\documentstyle{article}

\input epsf

\begin{document}

\title{Wetting and Capillary Condensation in Symmetric Polymer Blends:\\
A comparison between\\
Monte Carlo Simulations and Self-Consistent Field Calculations} 

\author{
M.\ M\"{u}ller and K.\ Binder
\\
{\small Institut f{\"u}r Physik, WA 331, Johannes Gutenberg Universit{\"a}t}
\\
{\small D-55099 Mainz, Germany}
}
\date{\today}
\maketitle

\begin{abstract}
We present a quantitative comparison between extensive Monte Carlo simulations and self-consistent field calculations
on the phase diagram and wetting behavior of a symmetric, binary (AB) polymer blend confined into a film. The flat walls 
attract one component via a short range interaction. The repulsion between monomers of different types leads to an upper
critical solution point in the bulk. The critical point of the confined blend is shifted to lower temperatures and 
higher concentrations of the component with the lower surface free energy. The binodals close the the critical point are 
flattened compared to the bulk and exhibit a convex curvature at intermediate temperatures -- a signature of the wetting 
transition in the semi-infinite system. We present detailed profiles of the two coexisting phases in the film and estimate 
the line tension between the laterally coexisting phases. Using the dependence of the thickness of the wetting layers and 
the shift of the chemical potential on the film width, we determine the effective interaction range between the wall and 
the $AB$ interface. Investigating the spectrum of capillary fluctuation of the interface bound to the wall, we find 
evidence for a position dependence of the interfacial tension. This goes along with a distortion of the interfacial
profile from its bulk shape. Using an extended ensemble in which the monomer-wall interaction is a stochastic variable, 
we accurately measure the difference between the surface energies of the components, and determine the location of the 
wetting transition via the Young equation. The Flory-Huggins parameter at which the strong first order wetting transition 
occurs is independent of chain length and grows quadratically with the integrated wall-monomer interaction strength. We 
estimate the location of the prewetting line. The prewetting manifests itself in a triple point in the phase diagram of 
very thick films and causes  spinodal dewetting of ultrathin layers slightly above the wetting transition. We investigate 
the early stage of dewetting via dynamic Monte Carlo simulations.  We compare our findings to phenomenological descriptions
and recent experiments.
\end{abstract}

\section{ Introduction }

The behavior of confined complex fluids is of practical importance for 
various applications (e.g.\ adhesives, coatings, lubricants, zeolites).
The confining surfaces give rise to packing effects and alter the conformations
of macromolecules in its vicinity. In general one component
(say A) may absorb preferentially at the surface, such that the wall is
coated with a layer of the component with the lower surface free energy. The structural and thermodynamic 
properties of these wetting layers are of practical importance and of fundamental interest in the statistical
mechanics of condensed matter.
At phase coexistence of the binary mixture, the surface free energy in the semi-infinite system undergoes a 
transition, at which the thickness of the adsorbed $A$-rich layer diverges.
This wetting transition may be continuous (second order wetting) or the thickness jumps from 
a finite value to infinity at the wetting transition temperature. The order of this transition and the 
temperature at which it occurs have attracted longstanding interest. \cite{CAHN,BREV,BINDERWET,SCHICK,LIP,PARRY3,PG}
The walls are wetted by the $A$ component, if the difference 
$\Delta \sigma_{\rm w}$ between the surface free energy of the wall with respect to 
a B-rich bulk $\sigma_{WB}$ and the surface free energy against an A-rich bulk $\sigma_{WA}$
exceeds the free energy cost of an $AB$ interface $\sigma_{AB}$ 
at an infinite distance from the wall.\cite{SCHICK,YOUNG}
\begin{equation}
\Delta \sigma_{\rm w} =  \sigma_{WB} - \sigma_{WA} > \sigma_{AB}  \qquad \mbox{(Young equation)}
\label{eqn:young}
\end{equation}
The spreading parameter $\Delta \sigma_{\rm w}-\sigma_{AB}$ controls the static and dynamic wetting 
behavior and is also accessible experimentally.\cite{BROCH} In structurally symmetric blends, the
surface free energy difference $\Delta \sigma_w$ is dominated by the different enthalpic interactions
of the monomers with the wall and thus $\Delta \sigma_w$ is largely independent of the molecular weight.
In the strong segregation limit, the interfacial tension $\sigma_{AB}$ is also independent of chain length, and
hence the wetting temperature is to leading order chain length independent. This is in marked contrast
to the critical temperature of the mixture, which increases linearly with molecular weight. Thus, wetting in polymeric
systems occurs far below the critical point,\cite{SCHMIDT} unlike the generic situation in mixtures of small 
molecules.

If the mixture is confined into a pore or a thin film, this wetting transition 
is rounded. Also the unmixing temperature is reduced and the coexistence curve
in the vicinity of the critical point is flattened compared to the bulk behavior.
Moreover, the preferential interactions at the surfaces shift the coexistence 
pressure or chemical potential away from its bulk coexistence value.\cite{NAKANISHI} At coexistence, 
the confined system phase separates {\em laterally} into A-rich domains which coexist with
regions in which there are A-rich layers at the surfaces and the
B-component prevails in the center of the film. The two phases are separated by interfaces,
which run perpendicular to the surfaces.
There is a delicate interplay between the wetting behavior of the semi-infinite system
and the phase behavior in a thin film.\cite{BREV,NAKANISHI}
In the temperature range between the critical temperature of the film
and the wetting temperature, the thickness of the wetting layer at
coexistence is determined by balancing the repulsion between the wall
and the AB-interface, which favors a thick wetting layer, against the
shift of the coexistence chemical potential, which suppresses the total
amount of A component in the film.\cite{PARRY}

A binary polymer blend between walls is a
suitable testing bed for these phenomenological ideas because the chain
length $N$ constitutes an additional control parameter which can be varied without changing enthalpic interactions.
Increasing the chain length $N$, we reduce bulk composition fluctuations, which are
neglected in most phenomenological approaches, and there are powerful
self-consistent field (SCF) techniques to describe the bulk and surface
behavior. In addition, the larger length scales of the occurring phenomena due to
the large size of the polymer coils facilitates applications of several experimental techniques.
Consequently, the behavior of polymer blends in 
thin films has attracted abiding theoretical,\cite{SCHMIDT,PINCUS,CN,FLEBBE,JONES}
experimental\cite{SCHEFFOLD,BUDKOWSKI,HARIHARAN,GENZER,ZHAO}
and simulational\cite{WANG,KUMAR,YR1} interest. We briefly summarize some of the findings 
pertinent to the present paper below.

Nakanishi and Pincus\cite{PINCUS} and Schmidt and Binder\cite{SCHMIDT} have explored the 
wetting behavior at a wall of a
binary polymer blend in the framework of a Cahn-Hilliard mean field
theory. Employing a quadratic form of the surface free energy, they
found first order wetting at low temperatures and second order
transitions close to the critical point. Flebbe et al.\ \cite{FLEBBE} studied the
confined system and revealed that pronounced differences between the
wetting behavior of the infinite system and the capillary condensation
in a film of thickness $D$ persist up to thicknesses which exceed the
radius of gyration $R_g$ by roughly two orders of magnitude. However,
both studies employed a square gradient (SG) approximation which is only
adequate for describing composition variations on length scales larger
than the coil extension -- a situation which occurs close to the critical point.
Moreover, estimating the parameters of the phenomenological surface free energy 
in the framework of a microscopic model is not straightforward. 
Self-consistent field techniques \cite{GENZER,NOOLANDI,COMPOSTO} overcome
the limitations of the square gradient approximation and more  microscopic 
treatments of the surface free energy contribution\cite{SCHMID2,NATH,FREED,MUTH,JERRY}
have also been explored.

The detailed composition profiles at surfaces of polymer blends are experimentally 
accessible via a variety of techniques. Investigating structurally symmetric
pairs of homopolymers via neutron reflectometry, Genzer et al.\ \cite{GENZER}
have compared the experimental results to the square gradient theory \cite{SCHMIDT}
and self-consistent field calculations. They found qualitative agreement between
experiments and theories. However, deviations from the quadratic 
dependence of the excess surface free energy on the surface composition are found. Similar deviations
were reported by Scheffold et al.\cite{SCHEFFOLD} and Budkowski et al.\cite{BUDKOWSKI} using nuclear reaction analysis
on random copolymers
with different microstructures. In these experiments the wetting transition occurs presumably
far below the critical point.

The reduction of the critical temperature and the crossover from  3D to
2D critical behavior have been studied in Monte Carlo (MC) simulations by
Kumar and co-workers\cite{KUMAR} and Rouault et al.\ \cite{YR1,YR2}
These simulations have also been compared to mean field calculations.\cite{SCHMID2,KUMAR2}
However, the simulations were restricted to symmetric blends and ``neutral'' walls (i.e., no
preferential interaction between the walls and any component). In this
limit, the coexistence chemical potential is not shifted
away from its trivial bulk value $\Delta \mu_{\rm coex} = 0$. Simulations of Wang and co-workers
\cite{WANG} extensively investigated the wetting behavior of a symmetric binary polymer blend at a 
hard wall which favors one component. Their simulation studies are closely related to the present
investigation, but they did not systematically explore the effects of finite film thickness.

In this work, we explore the combined effect of confined geometry
and preferential attraction of one component at the walls (``capillary
condensation'') and investigate the interplay between the wetting behavior 
and the phase diagram in the confined geometry. We restrict ourselves to 
perfectly flat walls which both attract the same component via a short 
range potential. We obtain profiles across the film and of the interface 
between the laterally segregated phases. Employing finite size scaling 
and reweighting techniques we are able to measure the  coexistence chemical potential
as a function of the wall separation $D$ and investigate corrections to the
Kelvin equation. We determine the phase diagram over a wide temperature range.
Using a novel Monte Carlo scheme, we measure the surface
free energy difference $\Delta \sigma_w$ and determine the wetting transition
via the Young equation (\ref{eqn:young}). We derive an approximate analytical 
expression of the wetting temperature in the strong segregation limit as a function 
of the interactions between the monomers and the wall and test this against our Monte Carlo 
simulation. At low temperatures the wetting transition is of first order and we
estimate the location of the concomitant prewetting transition. We suggest
an interpretation of recent experimental studies of spinodal
dewetting of ultrathin polymer films\cite{ZHAO} by Zhao and co-workers in terms
of prewetting--like phase separation.
This is illustrated via dynamic
Monte Carlo simulations of ultrathin films. We measure the effective interaction 
range between the wall and the $AB$ interface and investigate the capillary fluctuation
spectrum\cite{MW,M3} of the bound interface. We find evidence for a position dependence of the
interfacial tension $\sigma$. We compare our results with phenomenological approaches 
and detailed self-consistent field calculations. The latter take due account of the chain
conformations via a partial enumeration scheme\cite{MW,SZLEIFER,M2} and incorporate 
details of the interactions at the surface. We find qualitative (and sometimes even quantitative)
agreement between the simulations and the self-consistent field calculations without any 
adjustable parameter.

Our paper is arranged as follows: First, we give a phenomenological
description of capillary condensation\cite{PARRY} and discuss its limitations. Then
we introduce our coarse grained polymer model, give a brief synopsis of
the simulation techniques, and detail the salient features of our
self-consistent field scheme. In the main section we present a comparison between
the results of our Monte Carlo simulations and self-consistent field calculations. 
We close with a brief discussion and an outlook on future work.

\section{ Background }

We consider a binary polymer mixture confined into a thin film of
lateral extension $L \times L$ and wall separation $D$. Let $\Phi$ denote the
monomer volume fraction. Both types of structurally symmetric polymers consist of $N$
segments. The A-component of the blend is favored by both walls. The
monomer-wall interactions are taken to be short ranged and the parallel walls
are ideally flat.  Phase coexistence comprises two laterally segregated
phases. One of which is A-rich and exhibits only a minor compositional
variation across the film. Following Parry and Evans,\cite{PARRY} we
approximate its excess semi-grandcanonical potential $G$ per unit area with
respect to the A-rich branch of the bulk coexistence curve by:

\begin{equation}
\frac{\Delta G_A}{L^2} = 2 \sigma_{WA} - \frac{\Phi }{N}D\left( \mu_A\langle\rho\rangle +\mu_B(1-\langle\rho\rangle) -\frac{\mu_A+\mu_B}{2}\right)
		       = 2 \sigma_{WA} - \frac{\Phi }{2N}D \Delta \mu(2\langle\rho\rangle-1)
\end{equation}
where $\sigma_{WA}$ denotes the surface free energy of the wall,
$\mu_A$, $\mu_B$ the chemical potentials of A and B-polymers, $\Delta
\mu=\mu_A-\mu_B$ their chemical potential difference, and $\langle \rho \rangle$ the
composition of the A-rich phase at bulk coexistence.  In the second
phase the density of the  B-component comes up to its bulk value at the
center of the film, and this B-rich center region is separated from the
wall by A-rich layers of width $l$. Provided that the layer thickness $l$
is larger than the microscopic length scale but much smaller than the wall
separation $D$, the surface free energy of the wall $\sigma_{\rm WB}$ can be decomposed
into the surface free energy $\sigma_{\rm WA}$, the free energy of an AB interface $\sigma_{AB}$,
and the interaction $g(l)$ of this AB interface with the wall  (complete wetting).
Thus the excess potential takes the form:

\begin{equation}
\frac{\Delta G_B}{L^2} 
                       = \underbrace{2 \sigma_{WA} + 2 \sigma_{AB} + 2g(l) }_{2 \sigma_{WB}}
			    - \frac{\Phi}{N}   2l \Delta \mu (2\langle\rho\rangle-1)
			    + \frac{\Phi}{2N}   D \Delta \mu (2\langle\rho\rangle-1) 
\end{equation}
Above the wetting temperature, the wall repels the AB interface and we
take the short range interaction to be: $g(l)=A\exp(-\lambda l)$.  
Here $\lambda$ is a decay length which is of the same order as the correlation 
length of concentration fluctuations at bulk coexistence, as we shall discuss later.
The thickness of the wetting layers $l$ is determined by the condition
$\partial \Delta G_B/\partial l = 0$, which yields:

\begin{equation}
l(D) = \frac{1}{\lambda} \ln \frac{NA\lambda}{-\Phi(2\langle\rho\rangle-1)\Delta\mu}
\label{eqn:l}
\end{equation}
At phase coexistence the chemical potential difference is shifted such
that the two phases have the same semi-grandcanonical potential:
\begin{equation}
\frac{1}{\Delta \mu_{\rm coex}} = 
- \frac{\Phi(2\langle \rho \rangle-1)}{2N\sigma_{AB}}\left(D - 2l - \frac{2}{\lambda}\right)
\qquad \mbox{(Kelvin equation)}
\label{eqn:mcoex}
\end{equation}
To leading order the shift of the chemical potential $\Delta \mu$ from
the bulk coexistence value ($\Delta \mu=0$) is inversely proportional
to the wall separation $D$ and the coefficient involves the $AB$
interfacial tension and the bulk composition. This expression rigorously describes the $D \to \infty$ limit.\cite{PARRY3}
Thus the inverse wall separation $1/D$ plays a similar role as the chemical potential (or the bulk composition) 
in complete wetting. The thickness $l$ of the
wetting grows like $\ln D$, and the leading correction to the Kelvin
equation  $-\Delta \mu \sim 1/D$ are of relative order $\ln
D/D$.\cite{COMMENT1}

Though this phenomenological description is expected to capture the qualitative
behavior for very large wall separations and low temperatures, there are some
limitations which might prevent a quantitative agreement with our
Monte Carlo simulations.  For very small wall separations $D < O(R_g)$ ($R_g$: radius of gyration)
of the film, neutron reflexion experiments reveal that the two $AB$ interfaces interact with each other;\cite{HARIHARAN}
the coupling of the two interfaces across the film has been neglected in the treatment above.
More important, the detailed polymer conformations at the surface have been ignored. Polymers
orient and deform at the hard walls.\cite{YR1,BITSANIS} Since the length scale of these conformational changes is 
set by the radius of gyration $R_g$, one expects a distortion of
the intrinsic interfacial profile and a concomitant modification of the
effective interfacial tension for $l \sim \ln D < O(R_g)$. Furthermore the repulsion of the 
interface by the wall involves many length scales (e.g.\ range of the
wall-monomer interaction $d$, the width of the interface $w$, which controls the
composition profile at the center of the interface, the bulk correlation length $\xi$, which
sets the length scale in the wings of the interfacial profile.\cite{FRISCH})
Thus a simple single exponential decaying interaction is only an effective 
description. However, these effects can be described duely in a self-consistent field 
framework.

Moreover, the mean field treatment neglects fluctuations.  
In the vicinity of the critical point, one expects a critical behavior
belonging to the 2D Ising universality class.\cite{YR2}
Additionally, long wavelength fluctuations of the local
position of the interface (``capillary waves''\cite{AW}) from its most
probable value have been ignored. They can be described by a capillary fluctuation
Hamiltonian:\cite{HELFRICH}
\begin{equation}
\frac{{\cal H}_{\rm eff}}{k_BT} = \int dx\;dy\; \left\{  \frac{\sigma}{2} (\nabla \delta l(x,y))^2 + 
				  \frac{1}{2} \frac{\partial^2g}{\partial l^2} \delta l^2 \right\}
\label{eqn:helfrich}
\end{equation}
where $\sigma$ denotes the ``effective'' $AB$ interfacial tension in the
thin film geometry, and $\delta l(x,y)$ measures the deviation of the
local interfacial position from its average. Since we study a three dimensional system, which is at
the upper critical dimension of wetting phenomena, 
capillary fluctuations do not alter the functional dependence of the layer
thickness $l$ on the film size $D$.\cite{SCHICK} However, the prefactor $1/\lambda$
in eq.\ (\ref{eqn:l}) is multiplied by $(1+\omega/2)$,\cite{PARRY2} where  $\omega=
{k_BT}/({4\pi\xi^2\sigma_{AB}}) $ is the wetting parameter. 
Thus, the range of the wall-interface
interaction\cite{COMMENT2} is larger than in the self-consistent field
theory, and the wetting layers are thicker, respectively. 
The Fourier components $a_q$ of the local interfacial position $\delta l$ are Gaussian
distributed with width
\begin{equation}
\frac{2}{L^2\langle a_q^2\rangle} 
= \frac{\sigma}{k_BT} \left\{ q^2 + \frac{\partial^2g}{\sigma \partial l^2}\right\}
= \frac{\sigma}{k_BT} \left\{ q^2 + \left(\frac{2 \pi}{\xi_{\|}}\right)^2 \right\}
\label{eqn:fourier}
\end{equation}
Thus the spectrum\cite{MW,M3} yields information about the interfacial tension $\sigma$ and the 
parallel correlation length $\xi_{\|}$\cite{SCHICK}
\begin{equation}
\xi_{||} = 2\pi \sqrt{\sigma \left(\frac{\partial^2g}{\partial l^2}\right)^{-1}}
= \sqrt{\frac{4 \pi^2 \sigma}{A \lambda^2}} \exp\left( \frac{\lambda l}{2} \right) \sim \sqrt{D}
\sim \frac{1}{\sqrt{\Delta \mu}}   \qquad \mbox{(complete wetting)}
\label{eqn:xi}
\end{equation}
$\xi_{\|}$ acts as a long wavelength cut-off for the capillary fluctuation spectrum.\cite{SCHICK}
Additionally,  capillary fluctuations result in a broadening of the apparent interfacial
width which increases like $\ln(\xi_{\|}).$\cite{AW} Since  the parallel correlation length 
$\xi_{||} \sim \sqrt{D}$ growths with the wall separation $D$, 
we expect the squared interfacial width to depend logarithmically on the wall separation.
This behavior contrasts the behavior of mixtures confined into a thin film with asymmetric walls.
In this case a single interface occurs inside the thin film ($l \sim D$) and then a similar reasoning as above
yields $\ln (\xi_{||}) \sim D$.\cite{AW}

In the present study, we compare this phenomenological description to self-consistent field (SCF)
calculations and Monte Carlo (MC) simulations quantitatively. This allows us to examine the effects discussed
above and to obtain a detailed picture of the static structure and the thermodynamics of polymer blends in 
thin films.

\section{ Model and computational techniques }

\subsection{The bond fluctuation model and Monte Carlo (MC) technique}

Investigating the universal behavior of confined polymer blends, we
employ a coarse grained model that combines computational
tractability with the important qualitative features of real polymeric
materials: monomer excluded volume, monomer connectivity, and short
range interactions. We employ the bond fluctuation model.\cite{BFM}
Much is known about the phase behavior\cite{M3,M0} and interfacial properties
\cite{WANG,YR1,YR2,AW,MW,M3,M4,M5} of this polymer model and the
results have been compared to mean field calculations.\cite{SCHMID2,MW,SCHMID1}
Within the framework of this model, each monomer
blocks a whole unit cell of a 3D cubic lattice from further occupation.
Monomers along a chain are connected by bond vectors of length
$2,\sqrt{5},\sqrt{6},3$, and $\sqrt{10}$ lattice spacings. The blend comprises two
structurally symmetric polymer species - A and B - of the same chain
length $N=32$. This corresponds to roughly 150 repeat units in
chemically realistic models. At a monomer density $8\Phi=0.5$ the model
captures many features of a dense polymer melt. The size disparity between monomers
and (single site) vacancies results in a fluid-like structure with pronounced packing effects on
the monomer scale.\cite{M0} The structure of the monomer fluid is largely determined 
by the density and rather independent from the local composition or the 
temperature. Therefore we can lump the structure of the underlying bulk
fluid  into an effective coordination number {\tt z} or a Flory-Huggins 
parameter $\chi$, which is accessible in the simulations via the 
intermolecular paircorrelation function. We would like to emphasize that our
simulation techniques (e.g.\ extended ensemble technique to calculate
the spreading parameter or the analysis of the capillary fluctuation spectrum\cite{MW,M3})
and our self-consistent scheme can be readily applied to off-lattice models.

Thermal monomer-monomer interactions
are catered for by a short range square well potential extended over
54 neighboring lattice sites. This choice is motivated because it just includes 
all distances contributing to the first peak of the radial density correlation function.
The contact of monomers of the same type
lowers the energy by $\epsilon$, whereas contacts between unlike
species increase the energy by $\epsilon$. 
Similar to previous work by Wang et al.\ \cite{WANG}, we
consider a cuboidal system with a $L \times L \times D$ geometry. Periodic boundary
conditions are applied in $x$ and $y$ direction and there are 
impenetrable, flat walls at $z=-1$ and $z=D$. Every $A$ monomer, which is in the
d=2 layers adjacent to the walls reduces the
energy by $\epsilon_w$, whereas each B-monomer in this
wall interaction range increases the energy by the same amount. We keep
$\epsilon_w/k_BT=0.16$ fixed during our simulations, thus the monomer-wall
interaction is entropic in its character. This might be motivated by 
different packing behavior of the monomer species at the wall.
In the following all lengths are measured in units of the lattice spacing.

The polymer conformations are updated via a combination of local random 
monomer hopping and slithering snake-like moves.\cite{M0} The latter ones relax the
conformations roughly a factor $N$ faster than the local updates, and allow us to
investigate longer chain lengths than in previous work.\cite{WANG} We work
in the semi-grandcanonical ensemble, i.e.,\ at fixed temperature $k_BT/\epsilon$ and exchange
potential $\Delta \mu/k_B T$, and the composition is allowed to fluctuate. These
semi-grandcanonical moves consist for our symmetric blend of switching the
chain type $A \rightleftharpoons B.$\cite{M0,SARIBAN}

The ratios between the semi-grandcanonical Monte Carlo moves and the ones
which update the polymer conformations are adjusted to relax the composition of
the blend and single chain properties on the same time scale. We employ:
local hopping : slithering snake : semi-grandcanonical moves = 4:12:1

The coexistence curve in the confined geometry has been successfully
determined in mixtures of simple fluids via the peak in the order parameter
susceptibility or thermodynamic integration methods.\cite{BINDERWET} The former one is, however,
restricted to the vicinity of the critical point, whereas the latter one involves
the definition of a reference state in our polymer model.\cite{YRCOMMENT} In the present study,
we employ the semi-grandcanonical moves in junction with a reweighting\cite{VALEAU,BERG}
scheme to encourage the system to explore configurations in which both phases coexist 
in the simulation cell and the system ``tunnels'' often between the two coexisting phases. 
We obtain the reweighting factors via histogram analysis\cite{HISTO} of the joint composition-energy
probability distribution of previous runs at higher temperatures; the procedure starts 
around the critical temperature $k_BT/\epsilon=69.3$. E.g.\ to obtain the preweighting 
factors at $k_BT/\epsilon=50$, we employ 5-9 simulations at intermediate temperatures.
More technical details pertinent to the BFM can be found in Ref.\ \cite{M3}.
This scheme permits an accurate location of the coexistence chemical potential and yields 
additional information about the free energy as a function of the composition of the system,
the interfacial tension between coexisting phases\cite{BERG}, interactions between 
interfaces\cite{M3} and the wetting behavior.

\subsection{ Self-consistent field (SCF) calculations }
We compare our Monte Carlo simulations to self-consistent field calculations\cite{NOOLANDI,SZLEIFER,HT,SHULL,FLEER1,FLEER2,SZREV1,SZREV2}
which incorporate details of the polymer architecture\cite{MW,SZLEIFER,M2,SZREV1,SZREV2} and surface interactions.
The partition function of the binary polymer blend\cite{HT} containing $n_A$ A-polymers 
and $n_B$ B-polymers can be written in the form:

\begin{equation}
{\cal Z} \sim \frac{1}{n_A!n_B!} 
	      \int \Pi_{\alpha=1}^{n_A} {\cal D}[r_\alpha] {\cal P}_A[r_\alpha]
	           \Pi_{\beta=1}^{n_B} {\cal D}[r_\beta] {\cal P}_B[r_\beta]
	      \exp \left( -\frac{\Phi}{k_BT} \int d^3r\; {\cal E}[\hat{\rho}_A,\hat{\rho}_B]\right)
	      \label{eq:z}
\end{equation}
where the functional integrals ${\cal D}[r]$ sum over all polymer
conformations and ${\cal P}[r]$ denotes the probability distribution
characterizing the isolated (i.e. not mutually interacting) chain conformations in the confined geometry.  
${\cal P}$ includes intramolecular interactions and the interaction with the wall, but
not the pairwise interactions among different polymers.
${\cal E}$ represents a segmental free energy due to intermolecular interactions
which is specified below. The dimensionless monomer density takes the form\cite{HT}:

\begin{equation}
\hat{\rho}_A(r) = \frac{1}{\Phi} \sum_{\alpha=1}^{n_A}\sum_{i_A=1}^{N_A} \delta(r-r_{\alpha,i_A})
\label{eqn:dsz}
\end{equation}
The second sum runs over all monomers in the A-polymer $\alpha$, and a similar 
expression holds for $\hat{\rho}_B(r)$.

The probability distribution ${\cal P}[r]$ of an isolated single chain
conformations between two hard walls is the product of the bare
probability distribution ${\cal P}_0[r]$, which characterizes the
noninteracting, single chain conformations in the bulk, and the
Boltzmann weight ${\cal P}_w[r]$ of the interaction with the walls. The
latter vanishes if one of the segments is located outside the interval
$[0,D-2]$\cite{COMCORN}. Otherwise it takes the form $\exp(\pm \epsilon_w n_w[r])$,
where $n_w[r]$ is the number of segments in the wall interaction range
$[0,d-1]$ and $[D-d-1,D-2]$. The $+$ or $-$ sign holds for A-polymers or
$B$-polymers, respectively.

The segment free energy ${\cal E}$ comprises two contributions:  a free
volume term arising from hard core interactions and an energetic
contribution from the repulsion of unlike species.  Since the total
density fluctuations in the melt are small, we approximate the free
volume part by a simple quadratic expression introduced by
Helfand\cite{HT}, involving the knowledge of the inverse compressibility $\zeta$. 
This quantity has been measured
in simulations of the athermal model; $\zeta=4.1$\cite{WM1}. The
repulsion between unlike monomer species is incorporated via a
Flory-Huggins parameter $\chi = 2 {\tt z} \epsilon/k_BT$\cite{M0}, where ${\tt
z}$ denotes the number of monomers of {\em other} chains in the
range of the square well potential of depth $\pm \epsilon$. In principle the
number of intermolecular contacts can be evaluated for every chain conformation\cite{SZCONF}
as to include the coupling between chain conformations and energy. For simplicity we average
the number of intermolecular interactions over all chain conformations and employ the 
effective coordination number which has been extracted from the
intermolecular paircorrelation function $g(r)$ in the simulations of the bulk
system. Moreover, neglecting the slight temperature dependence\cite{COMMENT3}, we employ the
value {\tt z}=2.65 throughout our calculations. This value yields remarkably good agreement 
between SCF calculations and Monte Carlo simulations in our previous work on interfacial 
properties\cite{MW,M4,SCHMID1}. 
Hence, the segmental interaction free energy is taken to be\cite{HT}
\begin{equation}
\frac{{\cal E}[\rho_A,\rho_B]}{k_BT} =   \frac{\zeta}{2} \left( \rho_A + \rho_B - 1\right)^2 
-\frac{1}{2} {\tt z} \epsilon (\rho_A - \rho_B)\left(1+ \frac{1}{2}l_0^2\partial^2_\perp\right)(\rho_A - \rho_B)
\label{eqn:energy}
\end{equation}
$\partial_\perp $ denotes the spatial derivative perpendicular to
the wall. The spatial range of the monomer-monomer interactions 
$l_0^2=\int dz\; g(r) z^2 /\int dz\; g(r) \approx 16/9$, where the integration is extended over 
the range of the square well potential, has been estimated for the bond fluctuation model by Schmid\cite{SCHMID2}.
This non-local energy density mimics the influence of the reduction of the number of pairwise intermolecular interactions
due to the presence of the wall. The treatment of this ``missing neighbor'' effect is clearly an approximative one, and we
expect it to be only of limited validity in the presence of the large concentration gradient at the wall.

Introducing auxiliary fields, we rewrite the many chain problem in
terms of independent chains in external, fluctuating fields $W_A$ and
$W_B$.

\begin{equation}
{\cal Z} \sim \int {\cal D}[W_A,W_B,\Phi_A,\Phi_B] \exp \left(-{\cal F}[W_A,W_B,\Phi_A,\Phi_B]/k_BT\right)
\end{equation}
where the free energy functional is defined by
\begin{eqnarray}
f=\frac{{\cal F}[W_A,W_B,\Phi_A,\Phi_B]}{\Phi k_BT V} &=& 
					    \frac{\bar{\rho}_A}{N_A} \ln \bar{\rho}_A
					 +  \frac{\bar{\rho}_B}{N_B} \ln \bar{\rho}_B
					 +  \frac{1}{V} \int d^3r \; {\cal E}(\Phi_A,\Phi_B)  \nonumber \\
		&&   	                 -  \frac{1}{V} \int d^3r \left\{  W_A\Phi_A + W_B\Phi_B \right\}
					 -  \frac{\bar{\rho}_A}{N_A} \ln q_A[W_A]
					 -  \frac{\bar{\rho}_B}{N_B} \ln q_B[W_B]
					 \label{eqn:F}
\end{eqnarray}
$V$ is the volume of the system, 
$\bar{\rho}_A = \frac{n_A N_A}{\Phi V} = 1 -\bar{\rho}_B$ denotes the
average A-monomer density, and $q_A[W_A]$ the single chain partition
function of an $A$ polymer in the external field $W_A$. Using the definition of the monomer
densities (cf.\ (\ref{eqn:dsz}), we can write the latter as an explicit function
of the location $r_i$ of the monomers along the $A$ polymer:
\begin{equation}
q_A[W_A] = \frac{1}{V} \int {\cal D}[r] {\cal P}_A[r] \exp \left(- \sum_{i=1}^{N_A} W_A(r_i) \right)
\end{equation}
The functional integral for the partition function cannot be solved,
therefore we approximate it by the saddle point of the integrand. The
values of the collective variables, which extremize eq.\ (\ref{eqn:F}) are 
denoted by lower case letters. They are determined by the equations:

\begin{eqnarray}
\frac{\delta {\cal F}}{\delta \rho_A} = 0 &\Rightarrow &
w_A = \frac{\delta}{\delta \rho_A} \int d^3r\; {\cal E}(\rho_A,\rho_B) =
\zeta (\rho_A + \rho_B -1) - {\tt z} \epsilon \left(1+\frac{1}{2}l_0^2\partial_\perp^2\right)(\rho_A - \rho_B) \\
\frac{\delta {\cal F}}{\delta w_A} = 0    &\Rightarrow &
\rho_A = \frac{\bar{\rho}_A V}{N_A q_A} \frac{\delta q_A}{\delta w_A} \label{eq:d}
\end{eqnarray}
and similar expressions for $w_B$ and $\rho_B$. The saddle point
integration approximates the original problem of mutually interacting
chains by one of a single chain in an external field, which is
determined, in turn, by the monomer density.  The coupling between composition 
and polymer conformations is retained; however, composition fluctuations and the coupling
between the individual polymer conformations and the effective coordination number {\tt z}
are ignored. In inhomogeneous situations this coupling gives rise to a position-dependence of {\tt z} \cite{M4}.

To determine the phase
diagram we calculate the chemical potential difference $\Delta \mu$ and
the semi-grandcanonical potential $G$ according:
\begin{equation}
\frac{\Delta \mu}{k_BT} = \frac{\partial {\cal F}}{k_BT \partial n_A} = N \frac{\partial f}{\partial \bar{\rho}_{A}}
\end{equation}
\begin{equation}
\frac{G(\Delta \mu)}{kT\Phi V} = f -\frac{\Delta \mu}{k_BT}
\frac{\bar{\rho}_{A}(\Delta \mu)}{N} \qquad \mbox {+linear terms in $\Delta \mu$}
\end{equation}
Note that we have specialized here to the case $N_A=N_B=N$ while eq.\ (\ref{eq:z})-(\ref{eq:d})
still hold for the general situation of chain length asymmetry.
At coexistence, the phases have equal semi-grandcanonical potential $G(\Delta \mu)$.

We evaluate the single chain partition function via a partial
enumeration scheme\cite{SZLEIFER,M2,MW,SZREV1,SZREV2}. Using MC simulations of
the pure melt, we generated 81,920 independent polymer conformations
at temperature $\epsilon=0.02k_BT$ according to the distribution ${\cal
P}_0$. Since the chain conformations are almost temperature independent, 
we use the same sample to extend the SCF calculations to different temperatures. 
Rotating and translating those original conformations, we obtain between 3932160 and 15726640
polymer conformations. (Note that only the perpendicular coordinates of the chains are employed.) 
The position of the first monomer 
is chosen randomly with a  uniform distribution inside the interval $[0,D-2]$ along the $z$-axis.
The polymer conformation is discarded (${\cal P}_w=0$) if any segment is located outside this interval. 
Otherwise, the number of segments within the two nearest layers to the walls is properly counted and the Boltzmann 
factor of the interaction with the wall yields the weight ${\cal P}_w$.
Note that the procedure incorporates the coupling between chain
extensions parallel and perpendicular to the walls; an effect which is
ignored in the Gaussian chain model. Moreover, it incorporates the 
chain architecture on all length scales without any adjustable parameter\cite{MW}.
Within this framework, the
A-monomer density (c.f.\ eq.\ \ref{eq:d}) is the statistical average of
independent A-polymers with distribution ${\cal P}_w$ in the external
field $w_A$:

\begin{equation}
\rho_A(r) = \bar{\rho}_A \frac{   \sum_{\alpha=1}^{C} {\cal P}_w[r]\frac{1}{N_A} \sum_{i=1}^{N_A} V \delta(r-r_{\alpha,i}) 
			       \exp \left( -\sum_{i=1}^{N_A} w_A(r_{\alpha,i})  \right)                       }
                           {   \sum_{\alpha=1}^{C} {\cal P}_w[r]
			       \exp \left( -\sum_{i=1}^{N_A} w_A(r_{\alpha,i})  \right)                       }
\end{equation}
Other single chain quantities (e.g.\ orientations, chain end densities)
are given by corresponding averages over independent chains in the
fields $w_A$ and $w_B$.

We expand the spatial dependence of the densities and fields in a Fourier
series\cite{MATSEN} $\{f_1(z)=1$ and $f_k(z)=\sqrt{2}\cos(2\pi k (z+1)/D)$ for $k=2,(D+1)/2\}$;
{\em e.g.}, $w_A(r) = \sum_{k=1}^{(D+1)/2} w_{A,k} f_k(z)$.
This allows only for solutions which are symmetric around the middle of the film. 
We assume the breaking the symmetry in the $z$ direction to be thermodynamically less 
favorable than lateral phase separation\cite{FLEBBE}. Certainly, other decomposition schemes
can be chosen\cite{MARK} which allow for asymmetric profiles. However this requires a larger
number of basis functions and, hence, increases the computational effort. 
Defining the contribution of an $A$ polymer conformation $\alpha$ 
to the Fourier component $k$ according to $\phi_k(\alpha) = \sum_{i=1}^{N_A} f_k(r_{\alpha,i})$
we rewrite the above equation in the form:
\begin{equation}
\rho_{A,k} = \bar{\rho}_A \frac{   \sum_{\alpha=1}^{C} {\cal P}_w[r] \phi_k(\alpha)/N_A
			       \exp \left( -\sum_{k=1}^{(D+1)/2} w_{A,k} \phi_k(\alpha)  \right)                       }
                           {   \sum_{\alpha=1}^{C} {\cal P}_w[r]
			       \exp \left( -\sum_{k=1}^{(D+1)/2} w_{A,k} \phi_k(\alpha)  \right)                       }
\end{equation}
For a fast evaluation of the above average we keep all $\phi_k(\alpha)$ ($k=1,\cdots,(D+1)/2$ and
$\alpha = 1,\cdots,C$) in the computer memory. This poses rather high memory demands (several Gbytes) 
and we employ a massively parallel  CRAY T3E computer, assigning a subset of conformations to each
processing element. We use up to 512 processors in parallel. The resulting set of $D-1$ non-linear 
equations is solved {\em via} a Newton-Raphson like method. Usually we achieve convergence within 4-7 
iterations.

\section{ Results}

\subsection{Phase coexistence in a thin film}

\begin{figure}[htbp]
    \begin{minipage}[t]{160mm}%
       \setlength{\epsfxsize}{9cm}
       \mbox{
       \epsffile{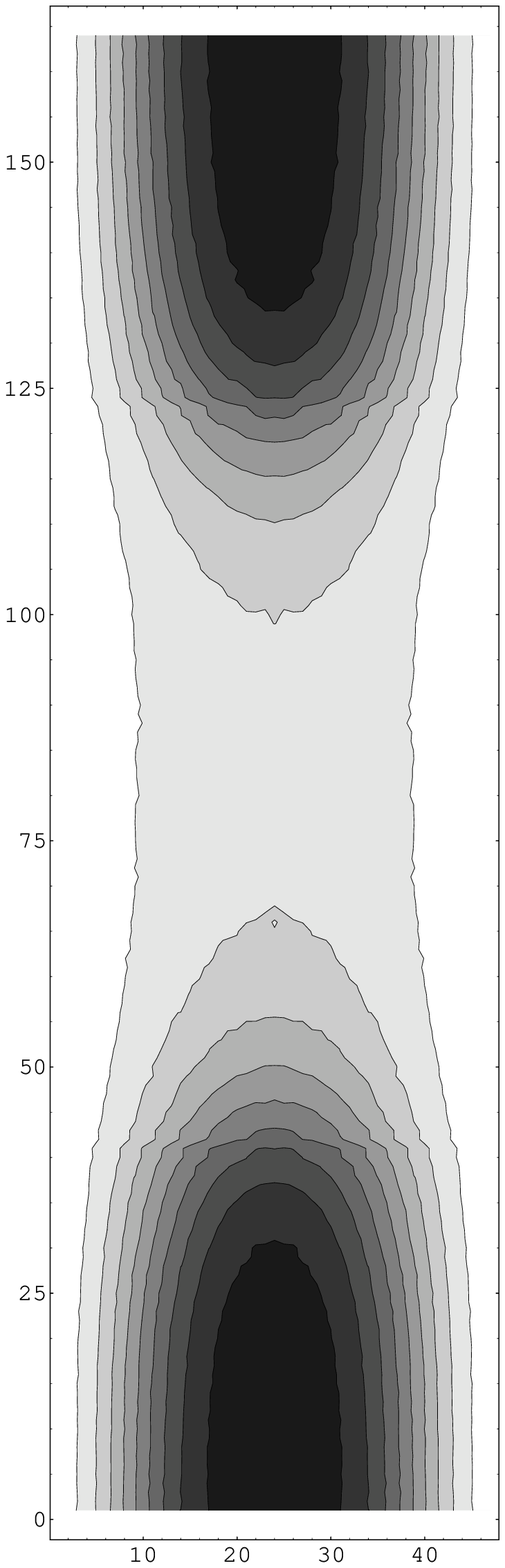}
       \hspace*{-1cm}
       \setlength{\epsfxsize}{9cm}
       \epsffile{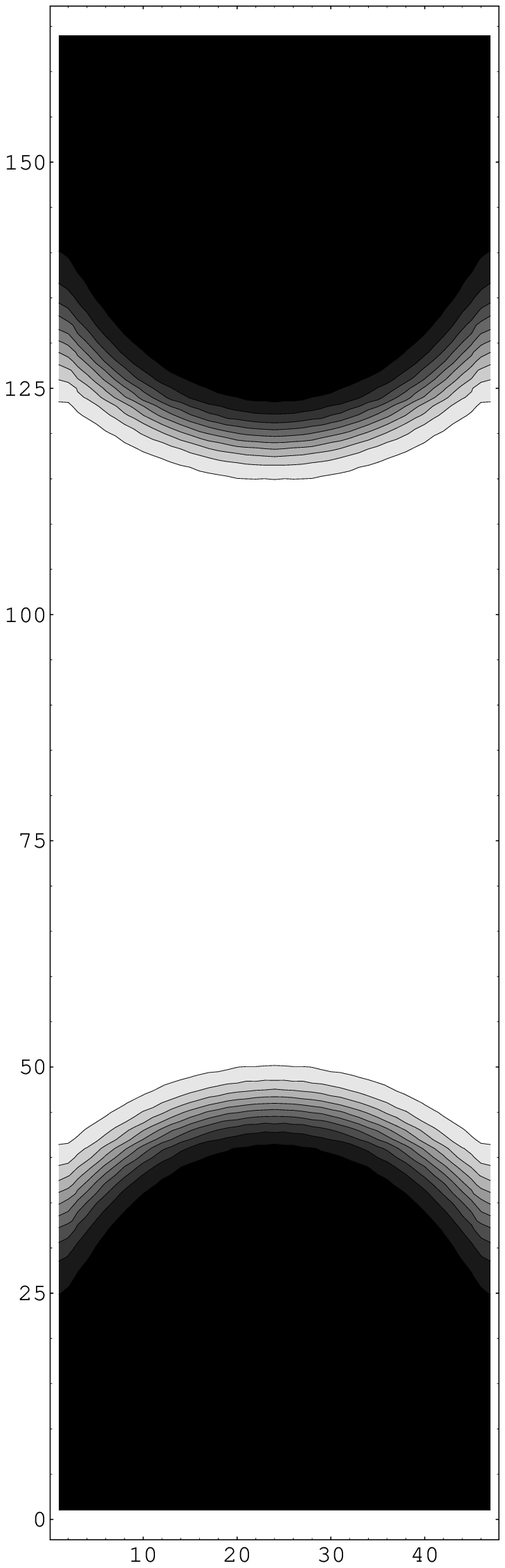}
       \vspace*{1cm}
       }
    \end{minipage}%
    \hfill%
    \begin{minipage}[b]{160mm}%
    \vspace*{2cm}
       \caption{
		2D composition profiles in a $164 \times 48 \times 48$ geometry. $B$ rich regions are shaded darker.
		The walls attract the $A$ component ($\epsilon_w/k_BT=0.16$). \newline
		({\bf a}) $\epsilon/k_BT=0.02$ (above the wetting temperature)  and
			  average A-monomer density $\langle \rho \rangle =0.67751$. \newline
		({\bf b}) $\epsilon/k_BT=0.08$ (below the wetting temperature)
			  and average A-monomer density $\langle \rho \rangle = 0.5$.
                }
       \label{fig:contour}
    \end{minipage}%
\end{figure}

We begin by exploring the qualitative features of phase coexistence and illustrating our simulation methodology.
The averaged 2D composition profiles of the laterally segregated, coexisting phases in a thin film of geometry
$L_x \times L_y \times D$ with $L_x=164$ and $L_y=D=48$
are presented in Figure \ref{fig:contour}. The simulations are performed at
constant composition, i.e.\ no semi-grandcanonical identity switches are employed. The grey scale indicates the 
composition; A-rich regions are lighter shaded than B-rich ones. The two interfaces are parallel to the
$yz$ plane and are free to move in $x$-direction. Only their distance is fixed by the overall composition.
Thus we average the profiles with respect to the instantaneous center of gravity of $B$ polymers in each 
configuration. The 2D profiles resemble qualitatively the results of 2D SCF calculations of Schlangen et al.\cite{LEERM};
however, our profiles are broadened by capillary waves\cite{AW}. The profiles at the two temperatures are qualitatively
different. At the higher temperature ({\bf a}) $\epsilon/k_BT=0.02$ the interface between the $A$-rich and $B$-rich
phases is quite broad and even in the $B$-rich phase, there is an $A$-rich layer at the wall. This corresponds
to the situation above the wetting transition. At the lower temperature ({\bf b}) $\epsilon/k_BT=0.08$, the profile is much 
sharper and there is no $A$-rich layer at the wall in the $B$-rich phase. Moreover, the interface between the
coexisting phases meets the wall at a finite angle. Thus the $A$ polymers do not wet the wall.
Unfortunately, the film width $D=48$ is not large enough to extract the contact angle $\Theta_e$ of 
a macroscopic droplet reliably, because the interface exhibits a pronounced curvature. Thus it is difficult 
to quantify a contact angle, and systems with opposing boundary fields might yield more reliable estimates. 
Using the independent determined interfacial tension $\sigma_{AB}=0.0436 k_BT$  and the difference in the 
surface free energy $\Delta \sigma_w=0.0397 k_BT$ (see below), we estimate the contact angle $\Theta_e$ for a thick film to:
$\cos \Theta_e = \Delta \sigma_{\rm w}/\sigma_{AB}\approx 0.91 =\cos 25^o$

\begin{figure}[htbp]
    \begin{minipage}[t]{160mm}%
       \setlength{\epsfxsize}{8cm}
       \mbox{\epsffile{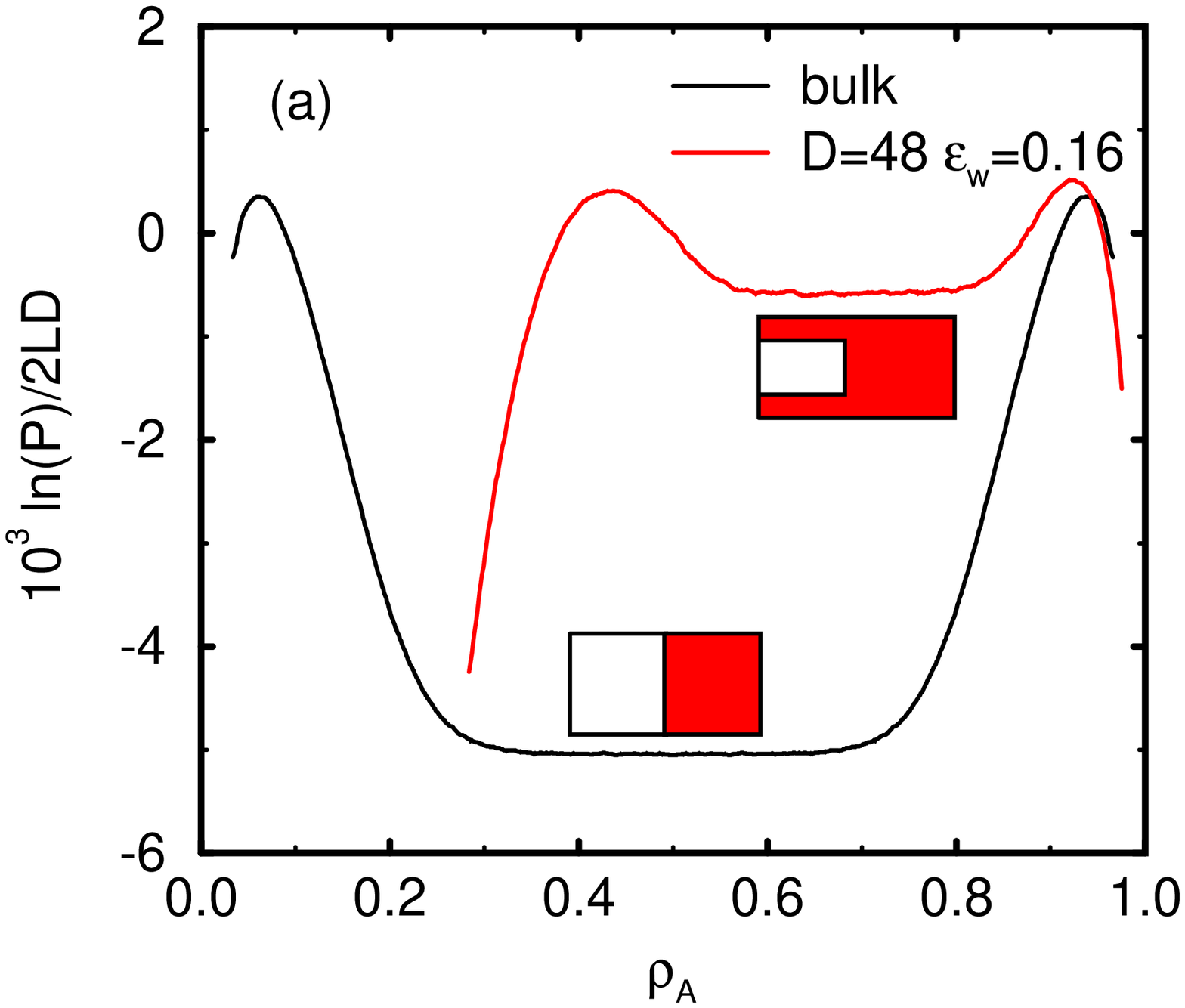}
       \setlength{\epsfxsize}{8cm}
       \hspace*{1cm}
       \epsffile{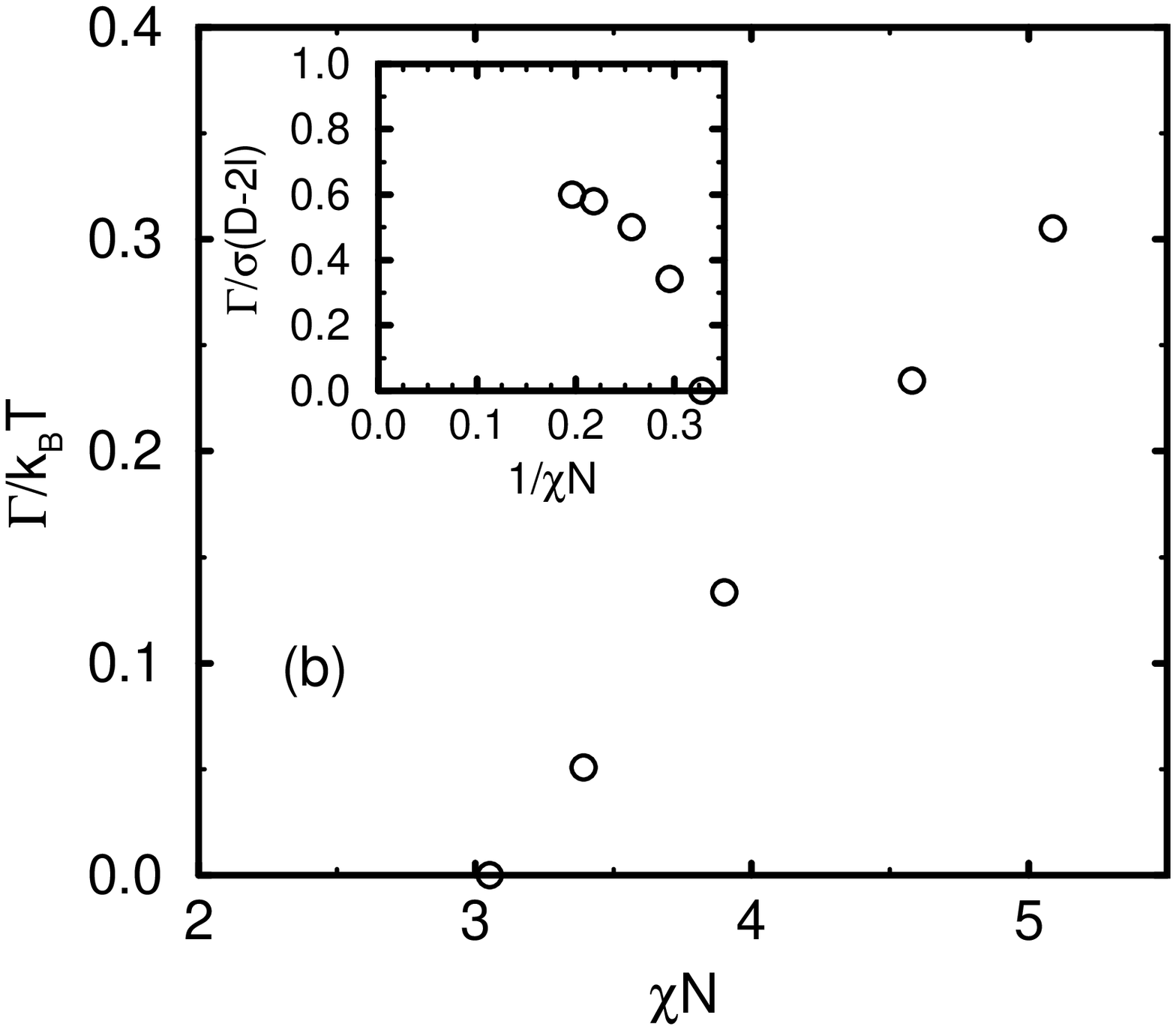}
       }
    \end{minipage}%
    \hfill%
    \begin{minipage}[b]{160mm}%
       \caption{({\bf a}) Probability distribution of the composition at $\epsilon/k_BT=0.02$ for the bulk system 
		($L_x=128 \times L=64 \times D=64$ and periodic boundary conditions in all directions)
		and the confined system ($L_x=164 \times L=48 \times D=48$, $\epsilon_{\rm w}/k_BT=0.16$).
		The peaks correspond to the coexisting phases and have equal probability weight.
		The typical configurations of the plateau between the two peaks contain two interfaces 
		parallel to the $yz$-plane (as sketched). The detailed composition profile of the
		confined blend is displayed in Fig.\ 1 ({\bf a}).\newline
		({\bf b}) Temperature dependence of the line tension $\Gamma$ between the coexisting phases 
		in a thin film $D=48$. The Flory--Huggins parameter is given by $\chi= {2 {\tt z} \epsilon}/{k_BT}$,
		with the intermolecular coordination number ${\tt z}=2.65$.\newline
		The inset shows the ratio between the line tension and the bulk interfacial tension as a function
		of the inverse temperature.
                }
       \label{fig:prob}
    \end{minipage}%
\end{figure}

Employing semi-grandcanonical identity changes in addition, we allow the overall composition of the system to fluctuate.
The reweighting technique\cite{VALEAU,BERG} permits us to explore a wide range of compositions $\rho$ and the probability
distribution $P(\rho)$ yields information about the free energy $F(\rho) = - k_BT \ln P(\rho)+\mbox{const}$.
The distributions in a thin film (in the same geometry and temperature as in Figure \ref{fig:contour}
({\bf a})) and in the bulk are displayed
in Figure \ref{fig:prob} ({\bf a}). The chemical potential difference $\Delta \mu$ has been adjusted to its coexistence 
value (cf.\ next section). The locations of the two peaks correspond to the composition of the
coexisting phases. The distribution of the bulk system is symmetric around $\rho=0.5$. In the thin film, however, 
the composition of the B-rich peak is shifted to higher values of $\rho$ due to the wetting layers at the walls.
Moreover, the $A$-poor peak is broader than the $A$-rich one.
The plateau in the probability distribution indicates that the two interfaces can change their distance (and thereby
alter the composition) at negligible free energy costs. Hence the interfaces do not interact, and we calculate the 
interfacial tension\cite{INTER} according to $\sigma/k_BT =  - \ln \left(P_{\rm max}/P_{\rm min}\right)/2L_yD$.
In the confined blends, $\sigma D$ rather corresponds to the line tension $\Gamma$ between the coexisting phases in the film.
The MC data in Figure \ref{fig:prob} ({\bf a}) yield $\sigma = 0.0054(1)k_BT$ for the bulk interfacial tension
and $\Gamma = 0.051(5) k_BT$ for the line tension in the film of width $D$. 
In a crude approximation, the line tension $\Gamma$ is proportional to the bulk interfacial tension $\sigma$, and the 
effective length of the $AB$ interface $D-2l$ across the film, where $l$ denotes the thickness of the $A$-rich
wetting layer in the $B$-rich phase. Of course, this is an upper bound to the free energy cost of the $AB$
interface in a film, because the system chooses rather a curved interface (cf.\ Figure \ref{fig:contour} ({\bf a}))
and even in the middle of the film there is no region where the interface is appropriately described by the
bulk behavior. Moreover, $\Gamma$ vanishes at the critical temperature of the thin film, whereas the
bulk interfacial tension becomes zero at the higher bulk critical temperature. Upon decreasing the temperature
the interfacial tension $\sigma$ and the line tension $\Gamma$ increase. Figure \ref{fig:prob} ({\bf b})
presents the temperature dependence of $\Gamma$ and the ratio $\Gamma/\sigma(D-2l)$. We estimate the thickness $l$
via  $l = \frac{\rho-(1-\langle\rho\rangle)}{2\langle\rho\rangle-1} \frac{D}{2}$, where $\rho$ denotes the composition 
of the $B$-rich phase and $\langle \rho \rangle$ the composition in the bulk. The ratio $\Gamma/\sigma(D-2l)$
is smaller than one and grows upon decreasing the temperature.

\subsection{The phase behavior of the bulk and the confined system}

To determine the phase behavior of the confined system in the semi-grandcanonical ensemble, we locate the 
coexistence curve in the two-dimensional parameter space of temperature $k_BT/\epsilon$ and chemical 
potential difference $\Delta \mu/k_BT$, employing various concepts of finite size scaling theory.
At coexistence, the two phases have the same
semi-grandcanonical potential $G$ and we employ the equal weight criterion\cite{BORGS} for
the probability distribution $P(\rho,\epsilon,\Delta\mu)$ of the density
at fixed temperature and chemical potential; i.e.,\ we adjust the chemical potential
such that\cite{M0}:
\begin{equation}
\int_0^{\rho^*}P(\rho,\epsilon,\Delta\mu) d\rho = 
\int_{\rho^*}^1 P(\rho,\epsilon,\Delta\mu) d\rho \qquad \mbox{with} \qquad
\rho^* = \int_0^1 \rho P(\rho,\epsilon,\Delta\mu) d\rho
\end{equation}

This definition is very accurate for locating the coexistence curve below the critical temperature
even for systems with moderate linear dimensions $L$.  However it entails corrections of the
order $L^{-(1-\alpha)/\nu}\sim 1/L$\cite{M5} in the value of the critical composition
due to field mixing effects, where we have used 2D Ising values for the critical 
exponents $\alpha$ of the specific heat and $\nu$ of the correlation length.
Remember that the ``field mixing'' arises from a coupling between order parameter and energy density, whose singular
part involves the critical exponent $1-\alpha$\cite{NIGEL}.
Below the critical point, we use the system size $96 \times 96 \times 48$ (for $\epsilon/k_BT\leq0.03$)
or $48 \times 48 \times 48$ (for $\epsilon/k_BT>0.03$) and $\epsilon_w/k_BT=0.16$. To locate the critical
point we use substantially larger lateral extensions $L \leq 264$.

\begin{figure}[htbp]
    \begin{minipage}[t]{160mm}%
       \setlength{\epsfxsize}{13cm}
       \mbox{\epsffile{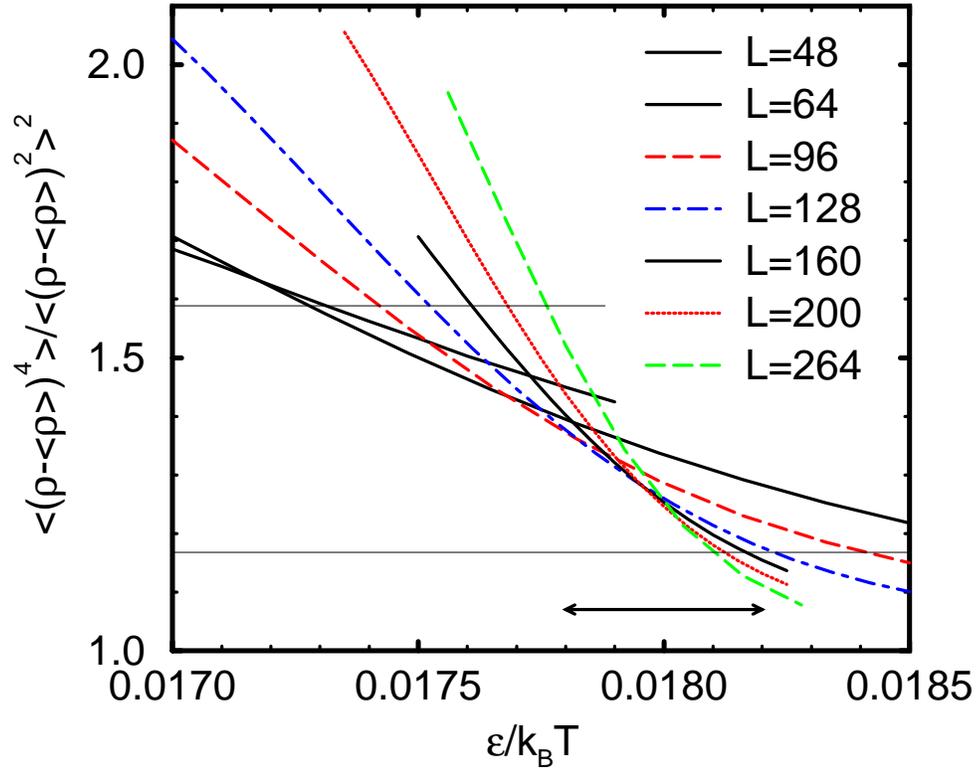}}
    \end{minipage}%
    \hfill%
    \begin{minipage}[b]{160mm}%
       \caption{
       Fourth order cumulant intersection along the coexistence curve to determine the critical temperature for 
       $D=48$ and $\epsilon_{\rm w}=0.16$. The error marks the uncertainty in the critical temperature 
       $\epsilon_c=0.0180(2)$. The horizontal lines mark the values of the cumulant of the 3D Ising (upper) and 2D Ising
       (lower) model.
                }
       \label{fig:kum2}
    \end{minipage}%
\end{figure}

Along the coexistence curve and its continuation that persists in finite-size systems,
we employ the cumulants of the composition probability distribution\cite{CUM} to locate 
the critical point. Finite size scaling theory implies that the moments of the order parameter
$\langle(\rho -\langle\rho\rangle)^l\rangle$ scale at criticality like $L^{-l\beta/\nu}$, where $\beta$ and $\nu$
are the critical exponents of the order parameter and the correlation length, respectively.
Therefore, the ratio $\langle(\rho -\langle\rho\rangle)^4\rangle/\langle(\rho -\langle\rho\rangle)^2\rangle^2$
becomes independent of the system size $L$ at criticality. Plotting this cumulant ratio vs.\ inverse temperature 
for different linear dimensions, one hence expects in the ideal case all curves to intersect in a common point,
which yields the critical temperature. Since this method involves
only even moments of the order parameter, it is rather insensitive to ``field mixing'' corrections\cite{NIGEL}.
Figure \ref{fig:kum2} displays the fourth order cumulant ratio of our simulations for wall separation
$D=48$. The cumulants for small lateral extension $L$ do not intersect at a common temperature, 
because the data fall into the crossover regime between three dimensional and two dimensional critical 
Ising-like behavior. The values of the cumulant for 3D and 2D Ising criticality\cite{BLOETE} are also shown in
the figure. Only, for aspect ratios $L/D>3$ a common intersection point gradually emerges
around $\epsilon_c/k_BT =0.0180(2)$, which is our estimate for the critical temperature in the film. 
The corresponding value of the critical composition is $\rho_{Ac}=0.686(4)$. Using the data for the three largest
lateral extensions $L=160,200,264$ and assuming 2D Ising critical behavior ($\beta=1/8$), we obtain
$\rho_\pm = \rho_c \pm 0.46(2) \left({(\epsilon-\epsilon_c)}/{k_BT}\right)^\beta$ for the binodal 
in the vicinity of the critical point. For a full understanding of the data in Figure \ref{fig:kum2}
combined analysis of crossover scaling and finite size scaling would be required, which is still a formidable
problem for the theory of critical phenomena in general.

\begin{figure}[htbp]
    \begin{minipage}[t]{160mm}%
       \setlength{\epsfxsize}{7cm}
       \mbox{\epsffile{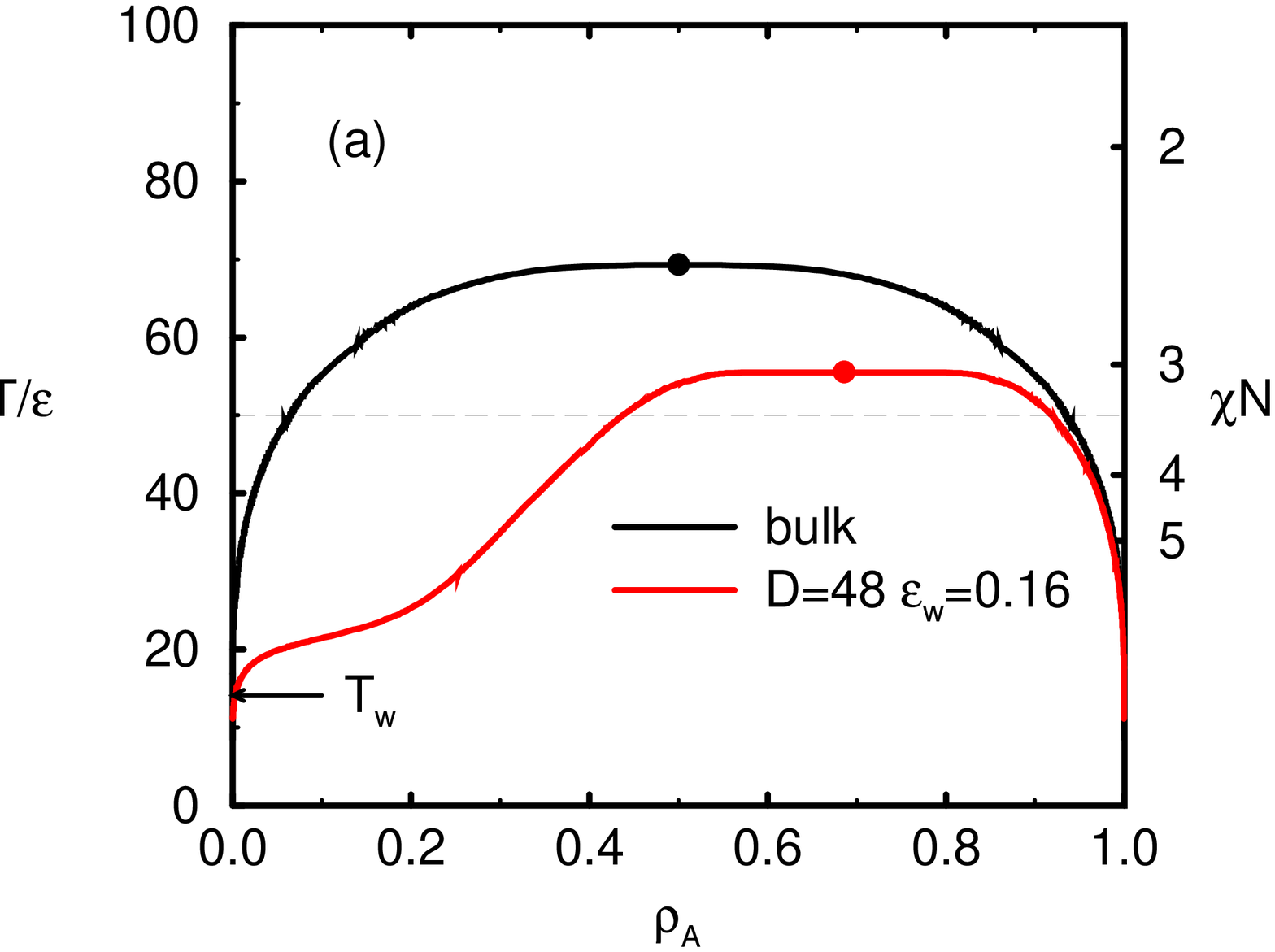}
       \setlength{\epsfxsize}{7cm}
       \hspace*{1cm}
       \epsffile{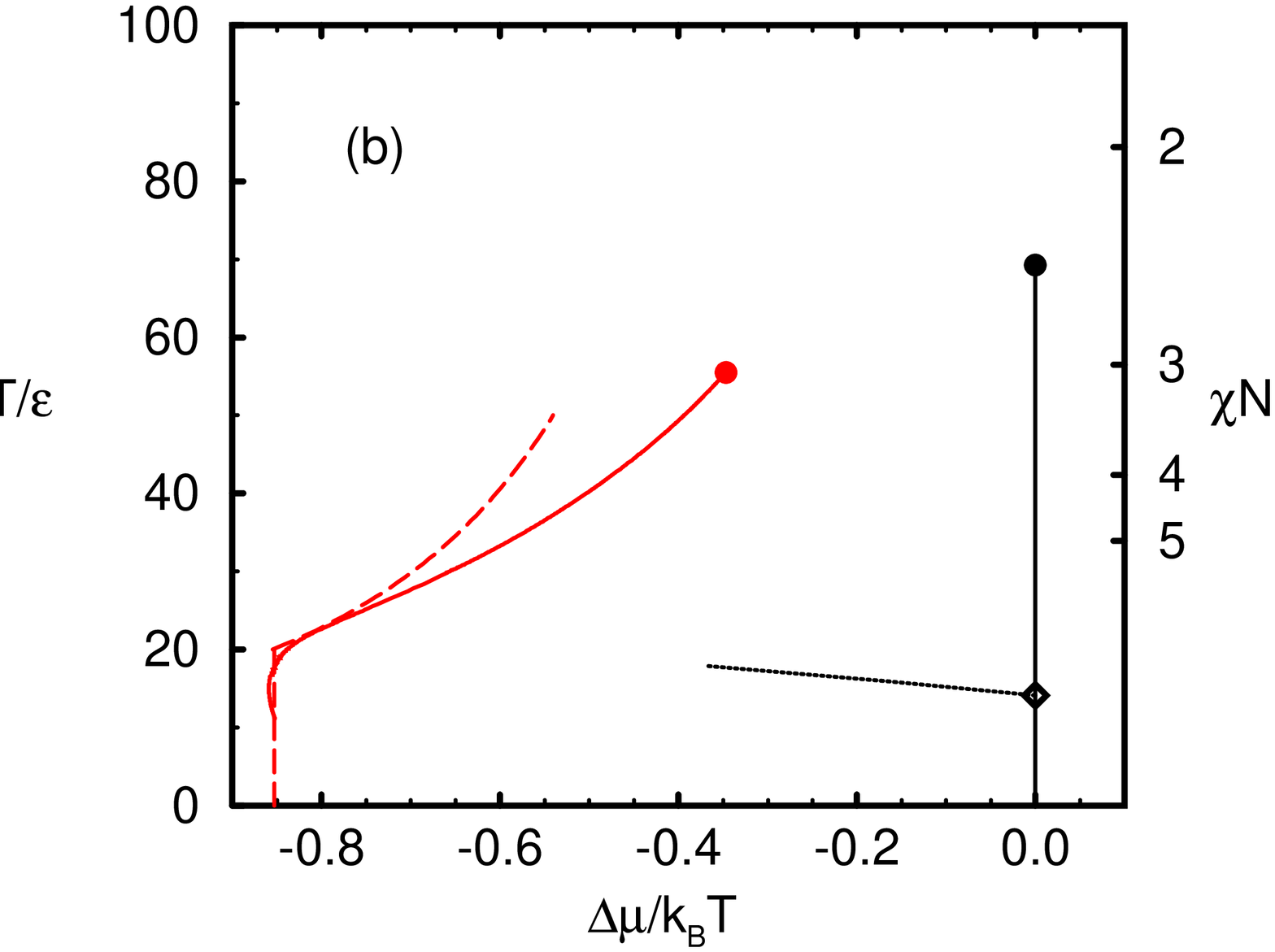}}\\
       \setlength{\epsfxsize}{7cm}
       \mbox{\epsffile{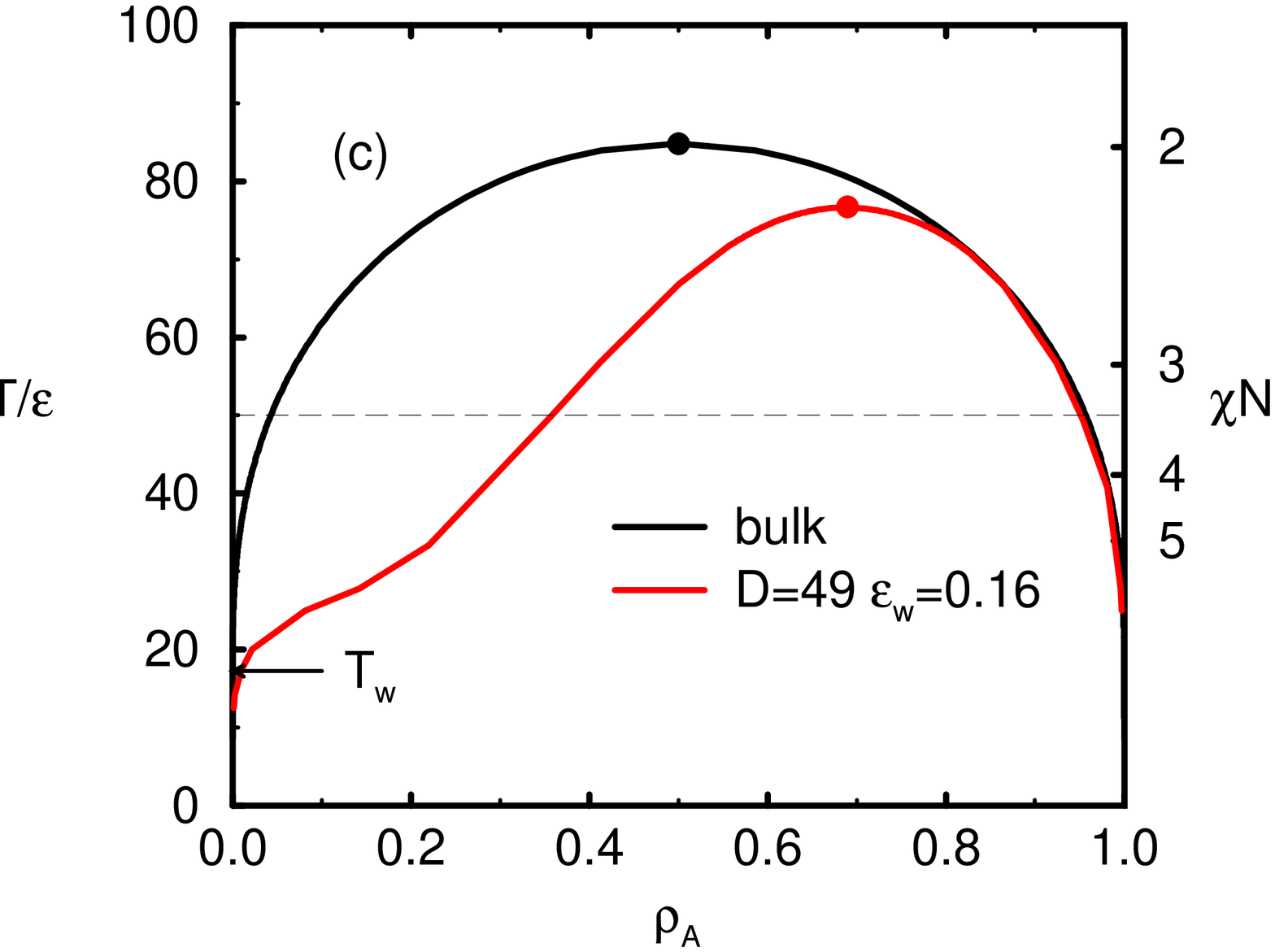}
       \hspace*{1cm}
       \setlength{\epsfxsize}{7cm}
       \epsffile{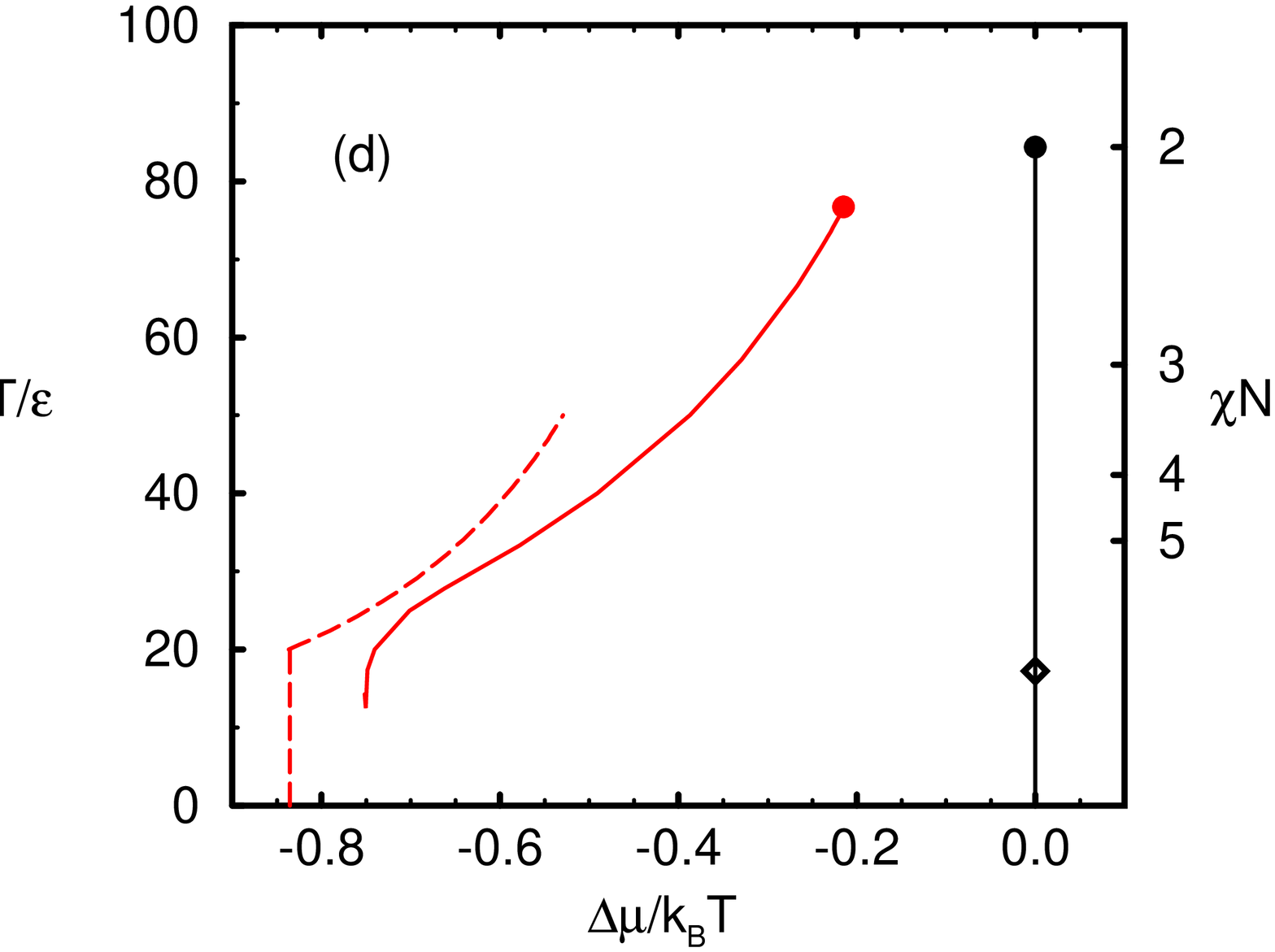}}\\
       \setlength{\epsfxsize}{7cm}
       \mbox{\epsffile{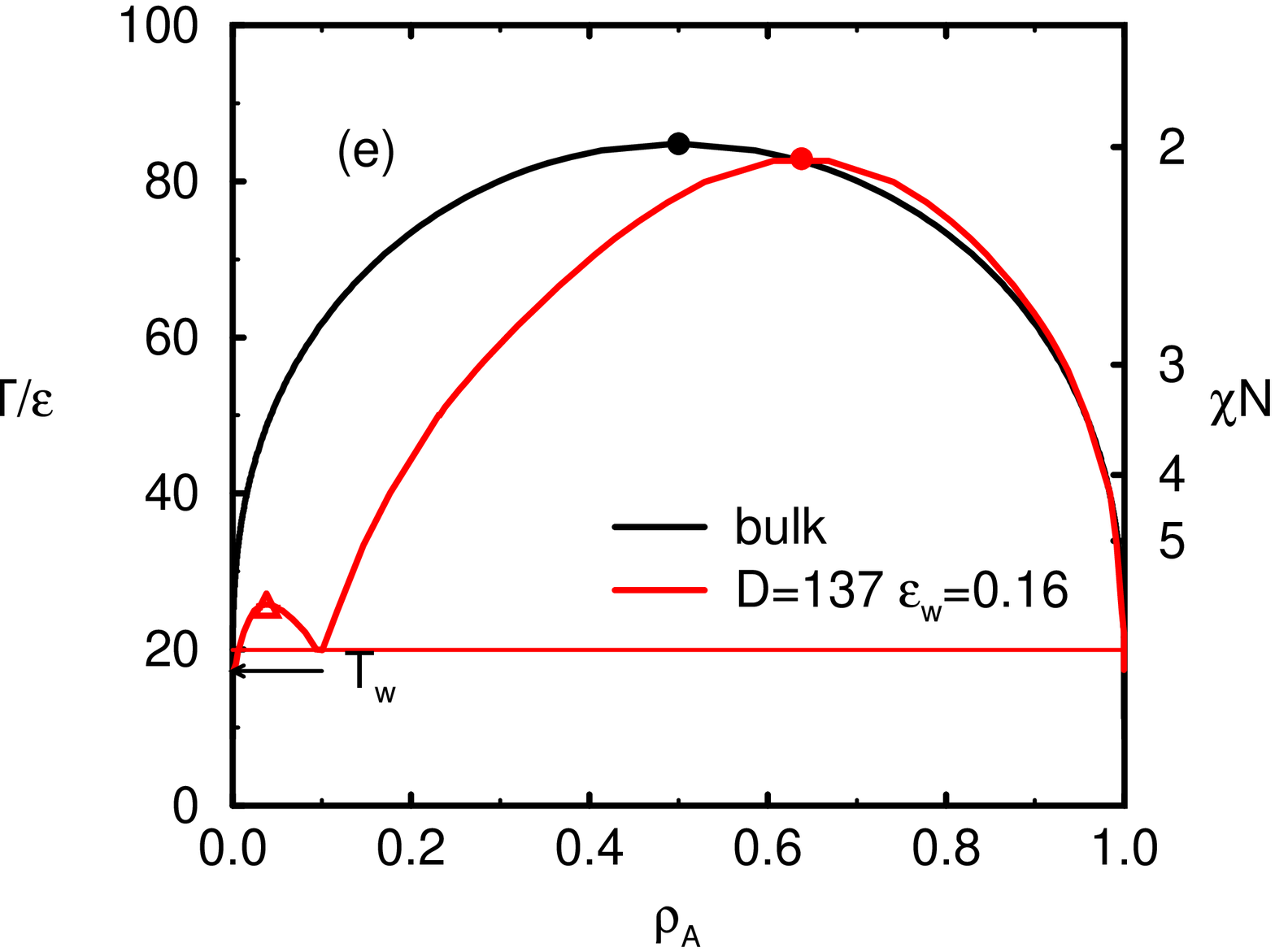}
       \hspace*{1cm}
       \setlength{\epsfxsize}{7cm}
       \epsffile{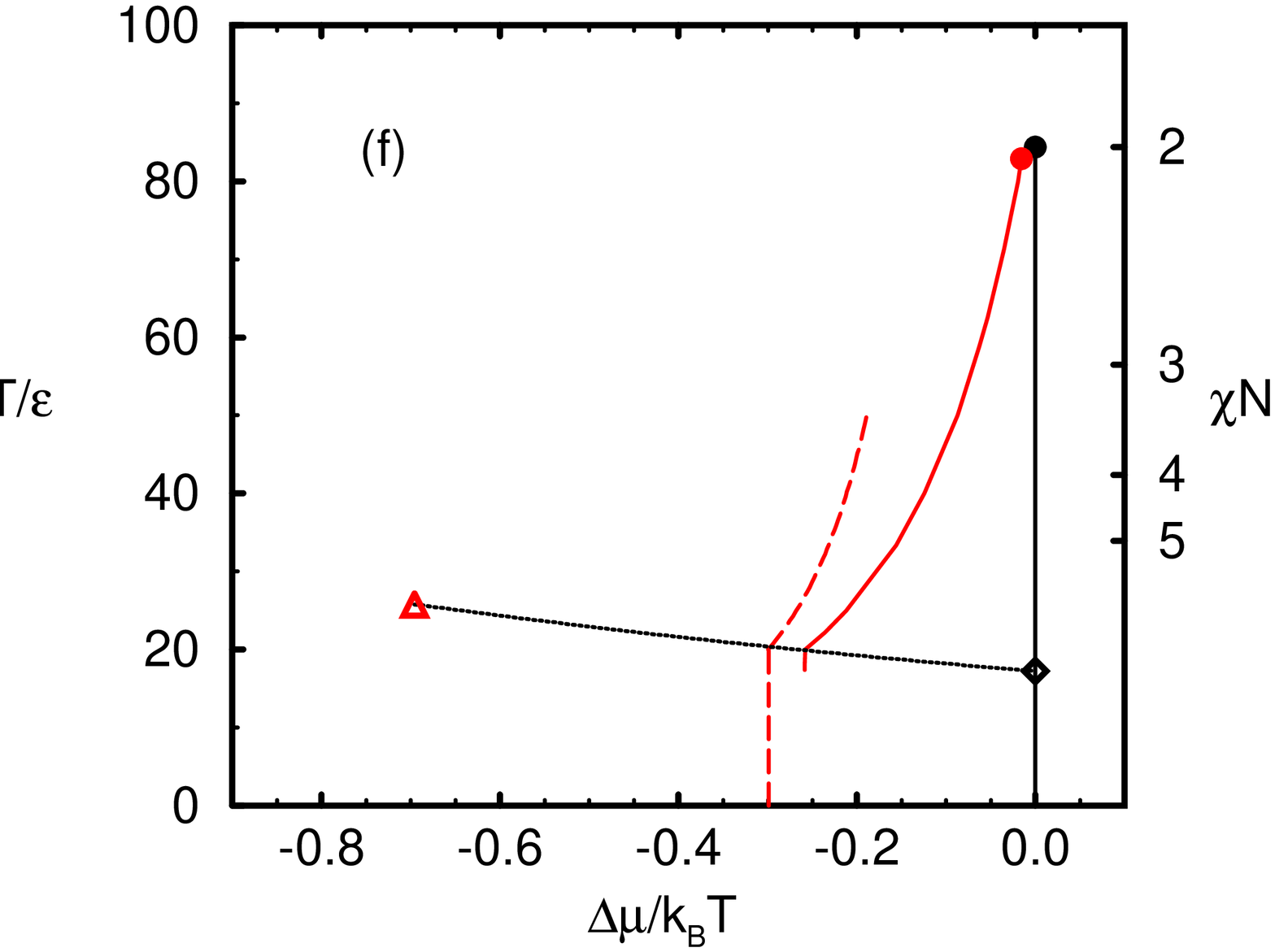}}
    \end{minipage}%
    \hfill%
    \begin{minipage}[b]{165mm}%
       \caption{ Comparison of the MC phase diagrams and the SCF results: \newline
       ({\bf a}) MC results for the phase diagram in the bulk and confined geometry 
		 ($D=48$ and $\epsilon_{ \rm w }/k_BT =0.16$). The critical points exhibit 3D
		 and 2D Ising critical behavior. The estimate for the wetting temperature 
		 is indicated by an arrow. The dashed line marks the temperature at which the 
		 thickness dependence is investigated.
		 \newline
       ({\bf b}) MC results for the coexistence chemical potential difference $\Delta \mu$. The dashed 
		 lines correspond to the low temperature estimates $-\Delta \mu D/k_BT = 2Nb({\chi/6})^{1/2} $
		 and  $-\Delta \mu D/k_BT = 4 d N\epsilon_w/k_BT$. The wetting transition in the semi-infinite
		 system is denoted by an diamond. The dotted line presents our estimate for the 
		 prewetting line (and its continuation above the prewetting critical temperature). \newline
       ({\bf c}) Phase diagram according to the SCF calculations (symbols as in ({\bf a})) for $D=49$.
		 \newline
       ({\bf d}) Phase diagram in the $T-\Delta\mu$ plane according to the SCF calculations 
		 (symbols as in ({\bf b})). \newline
       ({\bf e}) Phase diagram in the SCF calculations for a thick film $D=137$. Slightly above the wetting
		 temperature there is a triple point (horizontal solid line) at which a thin layer, a thick layer, 
		 and an $A$-rich phase coexist. The coexistence between layers of different thicknesses ends at a
		 tricritical point denoted by a triangle.\newline
       ({\bf f}) Phase diagram in the $T-\Delta\mu$ plane according to the SCF calculations for $D=137$
		 (symbols as in ({\bf b})) The diagram also includes the prewetting line (calculated from a film 
		 of width $D=137$).
                }
       \label{fig:phasen}
    \end{minipage}%
\end{figure}

The phase diagram of the bulk system and a film of width $D=48$ is presented in Figure \ref{fig:phasen} ({\bf a}).
The phase diagram of the bulk system is symmetric around $\rho=0.5$ and exhibits 3D Ising-like
critical behavior (i.e.\ $\beta = 0.325$). The binodal of the confined system near the critical point are flatter 
than in the bulk, 
indicating 2D Ising critical behavior\cite{YR1}. The critical temperature in the film 
($D=48$ and $\epsilon_w/k_BT=0.16$) is reduced by $20\%$ compared to the bulk. Note that this effect 
is much more pronounced than in the absence of preferential absorption of the A-component at the surfaces. 
Simulations of the symmetric system $D=48$ without preferential interactions $\epsilon_w=0$\cite{YR1} found 
a suppression of the critical temperature by only $4\%$.

There are two pronounced changes of curvature in the A-poor branch of the binodal $\rho_-(T)$.
The convex curvature around $k_BT/\epsilon\approx 20$ is the fingerprint of the wetting transition 
of the semi-infinite system. Around the wetting temperature, the wall-interface interaction $g(l)$ changes
from attractive (for $T<T_{\rm wet}$) to repulsive. Assuming that $A(T)$ grows upon increasing $T$ for $T>T_{\rm wet}$
and $\lambda \sim 1/\xi \sim 1/R_g$ is only weakly temperature dependent
we can combine eqs.\ (\ref{eqn:l}) and (\ref{eqn:mcoex}) to obtain for large $D$:
\begin{equation}
\rho_- \approx \frac{2l}{D} \approx \frac{2}{\lambda D} \ln \left( \frac{A(T)\lambda D}{2\sigma_{AB}}\right)
\end{equation}
which describes the convex portion qualitatively.
For temperatures far below the wetting temperature, the composition is given by: $\rho_- \sim \exp(-\chi N)$,
while we find critical behavior around $T_c$. In both temperature regimes the binodal is concave.
A similar shape of binodals is also observed in confined Lennard-Jones mixtures\cite{PETERSON}.

Figure \ref{fig:phasen} ({\bf c}) presents the results of our SCF calculations. Qualitatively similar to
the MC results, the critical point of the film is shifted to a lower temperature and a higher 
concentration of the A-component. SCF calculations and MC simulation agree nicely on the critical concentration.
The reduction of the critical temperature by
$10 \%$ is however somewhat smaller than in the MC simulations. Of course, both the bulk and
the thin film critical point exhibit mean field critical behavior with $\beta=1/2$, and the critical 
temperatures of the bulk and the film are overestimated by the SCF theory. This overestimation is
larger in 2D ($T_c^{\rm MF}/T_c^{\rm MC} = 1.38$) than in the bulk ($T_c^{\rm MF}/T_c^{\rm MC} = 1.22$).
At intermediate temperatures we also find a convex shape of the binodal in our SCF calculations.

The coexistence chemical potential is presented in Figure \ref{fig:phasen} ({\bf b}) and ({\bf d}) for the
MC simulations and for the SCF calculations, respectively. The chemical potential is shifted away from its
coexistence value. For temperatures above the wetting transition, eq. (\ref{eqn:mcoex}) can be simplified to
\begin{equation}
\frac{\Delta \mu_{\rm coex} D} {k_BT N} =  - 2 b \sqrt{\chi/6}  \qquad \mbox{(wet wall)}
\end{equation} 
This estimate is also displayed in the figures. It describes the qualitative behavior above $k_BT/\epsilon\approx 20$.
At lower temperatures, the coexistence value of the chemical potential difference becomes temperature independent.
As we shall explain below, this is again a signature of vicinity to the wetting transition in the semi-infinite system.

\subsection{The wetting transition of the semi-infinite system}

At the B-rich branch of the bulk coexistence curve ($\Delta \mu/k_BT \to 0^-$), there is a 
wetting transition at which the thickness $l$ of the A-rich layer at the wall of a semi-infinite system diverges
\cite{CAHN,BREV,SCHICK}. Since our 
monomer-wall interactions $\epsilon_w/k_BT=0.16$ are rather strong compared
to the monomer-monomer interactions at the critical temperature $\epsilon/k_BT_c=0.0180(2)$, we expect
the wetting transition to occur far below the critical point. As we shall see, the transition is first order at low 
temperatures, i.e. the thickness $l$ of the absorbed layer
jumps discontinuously from a finite value to infinity upon approaching the wetting temperature 
$k_BT_{\rm wet}/\epsilon$ from below.

Wetting occurs according to the Young equation (\ref{eqn:young}) when the surface free energy difference 
$\Delta \sigma_w$ exceeds the bulk interfacial tension.
By virtue of the symmetry of our polymer model with respect to exchanging the labels $A$ vs.\ $B$, the surface
free energy $\sigma_{WA}$ of a wall with interaction strength $\epsilon_{w}$ with respect to an A-rich bulk
equals the free energy cost of a wall with interaction strength $-\epsilon_{w}$ (i.e.\ favoring B-monomers)
with respect to a B-rich bulk. Thus the free energy difference can be obtained directly via thermodynamic integration 
at $\Delta \mu = 0^-$:
\begin{equation}
\Delta \sigma_{\rm w} = \frac{1}{2L^2} \Big\{G(\epsilon_{w})-G(-\epsilon_{w}) \Big\} 
= \frac{1}{2L^2} \int_{-\epsilon_{w}}^{\epsilon_{w}} de_{\rm w}\;\; \frac{\partial G}{\partial e_{\rm w}}
= -d \Phi \int_{-\epsilon_w}^{\epsilon_w} de_{\rm w}\;\; \langle \phi_{\rm wall}(e_{\rm w})\rangle
\qquad \mbox{(B-rich bulk)}
\label{eqn:integral}
\end{equation}
(for our choice of the Hamiltonian) where the surface composition $\phi_{\rm wall}$ denotes the difference of the number 
of A-monomers and B-monomers 
in the $d=2$ nearest layers at the wall normalized by $2d\Phi L^2$.  If the mixture was completely incompressible 
(i.e.\ in the absence of packing effects at the wall), the surface composition would range between $-1$ and $1$. 

For a B-rich bulk and temperatures below the wetting transition, the surface composition $\phi_{\rm wall}$ 
is rather independent of the monomer-wall interaction $\epsilon_w$ and close to -1.
If the wetting transition is a {\em strong} first order transition, the thickness of the absorbed layer will remain small,
as will the deviation of $\phi_{\rm wall}$ from -1. Thus eq.\ (\ref{eqn:integral}) yields the estimate 
$\Delta \sigma_{\rm w} \approx 2 d \Phi \epsilon_w$. $\Delta \sigma_{\rm w}$ is dominated by the monomer-wall
interaction; the entropy loss of the polymers at the wall is assumed to be the same for both species and
hence does not affect $\Delta \sigma_{\rm w}$ in our model.
Using  the expression $\sigma_{AB}/k_BT = b \Phi \sqrt{\chi/6}$
($b=3.05$ \cite{M4}: statistical segment length)
for the interfacial tension in the strong segregation limit (SSL)\cite{BREV}, 
we get the following estimate for first order wetting transition temperature in the strong segregation limit:
\begin{equation}
\chi_{\rm wet} \approx 24 \left(\frac{\epsilon_w d}{b^2 k_BT}\right)^2 \qquad \mbox{(SSL)}
\label{eqn:wettrans}
\end{equation}
which depends quadratically on the wall-monomer interaction strength $\epsilon_w d$ and is
independent of the chain length $N$. Using the parameters
of our simulations, the above equation yields $\epsilon/k_BT_{\rm wet} \approx 0.05$.
If the monomer-wall interaction $V_{WA(B)}(z)$ were not of square well type, we would replace $\epsilon_w d$ by the integrated
interaction strength.
\begin{equation}
\epsilon_w d = \int dz\; \Big[ V_{WB}(z)\rho_B(z)- V_{WA}(z)\rho_A(z) \Big]
\end{equation}
Using $\langle \rho \rangle = \exp(-\chi N)$ in the strong segregation limit (SSL)
we can rewrite the expression above in terms of the bulk composition:
\begin{equation}
\frac{\epsilon_w}{k_BT} = \sqrt{\frac{b^2}{24 N d^2} \ln \frac{1}{\langle \rho \rangle}} \qquad \mbox{(SSL)}
\label{eqn:wetr}
\end{equation}

However, for any finite film width $D$ the B-rich phase is  not stable for 
$ -\Delta \mu_{\rm coex}({\rm bulk}) = 0^- < -\Delta \mu < -\Delta \mu_{\rm coex}(D)$, 
and the thermodynamically stable phase at the bulk coexistence chemical potential is A-rich. 
Thus there is no wetting transition in a thin film in thermal equilibrium. We estimate the coexistence value
of the chemical potential difference below the wetting transition in the bulk to be:
\begin{equation}
-\frac{\Phi D \Delta \mu_{\rm coex}}{2N} = \Delta \sigma_w = 2 d \Phi \epsilon_w \qquad \Rightarrow  \qquad
 \frac{\Delta \mu_{\rm coex} D}{k_BT N} = -4 \frac{d \epsilon_w}{k_BT} \qquad \mbox {(dry wall)}
\end{equation}
by replacing $\sigma_{AB}$ by $\Delta \sigma_w$ in eq.\ (\ref{eqn:mcoex}).
Above the wetting transition in the bulk, the chemical potential difference depends on the strength of
the monomer-monomer interaction $\epsilon$ (or $\chi$), whereas below the wetting temperature it is
independent of $\chi$ but depends on the monomer-wall interaction $\epsilon_w$. Therefore we attribute the
change in the $\Delta \mu - T$ phase diagram (Figure \ref{fig:phasen} ({\bf b,d}))
around $k_BT/\epsilon \approx 14$ to the wetting transition in the semi-infinite system. If we had chosen 
an enthalpic surface interaction,
the chemical potential would depend linearly on the temperature below the wetting temperature, instead of
being temperature independent.

To investigate the influence of the wetting transition in the bulk on the behavior of the confined system
further, we calculate the composition dependence of the free energy close to the first order wetting transition 
in the  SCF scheme. At low concentration of the A-component, the A-rich 
wetting layers are bound to the wall (``dry'' state) and the free energy of this state with respect to the 
thermodynamically 
stable A-rich phase at $\rho \approx 1$ is $2\Delta \sigma_{\rm wall}L^2 $. Upon 
increasing the composition the thickness of the A-rich layers increases and for large layer thickness 
($\lambda D \gg \lambda l \gg 1$) the excess free energy is given by the interfacial tension $2\sigma_{AB}L^2$ (``wet'' state).
A plateau in the free energy indicates that the two interfaces are only weakly interacting. Both states are
separated by a free energy barrier of height $2\gamma L^2>0$; thus the wetting transition is first order.
For even higher concentrations the two $AB$ interfaces attract each other and finally annihilate to form the stable 
$A$-rich phase. The results of the SCF calculations for $\epsilon/k_BT=0.0575$ and $D=49,81$ and $137$ 
are presented in Figure \ref{fig:wet} ({\bf a}). One clearly identifies the ``dry state'' and the plateau, which corresponds
to the ``wet'' state. Both are separated by a free energy barrier $\gamma/k_BT \approx 0.0027(8)$.
Note that the data depend on the wall separation $D$ even for film widths as large as $D=137 \approx 20 R_g$.
The surface free energy difference $\Delta \sigma_w$ and the interfacial tension decrease both upon increasing the
film width. However, the effect is slightly more pronounced for the interfacial tension, thus 
using the data for thin films, we would systematically overestimate the wetting temperature.
Increasing the film width $D$ still further in our SCF calculations, exceeds our computational facilities.
The calculation of a single profile ($D=137$ and $7,863,320$ polymer conformations) requires about 10 minutes on 512 T3E 
processors and about 8 Gbyte of memory. Thus we use the width $D=137$ to explore the wetting behavior in the SCF 
calculations.

Though the wetting temperature can be estimated in the MC simulation via the scheme above (cf. Appendix),
it is limited to rather small system sizes, because the configurations relax via a  slow 
diffusion of the $AB$ interfaces across the film. Moreover, the connection between the composition of the system and
the thickness of the wetting layers is more involved. The probability distribution of the composition for 
$\epsilon/k_BT=0.0695$ is presented in Figure \ref{fig:wet} ({\bf b}) for system geometry $48 \times 48 \times 48$.
It shows qualitatively the same behavior as the
free energy in the SCF calculations. The equal probability of the ``dry'' state and the free interfaces
gives an estimate for the wetting temperature $\epsilon_{\rm wet}/k_BT \approx 0.0695$. Upon increasing the 
temperature, the ``dry'' state becomes metastable and ceases to exist even as a metastable state around 
$\epsilon/k_BT=0.063$ (cf.\ Figure \ref{fig:wet} ({\bf b} inset)). Thus the spinodal temperature is about $10\%$ higher than the
wetting temperature. In view of the  dependence of our SCF results on the wall separation $D$, we expect that
our simulation data for $D=48$ are affected in a similar manner. Indeed, the plateau value is slightly
higher than the interfacial tension $\sigma$ (displayed as a dashed horizontal line) obtained from the 
simulations of a system with periodic boundary conditions and $36 \times 36 \times 64$ geometry. Thus the 
wetting temperature is slightly lower than the estimate above.

In view of these difficulties, it is interesting to compare different methods for locating the wetting transition. 
Within the mean field framework, the 
wetting layers in a film are metastable for $\gamma>0$, 
and their lifetime $\tau$ in the MC simulations increases with the lateral extension $L$ 
like $\tau \sim \exp( \gamma L^2)$. Upon increasing the temperature $k_BT/\epsilon$ the coefficient 
$\gamma$ decreases and vanishes at the wetting {\em spinodal}. Of course, fluctuations cause a pronounced 
rounding of the spinodal when $\gamma L^2 \sim O(k_BT)$. The observation of this metastability has 
been used previously to determine the location of the wetting transition and its order by Wang et al.\
\cite{WANG}. To determine the spinodal point we measure the surface order parameter $\phi_{\rm wall}$ as a function 
of the temperature $k_BT/\epsilon$. We use rather large lateral extensions $L$ to increase the lifetime of the metastable
state and large film widths $D$ to avoid interactions between the two $AB$ interfaces. 
Using the SCF estimate, we obtain $\gamma L^2 \approx 25 k_BT$ for $L=96$ at the wetting transition.
The simulation data for the system geometry
$96 \times 96 \times 80$ and $128 \times 128 \times 128$ are presented as open symbols and filled symbols
in Figure \ref{fig:wet}({\bf c}), respectively. In the thermodynamically stable A-rich phase there are exclusively 
A-monomers near the wall and the surface composition is almost temperature independent.
In the metastable B-rich phase, the surface composition at low temperatures is B-rich and the surface order
parameter increases rapidly for $\epsilon/k_BT<0.7$. This increase goes along with a pronounced increase
of the surface composition fluctuations in the {\em metastable} phase, as shown in the inset.
We identify this change of the surface order parameter $\phi_{\rm wall}$ and the observation that no metastable 
B-rich phase could be detected in our simulations for $\epsilon/k_BT<0.065$ as the signature of the
spinodal and use $\epsilon/k_BT_{sp}=0.065(3)$ as our estimate for the spinodal temperature.

If the $AB$ interfacial tension has been measured independently (e.g.\ via reweighting techniques\cite{M4,BERG,INTER}
or the spectrum of capillary fluctuations\cite{MW,M3}), we can use the Young equation (\ref{eqn:young}) and
eq.\ (\ref{eqn:integral}) to determine the wetting temperature. However, rather than measuring the surface order 
parameter $\phi_{\rm wall}$ for many values of the wall interaction strength $e_{\rm w}$, we use an expanded 
ensemble\cite{VALEAU}
in which the wall-monomer interaction strength $e_{\rm w}$ is a stochastic degree of freedom which assumes 
values between $-\epsilon_w$ and $+\epsilon_w$. 
This permits us to calculate the free energy difference in a single simulation run:
\begin{equation}
{\cal Z} \sim \sum_{-\epsilon_w \leq e_{\rm w} \leq \epsilon_w} Z(e_{\rm w})/W(e_{\rm w}) = \sum_{e_{\rm w}} W^{-1}(e_{\rm w}) \exp(-G(e_{\rm w})/k_BT)
\end{equation}
where $Z(e_{\rm w})$ is the semi-grandcanonical partition function at fixed temperature, exchange potential and wall
interaction. We chose the preweighting factors $W(e_{\rm w}) \sim \exp(-G(e_{\rm w})/k_BT)$ as to 
achieve uniform sampling of all $e_{\rm w}$ states. A good initial estimate of the preweighting factors is given by
$W(e_{\rm w}) = \exp( -2 d \Phi L^2 e_{\rm w}/k_BT)$. 


Figure \ref{fig:wet} ({\bf d}) displays the surface composition $\langle\phi_{\rm wall}\rangle$ at 
$\epsilon/k_BT=0.07$,$\Delta \mu = 0^-$ and $-0.16 \leq \epsilon_w/k_BT \leq 0.16$. The bars in the figure
do not denote statistical errors but the variance of the distribution of the surface composition. Thus
the distributions of $\phi_{\rm wall}$ at the different values of the monomer-wall interaction overlap strongly.
The increase of the monomer-wall interaction $\epsilon_w$ shifts the surface composition to higher values 
and its fluctuations increase. The crude estimate $\langle \phi_{\rm wall}\rangle = -1$
is also shown in the figure. Values of $|\phi_{\rm wall}|>1$ can be attributed to compressibility effects:
Assuming that the system contains only B-chains, we can decrease the wall interaction energy by changing
the monomer density in the first $d$ layers at the wall. This deviation from the bulk density costs entropy
and we approximate it by a quadratic compressibility term (cf.\ eq.\ (\ref{eqn:energy})). Balancing
these two terms we obtain $\langle \phi_{\rm wall}\rangle = -(1-\epsilon_w/\zeta)$. The linear dependence of 
$\phi_{\rm wall}$ 
describes the simulational data qualitatively when the walls favor the majority phase in the bulk (i.e. $\epsilon_w<0$),
though the effect is somewhat more pronounced than the estimate above with the bulk value $\zeta=4.1$ of the 
compressibility at infinite temperature.

In the simulations we divide the interval [$-\epsilon_w,+\epsilon_w$] into 16 subintervals, and the MC scheme
incorporates moves which change the wall interaction  $e_{\rm w}$. Care has to be exerted at the interval boundaries
to fulfill detailed balance. We employ the ratios 10:4 and 1:4  between the grandcanonical moves and the attempts to 
alter the wall interaction. The results of both ratios agree within statistical accuracy.
The correlation time at $\epsilon/k_BT=0.07$ and $\epsilon_w/k_BT=0.16$ is about 250
attempts to change $\epsilon_w$. Since the surface order parameter increases more rapidly at higher $\epsilon_w$,
we chose smaller subintervals at the upper edge. Let $P(e_{\rm w})$ denote the probability with which the state 
$e_{\rm w}$ is populated in the simulation. Thus the excess wall free energy $\Delta \sigma_{\rm wall}$ is given by:
\begin{equation}
\frac{\Delta \sigma _{\rm w}}{k_BT} = \frac{1}{2L^2} \ln \left( \frac{P(-\epsilon_{\rm w}) W(-\epsilon_{\rm w})}
{P(\epsilon_{\rm w}) W(\epsilon_{\rm w})}\right)
\end{equation}
The inset of Figure \ref{fig:wet} ({\bf d}) presents the dependence of the excess wall free energy 
$\Delta \sigma_w$ on the wall separation $D=48 \cdots 196$ at $\epsilon/k_BT=0.07$, $\epsilon_w/k_BT=0.16$ and $L=96$.
The finite size effects are compatible with a $1/D$ dependence. The data for $D=48$ have a finite size error of
$3\%$ which compares well with the SCF calculations (cf.\ Figure \ref{fig:wet} ({\bf a})) and
the deviation of the plateau in Figure \ref{fig:wet} ({\bf b}) from the bulk value of the interfacial tension. 
For $D=80$ the value $\Delta \sigma_w$ is overestimated by $2\%$. This is of the same order of magnitude as our
uncertainties in the bulk interfacial tension and we use films of width $D=80$ for the calculation of 
$\Delta \sigma_w$ in the following.

The results of this measurement for several monomer-wall interactions $\epsilon_w$ and the independently measured 
bulk interfacial tension are presented in Figure \ref{fig:wet}({\bf e}). The figure also displays the interfacial tension 
in the strong segregation limit and our naive estimate for the excess wall free energy at low temperatures.
While the estimate for $\Delta \sigma_{\rm w}$ agrees well with the simulation data, the interfacial tension
shows pronounced deviations from the strong segregation behavior due to chain end effects.
From the intersection of $\Delta \sigma_{\rm w}$ and $\sigma_{AB}$ we estimate the wetting temperatures. 
For $\epsilon_w/k_BT=0.16$ we find a first order wetting transition at $\epsilon_{\rm wet}/k_BT=0.0709(15)$.
As anticipated, this is about $10\%$ below the spinodal temperature, extracted from the observation of the 
metastable wetting layers.
The MC result is $15\%$ lower than the wetting temperature predicted in the SCF calculations.

Similar to the SCF calculations of Carmesin and Noolandi\cite{CN}, we find only first order wetting transitions
for the parameters studied ($T_{\rm wet}<0.64T_c$). However, as we reduce the monomer-wall interaction $\epsilon_w$, 
the wetting temperature approaches the critical point and the strength of the first order transition becomes weaker. 
Thus we cannot rule out that there are second order transitions (and a concomitant tricritical point)
in the ultimate vicinity of $T_c$. In case of a second order transition, the thickness of the wetting layer 
grows continuously and there are serious finite size effects. Even for weak first order transitions
we anticipate finite size effects when the thickness $\langle l \rangle$ in the ``dry'' state is not much 
smaller than the wall separation $D$. For the highest temperature investigated ($\epsilon_{\rm wet}/k_BT = 0.0226$), 
however, we find $\langle l \rangle < 4 \ll 80 = D$.  Thus our data are not strongly affected by finite size effects.
However, approaching the tricritical wetting point even further calls for a careful analysis of the film width dependence.
This is not attempted in the present study.

Figure \ref{fig:wet} ({\bf f}) presents the inverse wetting temperature $\chi_{\rm wet}$ as a function of the 
monomer-wall interaction $\epsilon_w$ and compares the MC results with our simple estimate (\ref{eqn:wettrans}). 
The MC results confirm that the inverse wetting temperature depends quadratically on $\epsilon_w$
and the prefactor is in almost quantitative agreement. The horizontal shift between the two curves
is due to chain end effects in the interfacial tension. Semenov\cite{SEM} predicted that the interfacial tension
is reduced by a factor $1- 4 \ln 2/\chi N$ for finite chain lengths. These first order corrections in $1/\chi N$
to the interfacial tension increase $\chi_{\rm wet}$ by
an $\epsilon_w$-independent term $8 \ln 2/N$. This correction (dashed line in Figure \ref{fig:wet} ({\bf f}) )
accounts almost quantitatively for the deviations between the MC results and our simple estimate.

The inset presents the dependence of the monomer-wall interaction at the wetting transition on the bulk composition 
$\langle \rho \rangle$. The solid line represents our estimate in the strong segregation limit (without chain end effects)
according to eq.\ (\ref{eqn:wetr}). Near the critical point, there is a layer of finite thickness at the wall in 
the ``dry'' state of the first order transition and we find deviations from the low temperature behavior. Moreover, 
in the vicinity of the critical point second order wetting is expected to occur. In this regime the square gradient (SG)
theories\cite{PINCUS,SCHMIDT} give a qualitative description.
To a first approximation we identify the parameters of the bare surface free energy $f^{\rm bare}_{SG}$ in the mean field 
theory by estimating the energy in the $d$ layers next to the wall\cite{JONES,CIFRA}:
\begin{eqnarray}
f^{\rm bare}_{SG} &\equiv& - \mu_1 \frac{1 + \phi_{\rm wall}}{2} - \frac{1}{2} g_1 
		  \left(\frac{1+\phi_{\rm wall}}{2}\right)^2 + \mbox{const} \nonumber \\
	   &\approx&  -\epsilon_w d \phi_{\rm wall} + 2\Delta z \epsilon d (1+\phi_{\rm wall})(1-\phi_{\rm wall})/4 
		 \nonumber \\
\mu_1 \approx -2d(\Delta z \epsilon - \epsilon_w) \qquad &\mbox{and}& \qquad g_1 \approx 4 \Delta z \epsilon d
\label{eqn:SZW}
\end{eqnarray}
where $\Delta z$ denotes the reduction of the intermolecular contacts at the wall. From the profiles (presented in
Figure \ref{fig:profile} ({\bf e})) we estimate $\Delta z \approx -1$. Note that $\mu_1$ and $g_1$ are strongly
influenced by the specific packing structure of the monomer fluid at the wall. Moreover, $g_1$ is temperature
dependent in our model. In the SG theory second order wetting occurs close to criticality along
$\epsilon_w = \epsilon|\Delta z|(1-2\langle\rho\rangle)$. Using the square gradient expression for the composition
$N\chi (1-2\langle\rho\rangle) = \ln\left((1-\langle\rho\rangle)/\langle\rho\rangle\right)$ we can rewrite the temperature 
dependence of $g_1$ in terms of the bulk composition and obtain for second order wetting in the weak segregation 
limit (WSL):
\begin{equation}
\epsilon_w = \frac{1}{2N} \frac{\Delta z}{z}  \ln \left(\frac{1-\langle \rho \rangle}{\langle \rho \rangle} \right)
\qquad \mbox{(WSL)}
\end{equation}
which complements eq.\ (\ref{eqn:wetr}).
This prediction is presented in the inset of Figure \ref{fig:wet} ({\bf b}) as a dashed line. Due to the small chain 
length in our simulations,
the SG treatment is not accurate close to the critical point (cf.\ Figure \ref{fig:phasen}) and thus
the SG theory can only describe the qualitative behavior. However, the simulation data seem to 
crossover from the first order wetting at low temperatures to transitions qualitatively described by the equation 
above. This correlates with the observation that the strength of the first order wetting transition in the MC
simulation decreases at higher temperatures. If the coefficient $g_1$ is solely attributed to the ``missing neighbor''
effect it is proportional to the inverse chain length $N$. In this scenario\cite{PINCUS,JONES} critical wetting occurs only 
for very small monomer-wall interactions of the order $N^{-1}$ or short chain lengths. Note, that the 3D Ising model without
enhanced nearest neighbor interactions at the wall -- a model similar to the ultimate short chain length limit of our
polymer model -- exhibits second order wetting\cite{BINDERWET}.

In the presence of specific contributions to the monomeric interactions at the wall (i.e. $g_1$ independent
of $N$ at the critical point), as modeled in simulations by Wang and Pereira\cite{WANG}, second order
wetting has been observed for rather short chain lengths. However, for large $N$ the wetting transition occurs 
far below the critical point; thus the bulk composition is given by $\exp(-\chi_{\rm wet} N)$ (where $\chi_{\rm wet}$
is chain length independent). This behavior contrasts the bulk composition at which tricritical wetting occurs. 
In the SG theory it scales like $1/g_1\sqrt{N}$. Even in the case of chain length independent $g_1$, the bulk composition 
at the wetting transition is smaller than the tricritical value for sufficiently long chains and, hence, the transition 
is first order.

\subsection{Prewetting} 
If the wetting transition is first order, then there is a discontinuous jump in layer thickness above the
wetting temperature off coexistence i.e. at $\Delta \mu_{\rm pre}<0$\cite{SCHICK}.  At this prewetting line
a thin wetting layer coexists with a thick layer. As one increases the temperature, the difference in the 
thickness of the coexisting layers becomes smaller and the chemical potential moves away from its bulk 
coexistence value. At the prewetting critical point, the difference of the coexisting phases vanishes; 
the transition is believed to exhibit 2D Ising critical behavior. 

For small wall separation $D$, the coexistence chemical potential $\Delta \mu_{\rm coex}$ is smaller than 
the prewetting chemical potential $\Delta \mu_{\rm pre}$ and the system phase separates laterally before the thickness of the layer
$l$ reaches the lower coexistence value. This situation occurs at wall separation $D=48$ \cite{COMMENTPRE}.
The situation for larger $D=137$ is  exemplified with SCF calculations in Figure \ref{fig:phasen} ({\bf e,f}).
For large wall separation, the prewetting line crosses the coexistence curve. This triple point is located at 
$k_BT/\epsilon= 19.9 = 1.16 k_BT_{\rm wet}/\epsilon$. At the intersection point a $B$-rich phase with a
thin ($l_- \approx 0.065 R_g$) and a 
thick ($l_+ \approx R_g$) wetting layer coexist with an A-rich phase. The distance between the triple temperature in 
the film of width $D$ and the wetting temperature increases upon reducing the wall separation.
The prewetting critical temperature is located at $k_BT_{\rm c,pre}/\epsilon= 26 = 1.51 k_BT_{\rm wet}/\epsilon$.

The determination of the complete phase diagram of a thick film in the MC simulations is 
beyond our computational facilities. However, we expect that the SCF calculations capture the qualitative
behavior. To locate the prewetting line, we monitor the dependence of the layer thickness on the chemical potential
for $k_BT/\epsilon = 15.87 = 1.13 k_BT_{\rm wet}/\epsilon$ and $k_BT/\epsilon = 17.86 = 1.27 k_BT_{\rm wet}/\epsilon$.
We employ the system size $L=96$ and $D=80$. This technique has been employed previously by Pereira\cite{WANG} and
we
use it in junction with a multi-histogram analysis\cite{HISTO} of our MC data. 
For large lateral extensions $L$, we expect a jump in the layer thickness $l$ and hysteresis at the (first order) 
prewetting transition. For $L=96$ however, a layer of thickness $R_g/2$ comprises only $\Phi L^2 R_g/2N \approx 63$ 
polymers. Hence the prewetting transition is strongly rounded by finite size effects. The MC data are presented in 
Figure \ref{fig:pre}. The dependence of the layer thickness exhibits a turning point and from the corresponding
maximum of the susceptibility we estimate the location of the prewetting line. Upon increasing the temperature
the jump of the layer thickness decreases and the peak in the susceptibility becomes less pronounced. 
An accurate estimation 
of the prewetting critical point calls for a thorough finite size analysis, which is not attempted here.
However, the data for the highest temperature are close to or above the prewetting critical temperature.
Hence, the prewetting line is presumably less extended in the MC simulations than in the SCF calculations.

In a recent experiment, Zhao et al.\cite{ZHAO} investigated the wetting properties of ultrathin 
polyethylene polypropylene (PEP) films on polished silicon wafers above the wetting temperature.
These experiments reveal that slightly above the wetting temperature thick layers ($l \gg R_g$)
wet the substrate, while ultrathin layers ($l < R_g$) break up into droplets and form pattern ``analogous
to those produced by spinodal decomposition''\cite{ZHAO}. The layer thickness below which the layer
dewets scales like the radius of gyration $R_g$. These findings were rationalized\cite{ZHAO,AUBOUY} by the 
entropy costs of confining a chain into a layer which is thinner than the unperturbed chain extension.
At higher temperatures, however, even ultrathin films of PEP wet the silicon waver.

This situation resembles the prewetting behavior: Slightly above the wetting transition two
layers of different thickness $l_-$ and $l_+$ coexist. Since the length scale of the
effective wall-interface interaction $g(l)$ scales like $R_g$, which is proportional to the
correlation length of concentration fluctuations in the bulk and the characteristic decay lengths in the
wings of the concentration profiles across an interface, so do the thicknesses 
of the coexisting layers. The reduction of the conformational entropy 
of the $A$ chains confined into an ultrathin layer is an important (though not unique)
contribution to the effective wall-interface interaction $g(l)$ for $l<R_g$. 
A layer of thickness $l_-<l<l_+$  separates laterally into regions of thickness $l_-$ and $l_+$. 
If $\partial^2 g/\partial l^2<0$ (cf.\ Figure \ref{fig:wet} ({\bf a}) inset) spinodal phase separation
occurs, i.e.\ the layer is unstable against capillary fluctuations, which grow exponentially
for large wavevectors in the early stage. Using the capillary fluctuation Hamiltonian (\ref{eqn:helfrich}), 
the fastest growing mode of the spinodal dewetting has the wavevector\cite{BROCHARD}
\begin{equation}
q^*= \sqrt{ - \frac{1}{2\sigma} \frac{\partial^2 g}{\partial l^2}}
\end{equation}
Using the SCF results for our model in Figure \ref{fig:wet} ({\bf a} inset), we estimate 
$\partial^2 g/\partial l^2 \approx \gamma/2R_g 
\approx -0.00062 k_BT$, $\sigma \approx 0.034 k_BT$ and $q^* \approx 0.09 = 0.1 \;\; 2\pi/R_g$ close to the wetting 
transition. At temperature higher than the prewetting critical point however, there is no coexistence between thick and 
thin layers. Thus, $g(l)$ is convex for all thicknesses $l$ and ultrathin layers are stable. This temperature, chain length
and layer thickness dependence and the concave curvature of the free energy, which leads to spinodal character
of the dewetting, are compatible with the experimental findings\cite{ZHAO}.

To illustrate the dewetting of ultrathin polymer layers further, we perform MC simulations in the canonical ensemble,
i.e.\ we let the chain conformations evolve via individual monomer hopping (LM) and slithering snake (SS) moves, however,
identity switches $A \rightleftharpoons B$ are not allowed. Due to the slithering snake moves, we do not observe Rouse 
dynamics on short time and length scales, but the number of $A$-chains is conserved and we recover a purely diffusive 
dynamics on length scales larger than $R_g$. Of course, our MC simulation cannot reproduce fluid-like
flow which is important for the late stage dynamics of spinodal decomposition\cite{BMAT}. Hence we do not attempt to 
relate our 
``MC time'' to physical time units. Moreover, unlike the experimental situation, we consider a binary polymer melt 
in contact with a wall rather than a polymer solution. Long-ranged dispersion forces at the wall are not incorporated 
in our model. Due to the diffusive dynamics with conserved composition, the characteristic length scale $1/q^*$
being larger than the chain extension, and the universality of the wetting behavior we do expect, however, that 
the salient features of the early stage of phase separation are qualitatively captured in our MC simulations.

We study a cubic $123^3$ system. Initially both walls are covered with a flat, pure $A$-polymer layer of thickness
$\langle l \rangle=4,8,16$, whereas the central portion of the  film contains only B-polymers. In Figure \ref{fig:dewet1}
we display snapshots of the A monomers in the lower half of the container. Each $A$ monomer is 
presented as a sphere. The left row shows the time evolution of an ultrathin layer $\langle l \rangle = 4 = 0.577 R_g$ at
$T=1.02 T_{\rm wet}$. Initial concentration fluctuations grow rapidly. Later we observe A-rich domains which coarsen
in time and in the last snapshots there is only one cluster which spans the whole system via the periodic boundaries
in the lateral directions. The domain size is comparable with the  extension of the container and no further domain
growth takes place. Thus, we find clear evidence for dewetting in an ultrathin layer above the wetting temperature.

The time evolution of a thicker layer $\langle l \rangle=8=1.153 R_g$ at the same temperature is presented in the  
middle row. Though we do observe some local thermal fluctuations of the $A$ concentration, the layer does not break 
up into domains. Thus, a thick layer does not dewet the walls at the same temperature\cite{COMMOVIE}. 
To complete the analogy to the experiments we display the time sequence for a thin layer $\langle l \rangle = 0.577 R_g$
at $T = 2.363 T_{\rm wet}$. This temperature is above the SCF estimate of the prewetting critical temperature.
Again we observe quite pronounced thermal lateral composition fluctuations on the length scale $R_g$. In the
last snapshot an $A$-polymer has even escaped the layer. Note that at this temperature ($T=0.48 T_c$) there is a small 
solubility of the $A$ component in the $B$ rich bulk. However, the length scale of composition fluctuations remains 
smaller than the box size. This indicates that the ultrathin layer is stable against spinodal dewetting at high 
temperatures.

This behavior can be quantified via a subbox analysis. We monitor the probability distribution $P(\nu)$ of the 
local $A$-monomer density $\nu$. For times larger 
than those displayed in the figures $P(\nu)$ is stationary, because if there is lateral phase separation the domain
size has become comparable with the lateral system size. We average the lateral A-monomer density over square blocks of 
linear extension $B=1.154 R_g$. The results of this analysis are presented in Figure \ref{fig:dewet2}.
In ({\bf a}) we study the dependence on the layer thickness $\langle l \rangle = \langle \nu \rangle/\Phi$ 
slightly above the wetting temperature $T=1.02 T_{\rm wet}$. The layers of thickness $\langle l \rangle/R_g=1.154$
and $2.308$ exhibit a single peaked distribution centered around the initially homogeneous density.
However, the thin layer $\langle l \rangle/R_g=0.577$ exhibits a bimodal distribution; one maximum is at
$\nu=0$ and the other is located at $\nu/\Phi R_g \approx 0.8$. This double peak structure indicates the
dewetting. The distribution for a thin layer at higher temperature $T=2.363 T_{\rm wet}$ is also shown for comparison.
It resembles the distribution of the thicker layers, just shifted to lower densities. Thus, the thin layer at
higher temperatures is stable against spinodal dewetting. In Figure \ref{fig:dewet2} ({\bf b}) we present the
temperature dependence of the subbox distribution as a function of the temperature. Upon reducing the temperature
the distributions change very gradually from single peak to bimodal. The inset shows schematically the path
along which we approach the coexistence between thick and thin layers. The solid curve is the result
of our SCF calculations.

 \subsection{Interfacial profiles across the film}

 We proceed by exploring the detailed profiles across the film (i.e. perpendicular to the wall)
 and its dependence on the film width $D$ in
 the temperature regime between the critical temperature of the film and the wetting transition in the bulk.
 We choose $k_BT/\epsilon=50 \approx 0.72 k_BT_c/\epsilon$, which is far enough below the critical point to limit the 
 influence
 of the shift of the critical temperature upon confinement and close enough to the critical point to obtain the
 preweighting factors of the composition within a small number of auxiliary simulations at intermediate temperatures.
 We study the wall separations $D=36,48,60,80,112$, which correspond to $D/R_g = 5.2 \cdots 16.2$.
 We employ a $96 \times 96 \times D$ geometry and for $D=48$ we gather some data for different lateral system sizes.
 The MC results are compiled in Tab.\ \ref{tab:resmc}; effects of the varying the system size are small at this temperature.
 
 Figure \ref{fig:profile} presents the composition profiles $c(z)=\rho_A(z)/(\rho_A(z)+\rho_B(z))$ at coexistence, 
 both in the MC
 simulations ({\bf a}) and the SCF calculations ({\bf b}). The profiles are symmetric about the middle of the 
 film 
 and only one half is displayed. Both methods yield qualitative similar results:
 Upon increasing the film width $D$, the coexistence potential approaches the bulk value and the thickness $l$ 
 growths. This is qualitatively similar to complete wetting\cite{SCHICK}. The surface order parameter and the width 
 of the $AB$ interface increases with growing $D$ too. The profiles at the wall are 
 flattened about the first $d$ lattice units in the simulations as well as in the SCF calculations.
 
 The inset of Figure \ref{fig:profile}({\bf a}) displays the density profiles $\rho(z)$ for $D=60$, which exhibit 
 pronounced packing effects. Also the results of the SCF calculations (not shown) exhibit some structure near the 
 walls, however, the effect is much less pronounced than in the Monte Carlo simulations and the detailed packing 
 structure is not reproduced by our SCF calculations. However, related SCF calculations of the surface segregation of a binary 
 blend at a hard wall in the framework of an off-lattice model\cite{SZWALL} achieve somewhat better agreement with MC 
 simulations.
 Moreover, the width of the $AB$ interface is larger in the simulations than in the
 SCF calculations. This is partially due to broadening of capillary fluctuations and also
 expected because the distance to the critical point is smaller in the simulations than in the SCF framework.
 Moreover, the segregation in the middle of the film increases upon increasing the width of the film, whereas
 the opposite trend is observed in the SCF calculations.
 
 The Figure \ref{fig:profile} also shows the behavior of the parallel and perpendicular components of the
 end-to-end vector $\vec{R}$ as a function of its midpoint from the wall for $D=60$.
 The simulation data are presented in ({\bf c}) and the SCF calculations in ({\bf d}). At the wall, 
 the perpendicular component $R_{\perp}$ vanishes for both components. Simulations and SCF calculations
 reveal that the parallel chain extension $R_{\|}$ of the A-component (majority) at the wall is larger 
 than in the bulk.
 This transpires that the chains are not only deformed by the presence of the walls, but orient
 the long axis of their instantaneous shape parallel to the wall. Note that such an effect cannot be
 observed for Gaussian chains because the parallel and perpendicular extensions of Gaussian chains decouple
 completely and, hence, the parallel components of the chain extension are independent from the distance to the wall.
 The B-component (minority at the wall) shows hardly any deviation from its bulk value across the film in the MC simulations
 and in the SCF calculations.
 
 Upon approaching the middle of the film simulations and SCF calculations show that the perpendicular extensions 
 of A-chains attain their unperturbed value sooner than B-chains. The A-chains which are close to the 
 $AB$ interface and in the minority (i.e.\ around z=20 and z=40) are stretched perpendicular to the $AB$ interface,
 as to reach with one end the corresponding A-rich layer close to the wall. Similar behavior is predicted
 for polymer/polymer interfaces\cite{SCHMID1}, however, this effect cannot be resolved within the
 scatter of the simulation data. (Note that the concentration of A-chains is only $6\%$ in the middle of the film.)
 Furthermore we observe that the chain conformations at the AB interface are strongly affected by the presence of the
 wall. Thus, a film width of $D=48\approx 6.9 R_g$ is not large enough to approximate the properties of the
 interfaces by their bulk behavior. At lower temperature, these deviations will become even more pronounced
 because the A-rich layer thickness decreases upon reducing the temperature.

 We characterize the local structure of the polymeric fluid further by profiles of the intermolecular contacts
 across the film ($D=60$). The number of inter- and intramolecular contacts, without discrimination of the monomer
 species is presented in Figure \ref{fig:profile} ({\bf e}).
 In the middle of the film the number of intermolecular contacts $z$ is close to 2.65, the value used in the 
 SCF calculations. In the vicinity of the wall the value decreases and exhibits pronounced oscillations. These 
 characterize the local structure of the monomer fluid. As discussed in the previous section, they are 
 indispensable for a quantitative prediction of the wetting behavior and surface thermodynamics. In the framework of 
 our SCF calculations these ``missing neighbors'' at the wall are accounted for via a gradient expansion of the composition
 (cf.\ eq.\ (\ref{eqn:energy}))\cite{HT}. 
Though the SCF
 treatment captures the qualitative effects it cannot reproduce the detailed structure of the monomer fluid.
 The figure also displays the number of intramolecular contacts, which are assumed to be independent of the position 
 in the SCF calculations. The number of intermolecular 
 contacts of A-chains increases at the wall. A-chains try to bring many monomers close to the wall and adopt a flat 
 (pancake-like) conformation which has a larger number of intermolecular contacts. This correlates 
 with the increase of the perpendicular extension at the wall. The number of self-contacts of B-chains is reduced at the 
 wall; B-chains try to escape the unfavorable monomer-wall interactions.

 Measuring the number of intermolecular contacts between the same ($n_{AA},n_{BB}$)  and unlike ($n_{AB}$) species, we 
 can assess the validity of the random mixing assumption inherent in the SCF treatment.
 \begin{equation}
  \frac{2n_{AA}}{\Phi\rho_A^2} = \Phi \int_{r\leq\sqrt{6}} d^3r\;\; g_{AA}(r) \equiv z_{AA} \qquad  \mbox{and} \qquad
  \frac{n_{AB}}{\Phi\rho_A\rho_B} = \Phi \int_{r\leq\sqrt{6}} d^3r\;\; g_{AB}(r) \equiv z_{AB} 
 \end{equation}
 $g_{IJ}$ denotes the intermolecular paircorrelation function,
 which is normalized such that $g_{IJ}(r \to \infty) = 1$. The integration is extended over the spatial extension
 of the square well potential. For the temperature studied, we find quite pronounce deviations of 
 our MC results from the random mixing assumption. The number of neighbors of the same species in the minority
 phase is strongly enhanced. This indicates a clustering of chains in the minority phase, similar to the observation 
 of composition fluctuations in the ultrathin film at $\epsilon/k_BT=0.03$ shown in Figure \ref{fig:dewet1}. This 
 non-random mixing also correlates with the underestimation of the composition of the minority component (cf.\
 Figures \ref{fig:phasen} ({\bf a,c})). For the temperature studied, the mean field approximation underestimates 
 the bulk minority composition $\langle \rho \rangle$ by a factor 0.68. 
 Increasing the chain length, however, we can reduce these local composition fluctuations
 and we find random mixing behavior for very long chains\cite{M0} in accord with the Ginzburg criterion\cite{GINZBURG}.

 \subsection{Dependence on the film width: Kelvin equation}

 We analyze the dependence of the layer thickness and the coexistence chemical potential on the film thickness.
 This yields information about the strength and spatial range $1/\lambda$ of the interaction $g(l)$ between the 
 wall and the $AB$ interface at a distance $l$ from the wall. 
 The shift of the coexistence chemical potential $\Delta \mu_{\rm coex}$ with the film width $D$ is displayed in
 Figure \ref{fig:kelvin} ({\bf a}) for the simulations and the SCF calculations. The straight line
 depicts the leading finite size behavior according to eq.\ (\ref{eqn:mcoex}), where we have used the {\em independently}
 determined interfacial tension\cite{M4} and have estimated the width $l$ of the A-rich layers as in section ({\bf A}).
 The phenomenological treatment
 describes the data very well. Only for the smallest system sizes there are some deviations, which are
 more pronounced for the simulation data. From the next-to-leading order corrections we can roughly estimate the
 spatial range of the effective interaction between the wall and the $AB$ interface, albeit with large uncertainties:
 $1/\lambda_{\rm MC}\approx 7$ and $1/\lambda_{\rm SCF} \approx 6.2$. Note that these length scales are much larger 
 than the range $d=2$ over which microscopic monomer-wall interactions $\epsilon_w$ are extended in our model.
 
 The increase of the width $l$ of the A-rich layers upon approaching the bulk coexistence chemical potential 
 is presented in Figure \ref{fig:kelvin} ({\bf b})  for the simulations and the SCF calculations.
 The behavior for large wall separations $D$ is well described by eq.\ (\ref{eqn:l}) and the
 slope of $l$ vs.\ $\ln (-\Delta \mu_{\rm coex})$ yields an estimate for the range $1/\lambda$ of the
 wall-interface potential. Again, we find deviations for small values of $D$ from the anticipated behavior.
 
 To explore the effect of the confinement in more detail, we investigate the profiles in the SCF
 framework. Figure \ref{fig:mfpot} displays the  composition profiles of the unconfined interface and the interface
 in a film of width $D=49$ on a logarithmic scale. The solid line  represents a tanh profile with width $w=6.91$
 which describes the SCF result at the center of the profiles. However, there are deviations in the wings of the profiles.
 There the length scale of the exponential decay is set by the correlation length $\xi$\cite{FRISCH}. 
 The interfacial profile
 in the thin film is narrower and decays more rapidly. To quantify this effect, we use the gradient
 of the composition profile and define an effective tension via
 $\Sigma/k_BT \sim \Phi b^2 \int dz \left(\frac{\partial c}{\partial z}\right)^2$. 
This expression yields the correct interfacial tension in the weak segregation limit, and we expect
to obtain qualitatively reasonable results also for the temperature studied here. The ratio between this
tension and the corresponding bulk quantity is shown in the inset as a function of the thickness of the
wetting layer $l$. The effective tension is clearly position dependent, and increases upon confinement.
The right inset presents the wall-interface potential g(l) as a function of the thickness of the wetting layer
$l$ for $D=49$. The data are describable by an exponential decay with length scale $1/\lambda=4.33$.
Using eq.\ (\ref{eqn:xi}) we extract a parallel correlation length  $\xi_{\|} \sqrt{\sigma_{AB}/\sigma} \approx 43$
($\sigma_{AB}$: bulk interfacial tension, $\sigma$ ``effective'' interfacial tension in the capillary fluctuation 
Hamiltonian).

As noted in Figure \ref{fig:prob} ({\bf a}), the composition fluctuations in the A-rich phase are larger than in the
phase without interfaces. This is caused by fluctuations of the average interfacial position $l$.
Assuming that the fluctuations at both walls are independent, we can relate the excess composition 
fluctuations to the fluctuations of the layer thickness:
\begin{equation}
\langle \Delta \rho^2\rangle_{\rm excess} = \frac{\chi_--\chi_+}{L^2D} = \frac{2}{D^2}(2\langle\rho\rangle-1)^2 \langle \delta l^2\rangle
\label{eqn:excess}
\end{equation}
where $\chi_\pm$ are the susceptibility in the A-rich and A-poor phases, respectively.
Using $L^2\langle \delta l^2 \rangle = \left(\frac{\partial^2 g}{\partial l^2}\right)^{-1}$ and
eq.\ (\ref{eqn:l}), we can estimate the interaction range $1/\lambda$ and parallel correlation length 
$\xi_{\|}$ from a single simulation via:
\begin{equation}
\frac{1}{\lambda} = \frac{\Phi}{2N(2\langle\rho\rangle-1)} D\Delta \mu_{\rm coex} \;DL^2\langle \Delta \rho^2\rangle_{\rm excess}
\qquad \mbox{and} \qquad \frac{\xi_{\|}^2}{\sigma}= \frac{\pi L^2 D^2}{(2\langle\rho\rangle-1)^2} 
\langle \Delta \rho^2\rangle_{\rm excess}
\end{equation}
The estimates for the parallel correlation length are compiled in Tab.\ \ref{tab:resmc}. $\xi_{\|}$ increases with
the film width $D$ and is always smaller than our lateral system extension $L$. However, this method entails
rather large uncertainties and a direct measurement\cite{AW} would yield more reliable estimates.

The dependence of the effective interaction range on the width of the film is presented in Figure \ref{fig:lambda}.
The data are compared with the local slope $dl/d \ln (-\Delta \mu_{\rm coex})$.
Both estimates are consistent, which reveals that the excess composition fluctuations are indeed due to
fluctuations of the average wetting layer thickness. Upon increasing the film width $D$,
the effective interaction range increases. Therefore, the wall-interface interaction is not a simple 
exponential repulsion. Assuming that the deviations from the large $D$ behavior are of the form 
$\ln D/D \sim l \exp(-\lambda l)$, the interaction range approaches a value 
$1/\lambda \approx 7.4(6) = 6.2(5) (1+\omega/2)$, where we have used the MC results for the
interfacial tension and the bulk correlation length (cf.\ Tab.\ \ref{tab:resmc}) at $k_BT/\epsilon=0.02$
to estimate $\omega \approx 0.382$. 
Assuming a $1/D$ correction (not shown), we obtain a slightly smaller value $\lambda \approx 6.8$ as 
estimate for large $D$. Thus $1/\lambda(1+\omega/2)$ is of the same order as the correlation length of 
composition fluctuations $\xi_{\rm MC}=6.21$ in the bulk, but somewhat smaller than the intrinsic interfacial 
width $w_{\rm SCF}/2 \approx 3.5$ (as extracted from the SCF calculations). Thus Figure \ref{fig:lambda} yields 
tentative evidence for the suggestion \cite{PARRY2} that $1/\lambda=\xi(1+\omega/2)$ indeed is relevant, corroborating
recent studies of Ising models\cite{BAGAIN}.

\subsection{Interfacial fluctuations}
The behavior of the fluctuating $AB$ interface, which is bound to the wall, can be described via an
effective Hamiltonian. Much theoretical investigations have been directed towards the detailed form
of this Hamiltonian\cite{PARRY2,FISHER}. In its simplest form (cf.\ eq.\ (\ref{eqn:helfrich})) this Hamiltonian 
comprises a contribution from the increase of the interfacial area due to deviations $\delta l$ of the local
interface position from its average value $\langle l \rangle$ and an effective wall-interface potential $g(l)$.
Since the $AB$ interface is distorted by the presence of the wall (cf.\ Figure \ref{fig:mfpot}), the range of 
the wall-interfacial interaction depends on the thickness of the wetting layer $\langle l \rangle$. A similar
dependence is expected for the effective interfacial tension $\sigma$\cite{PARRY2,FISHER}.

The effective interfacial tension $\sigma$ can be accurately measured in the MC simulations
via the spectrum of capillary fluctuations\cite{MW,M3}. Unlike previous studies\cite{MW,M3} we use an
integral criterion\cite{AW} for the local y-averaged interfacial position:
\begin{equation}
l(x) = \frac{\int_0^{D/2} dz \;\int dy \;\; \rho_A(x,y,z) }{\frac{2}{D} (2 \langle \rho\rangle -1) 
\int_0^{D/2} dz \;\int dy \;\; \left(\rho_A(x,y,z)+\rho_B(x,y,z)\right)} + \mbox {const}
\end{equation}
The local interfacial position can be Fourier decomposed according to
\begin{equation}
l(x) \sim  \frac{a_0}{2} + \sum_{k=0}^{L/2-1} \Big [ a(q_k) \cos(q_k x) + b(q_k) \sin(q_k x) \Big ]
\end {equation}
with $q_k = 2\pi k/L$. Using the equipartition theorem, we find that the Fourier amplitudes
are Gaussian distributed and their variances are given by eq.\ (\ref{eqn:fourier}). 
Previous simulations of unconfined interfaces\cite{MW,M3} have shown that the Fourier amplitudes are
indeed Gaussian distributed and that $\sigma$ can be identified with the interfacial tension 
$\sigma_{AB}$, which has been measured independently. In the present simulation
we verify this again for temperatures closer to the critical point. The full circles in 
Figure \ref{fig:cap} display the spectrum 
of interfacial fluctuations of the unconfined interface (using a local criterion for the
interfacial position\cite{M3}), whereas the dashed curve corresponds to the prediction
(\ref{eqn:fourier}) with $g=0$ and $\sigma=\sigma_{AB}$ as determined independently via
a reweighting technique.

The results for the confined system are presented in Figure \ref{fig:cap}. The open symbols
denote the simulation data, whereas the solid lines present linear regressions according to
eq.\ (\ref{eqn:fourier}). The fit yields the parallel correlation length $\xi_{\|}$ and the effective
interfacial tension $\sigma$. Unfortunately, the integral definition of the local interface position
is affected by bulk composition fluctuations, which result in an overestimation of the 
fluctuations of the local interfacial position.
However, assuming that these bulk composition fluctuations are laterally uncorrelated on length scales $2\pi/q$ larger 
than the bulk correlation length $\xi$, we can taken them into account by the substitution:
\begin{equation}
\left(\frac{\partial^2 g}{\partial l^2}\right)_{\rm eff}^{-1} \approx \left(\frac{\partial^2 g}{\partial l^2}\right)^{-1}+
\frac{D\chi_+}{2(2\langle \rho \rangle-1)^2}
\end{equation}
where $\chi_+$ is the susceptibility of the A-rich phase.
Using $\langle a_0^2\rangle = \frac{4}{L^2} (\frac{\partial^2 g}{\partial l^2})^{-1}$ and 
$a_0 = \frac{2}{L} \int dx\; l(x)$, we rederive eq.\ (\ref{eqn:excess}).
Thus, bulk composition fluctuations mainly influence our estimate of $\xi_{\|}$. However, we can still
use eq.\ (\ref{eqn:fourier}) to extract an effective interfacial tension $\sigma$ for $0< q < 2\pi / \xi$.

Most notably, the effective interfacial tension for width $D=48 \approx6.9 R_g$
is more than twice as large as in the unconfined system.
The thickness dependence of the effective interfacial tension is presented in
the inset of Figure \ref{fig:cap}. The additional contribution to the interfacial tension is
well describable by the predicted form\cite{PARRY2,FISHER} $\Delta \sigma \sim l \exp(-\lambda l)$ and the
fitting parameter $1/\lambda \approx 3.7$ is roughly compatible with the previous estimates for the
effective interaction range for small $l$. Moreover, this effect is also born out in the SCF field calculations.
As presented in Figure \ref{fig:profile}({\bf b}) and Figure \ref{fig:mfpot}, the interfacial width in the 
confined geometry is slightly smaller and the effective interfacial tension is larger than in the bulk.
The absolute magnitude of this effect in the SCF calculations is however somewhat smaller than in the 
MC simulations. 

\section{ Summary and discussion} 
We have presented a comparison between extensive MC simulations 
and SCF calculations on the phase behavior of a binary polymer blend which is confined 
into a film of width $D$. The hard walls at the film boundaries are ideally flat and preferentially absorb the 
A-component of the mixture via a short range interaction. Combining simulations in the semi-grandcanonical ensemble with 
reweighting techniques\cite{VALEAU,BERG},
we accurately determine the phase diagram of the confined blend and the shift of the coexistence chemical
potential. The critical point is shifted to lower temperatures and higher concentrations of the A-component. 
At the vicinity of the critical point 2D Ising-like critical behavior is observed in the MC simulations, 
in agreement with previous simulations by Rouault et al.\ \cite{YR1} on confined blends between ``neutral'' walls.
The binodals of the confined blend are asymmetric. The A-poor binodal is convex in an
intermediate temperature regime. This curvature is the signature of the wetting transition
in the semi-infinite system. Also the temperature dependence of the coexistence chemical potential changes
around the wetting temperature.

Using a novel extended ensemble which allows the monomer-wall interaction to fluctuate we accurately
locate the wetting transition in the MC simulations according to the Young equation.
Simulations and SCF calculations reveal that the film width $D$ 
gives rise to a pronounced rounding and shift of the wetting transition.
We find evidence for a first order wetting transition at rather low temperatures $T_{\rm wet} \approx 0.2 T_c$. 
This is also in agreement with our
SCF field calculations. In the limit of a strong first order transition the wetting
temperature is inversely proportional to the square of the integrated monomer-wall interaction 
strength and independent of the chain length $N$. We cannot rule out that critical wetting occurs 
in the ultimate vicinity of the critical point. However, for all parameters investigated, we find only 
first order wetting transitions in our simulations.

We determine the prewetting line, at which a thin and a thick wetting layer coexist, in the SCF calculations
and estimate its location in our MC simulations. We suggest that
recent experiments of Zhao and co-workers\cite{ZHAO} on the spinodal dewetting of ultrathin polymer films
can be analyzed in terms of spinodal phase separation into a thin and a thick wetting layer. The chain length
and temperature dependence of the prewetting is in agreement with the experimental observations. The results are
further illustrated by a MC study of the early stage of spinodal dewetting.

The dependence of the coexistence chemical potential on the film width is well describable by a Kelvin
equation\cite{PARRY} for large $D$. The leading behavior is quantitatively predicted by the phenomenological description.
The increase of the thickness $l$ of the wetting layer is compatible with a logarithmic dependence on the film
width or the coexistence chemical potential. Using the size dependence of the wetting layer thickness or
the excess composition fluctuations of the A-poor phase, we measure the effective wall-interface interaction
range $1/\lambda$.  Our results are compatible with the prediction of Ref.\ \cite{PARRY2} $1/\lambda = \xi(1+\omega/2)$, 
where $\xi$ is the bulk
correlation length which set the length scale in the wings of the interfacial profile\cite{FRISCH} and $\omega$ is
the wetting parameter.

The surfaces give rise to a pronounced orientation of the end-to-end vector. Both our SCF calculations, 
and MC simulations
show an increase of the lateral chain extension at the wall, whereas the z-extension vanishes at the wall.
The length scale of orientation is set by the radius of gyration. Even for large wall separations $D\approx 20 R_g$, the
conformations at the $AB$ interface are perturbed. We find that the interfacial width of the confined $AB$ interface
is smaller than in the bulk and approaches gradually the bulk value for very large $D$.

We determine the local (laterally resolved) position of the $AB$ interface bound to 
the wall. Analyzing the spectrum of capillary fluctuations\cite{MW,M3} of the bound interface, we determine the position 
dependence of the interfacial tension. The effective interfacial tension is compatible with the
predicted form $\sigma - \sigma_{AB} \sim l \exp(-\lambda l)$\cite{PARRY2,FISHER}. In the film of 
width $D=48 \approx 6.9 R_g$, the effective tension $\sigma$ exceeds the bulk value $\sigma_{AB}$ by more than a 
factor of 2.
This increase is due to a deformation of the interfacial profile near the wall, which is also observed in the 
SCF calculations.

The chain length dependence of our results is not addressed in the simulations; we
consider only polymers consisting of N=32 coarse grained monomers. This corresponds to
rather short polymers of about 150 chemical repeat units. Let us briefly comment on the behavior
of longer chains: The length scale of the wetting layer, the range of the effective wall-interfacial
interaction and the correlation length in the bulk is set by the radius of gyration $R_g \sim \sqrt{N}$.
The phase behavior in the bulk is characterized by the interaction strength per polymer $\chi N$.
However, if the preferential interactions of the monomers with the wall $\epsilon_w$ are chain length independent,
the wetting temperature is also independent of the molecular weight. Thus wetting in blends of high
molecular weight occurs far away from criticality and is generically of first order.

Our SCF calculations incorporates the detailed chain conformations on all length scales without any 
adjustable parameter. They qualitatively describe a variety of different MC results on polymer blends in thin films; 
for some properties nearly quantitative agreement is achieved. Moreover, the SCF calculations are at least an order of 
magnitude less computational intense than the MC simulations. They are useful to extend the MC results to larger film width, where we
find a triple point at which a thin and a thick wetting layer coexist with an $A$-rich bulk.
However, there are qualitative deviations in the behavior around the critical point, and the SCF calculations 
quantitatively underestimate the thickness of the wetting layer and overestimates the wetting transition.
We discuss them in turn:

{\bf A})
In the vicinity of the critical temperature, the mean field treatment fails to describe the
shape of the binodal which is characterized by Ising critical behavior. Thus the binodals are flatter in
the MC simulations than in the SCF calculations and the critical temperature is overestimated by the SCF calculations.
According to the Ginzburg criterion\cite{GINZBURG} we expect that in 3D the mean field behavior extends the closer to 
the critical point the longer the chain length is. Simulations\cite{M0} in 3D confirm 
that the Flory-Huggins theory describes the behavior correctly in the limit $N \to \infty$. 
Hence SCF calculations and MC simulations agree better for higher molecular weight blends.
We expect this also to hold true in the confined geometry. Though the system exhibits 
2D Ising critical behavior, the polymer conformations are not flat pancakes but interpenetrate and 
hence interact with many neighbors. The asymptotic critical behavior of Ising type is observed if the
correlation length $\xi$ exceeds a crossover length scale $\xi_{\rm cross} \sim N$ \cite{YR2}, and
depending whether $D>\xi_{\rm cross}$ or $D<\xi_{\rm cross}$ different scenarios apply.

{\bf B})
Moreover, the mean field treatment neglects capillary fluctuations. These lead to a pronounced
broadening of the apparent interfacial width\cite{AW} that persists even for very long chain lengths. They also
increase the effective wall-interface interaction range. Our present simulations and previous 
studies\cite{AW} are consistent with the prediction\cite{PARRY2} that the interaction range is 
amplified by a factor $(1+\omega/2)$, where $\omega=k_BT/4\pi\xi^2\sigma_{AB}$ denotes the wetting 
parameter. Thus the SCF calculation underestimates the thickness of the wetting layers, in agreement 
with our comparison. However, at constant $\chi N$, the interfacial
tension decreases like $1/\sqrt{N}$ whereas the correlation length $\xi$ is of the size of the polymer
extension $\sqrt{N}$. Thus the wetting parameter $\omega$ decreases like $1/\sqrt{N}$ at fixed $\chi N$
and we expect that the concomitant underestimation of the wetting layer thickness becomes smaller upon
increasing the chain length, at least for temperatures not too far from the critical point where the 
above formula for $\omega$ applies.

{\bf C})
Another important source of discrepancies between SCF calculations and the MC results in the long chain limit
is the treatment of the local structure of the fluid at the wall. Our SCF calculations incorporates the detailed
chain structure on all length scales, uses the same monomer-wall interactions as the simulations,
and accounts of the finite compressibility of the melt. 
In the present calculations, however, the reduction of the number of intermolecular interactions due to the presence 
of the wall (``missing neighbor effect'') is only qualitatively treated via a gradient expansion of the segmental
energy density\cite{HT}. The pronounced packing effects at the wall are not quantitatively captured by our SCF calculations. 
These effects plays a crucial role for a quantitative prediction of the wetting behavior 
(cf.\ eq.\ (\ref{eqn:SZW}) and (\ref{eqn:wettrans})). 
There are attempts to incorporate the local structure of the fluid at the wall into the SCF framework. 
Nath et al.\cite{NATH} incorporated fluid packing via the direct correlation function of the liquid. 
This quantity can be determined independently via P-RISM integral and/or density functional theory. 
In addition the polymers deform at the vicinity of the walls. The conformational changes alter the number of 
intermolecular and intramolecular contacts. These effects depend on the molecular architecture of 
the polymers and on the detailed structure (e.g.\ corrugation potential, microscopic roughness) of 
the wall. E.g.\ structural asymmetries between the components may give rise to entropic contributions to
the spreading parameter. We anticipate that these will be important for polymer-solvent systems in contact 
with a wall.  A proper treatment of these effects is out of the scope of the present study
which is focussed on the universal polymeric behavior of confined binary melts.

\subsection*{Acknowledgment}
It is a great pleasure to thank J.\ Baschnagel, F.\ Schmid, Y.\ Rouault, and A.\ Werner for 
helpful discussions. The generous allocation of about $10^5$ hours of single processor CPU 
time on the CRAY T3E at the HLRZ J\"ulich and the CONVEX SPP2000 at the computing center Mainz 
is grateful acknowledged. This work was supported by BMBF grant No.\ 03N8008C and DFG grant Bi314/17.

\section*{Appendix: Composition dependence of the free energy in the thin film near the wetting transition}

\begin{figure}[htbp]
    \begin{minipage}[t]{160mm}%
       \setlength{\epsfxsize}{7cm}
       \mbox{\epsffile{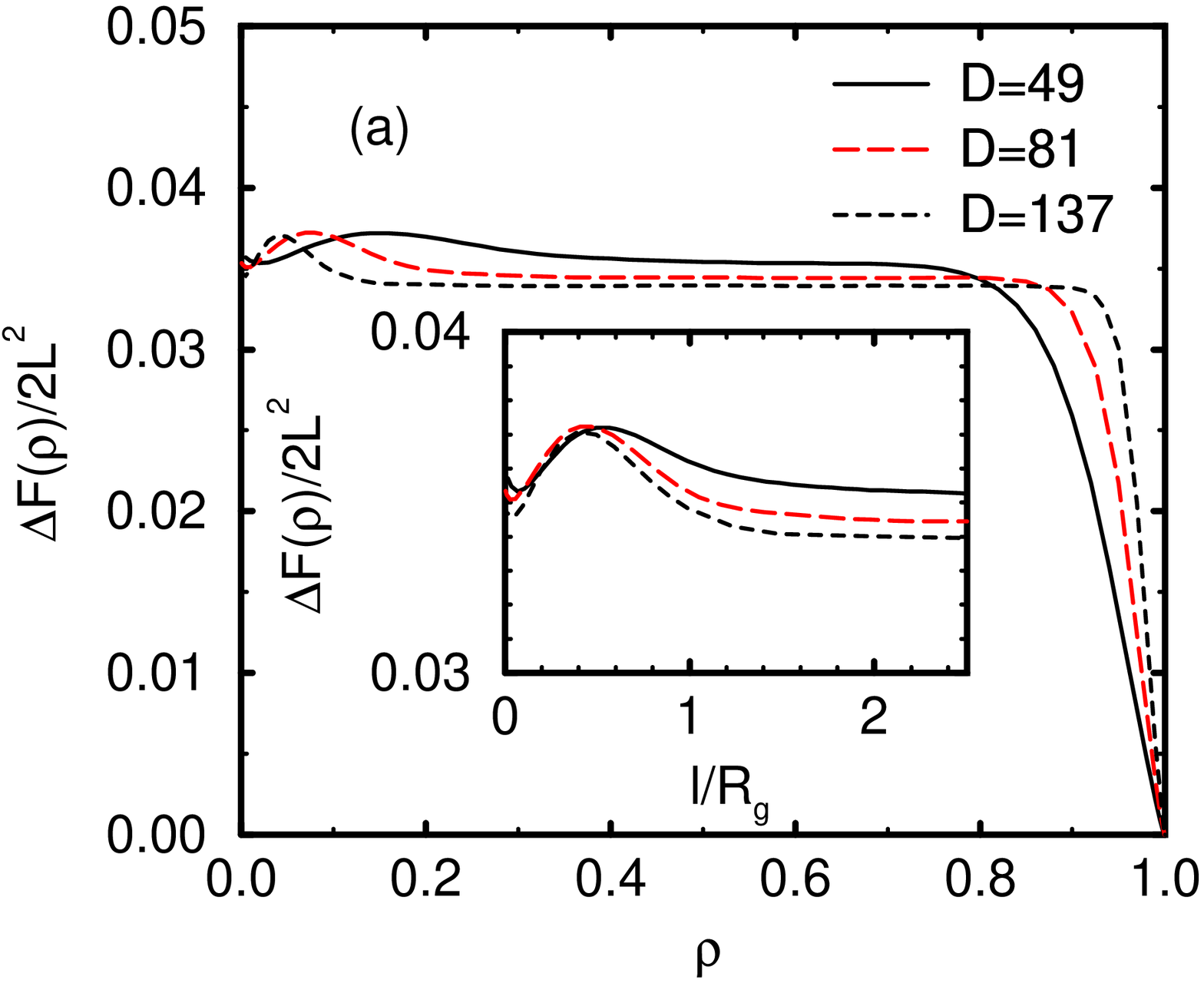}
       \setlength{\epsfxsize}{7cm}
       \hspace*{1cm}
       \epsffile{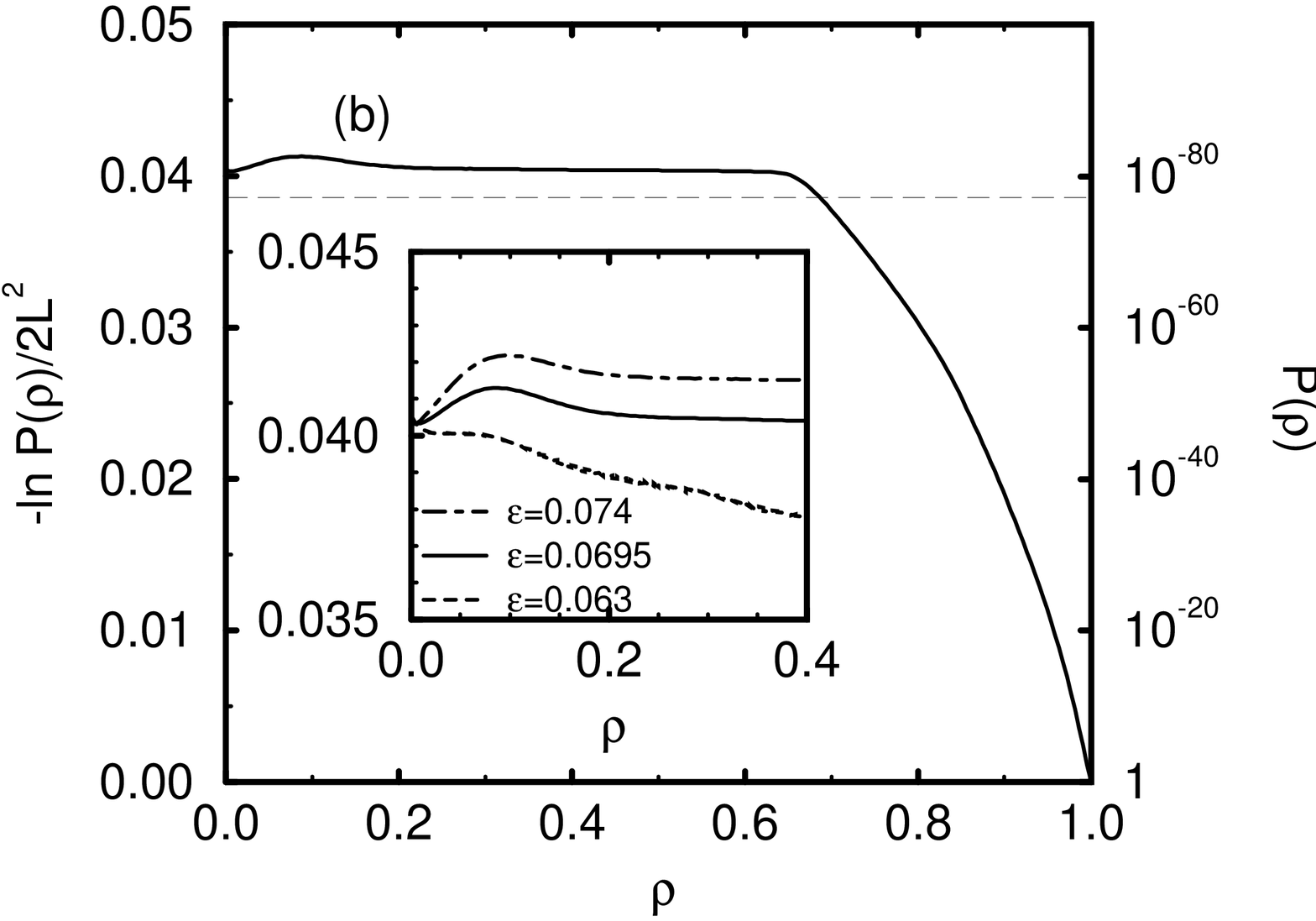}}
       \setlength{\epsfxsize}{7cm}
       \mbox{\epsffile{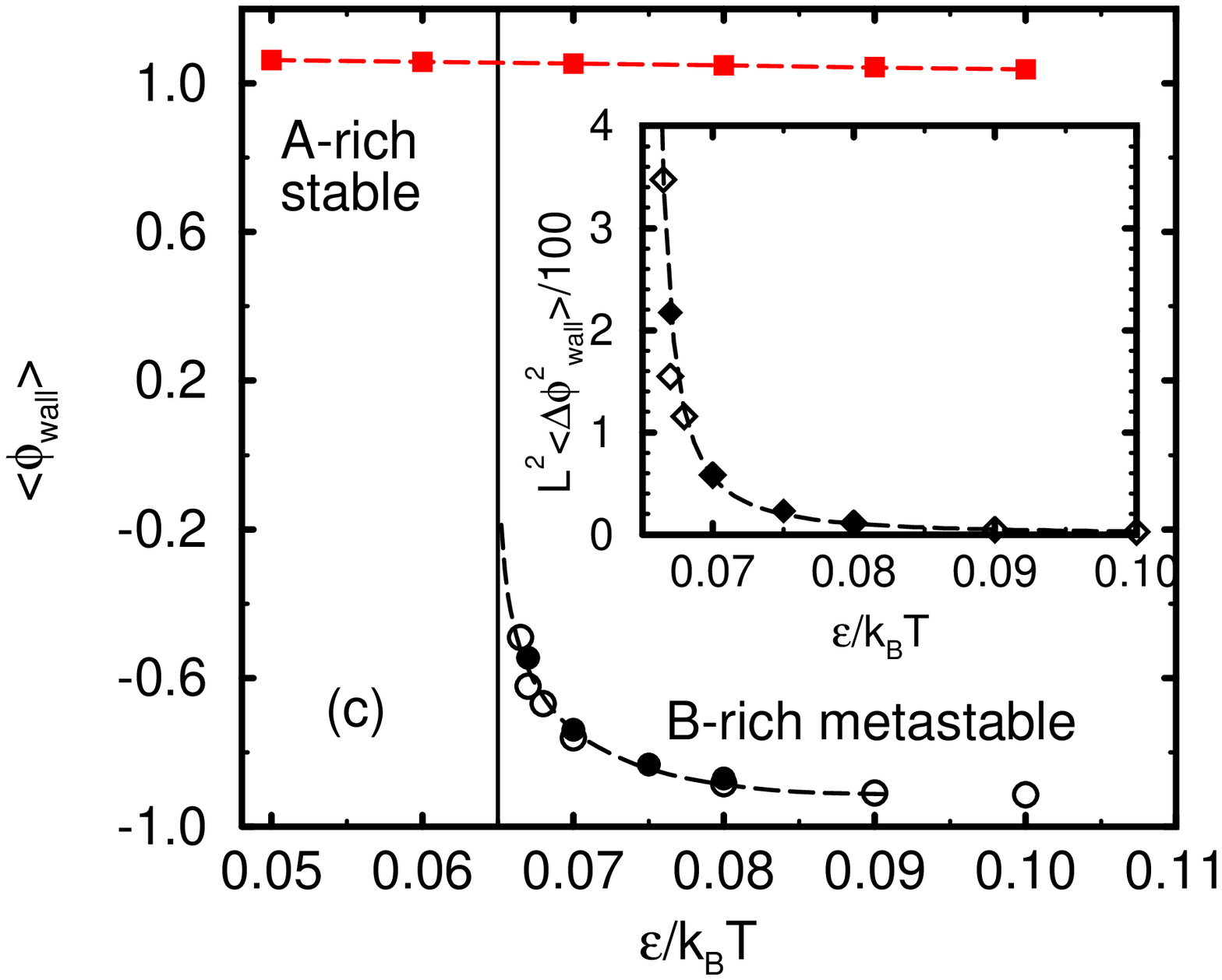}
       \setlength{\epsfxsize}{7cm}
       \hspace*{1cm}
       \epsffile{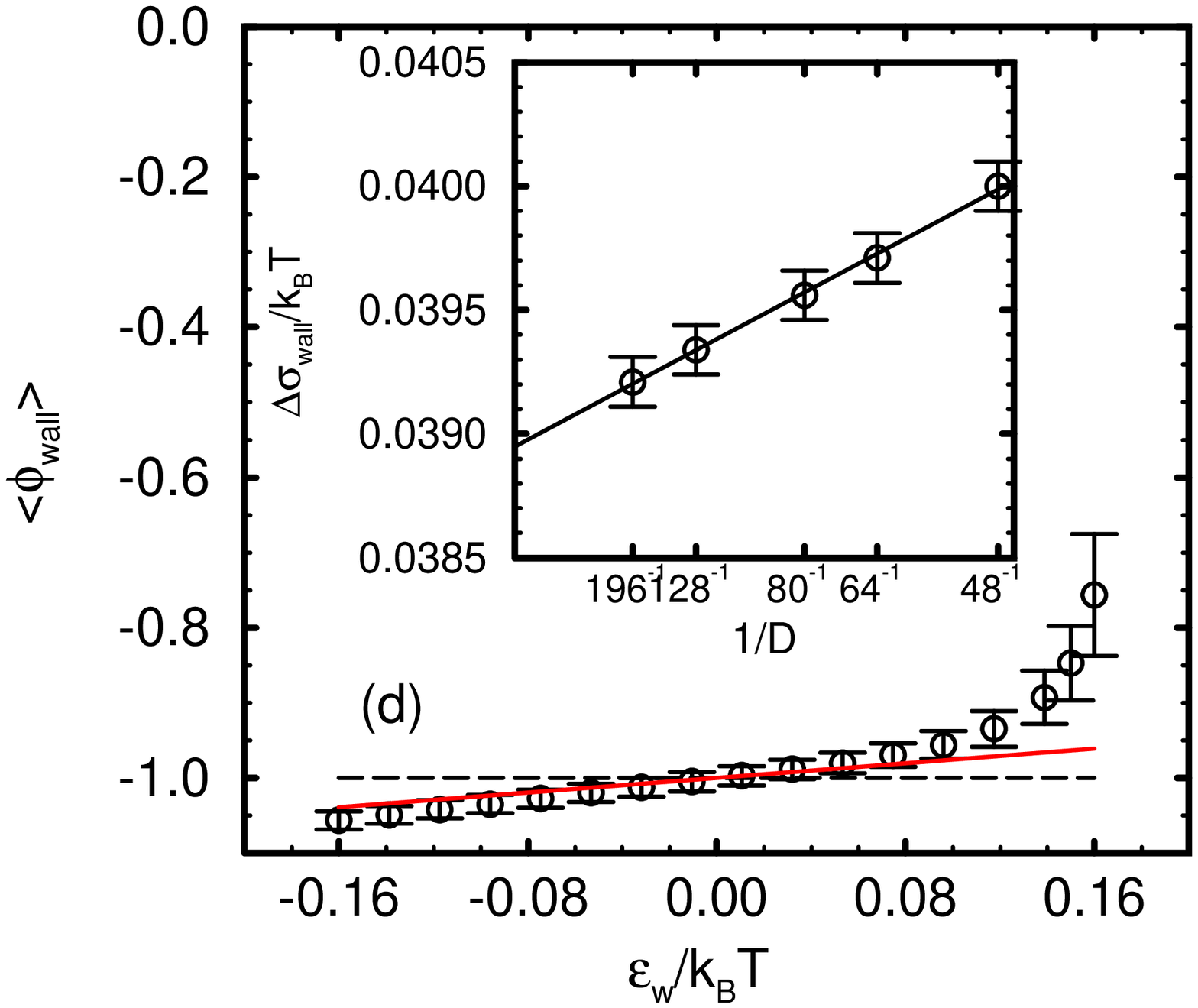}}
       \setlength{\epsfxsize}{7cm}
       \mbox{\epsffile{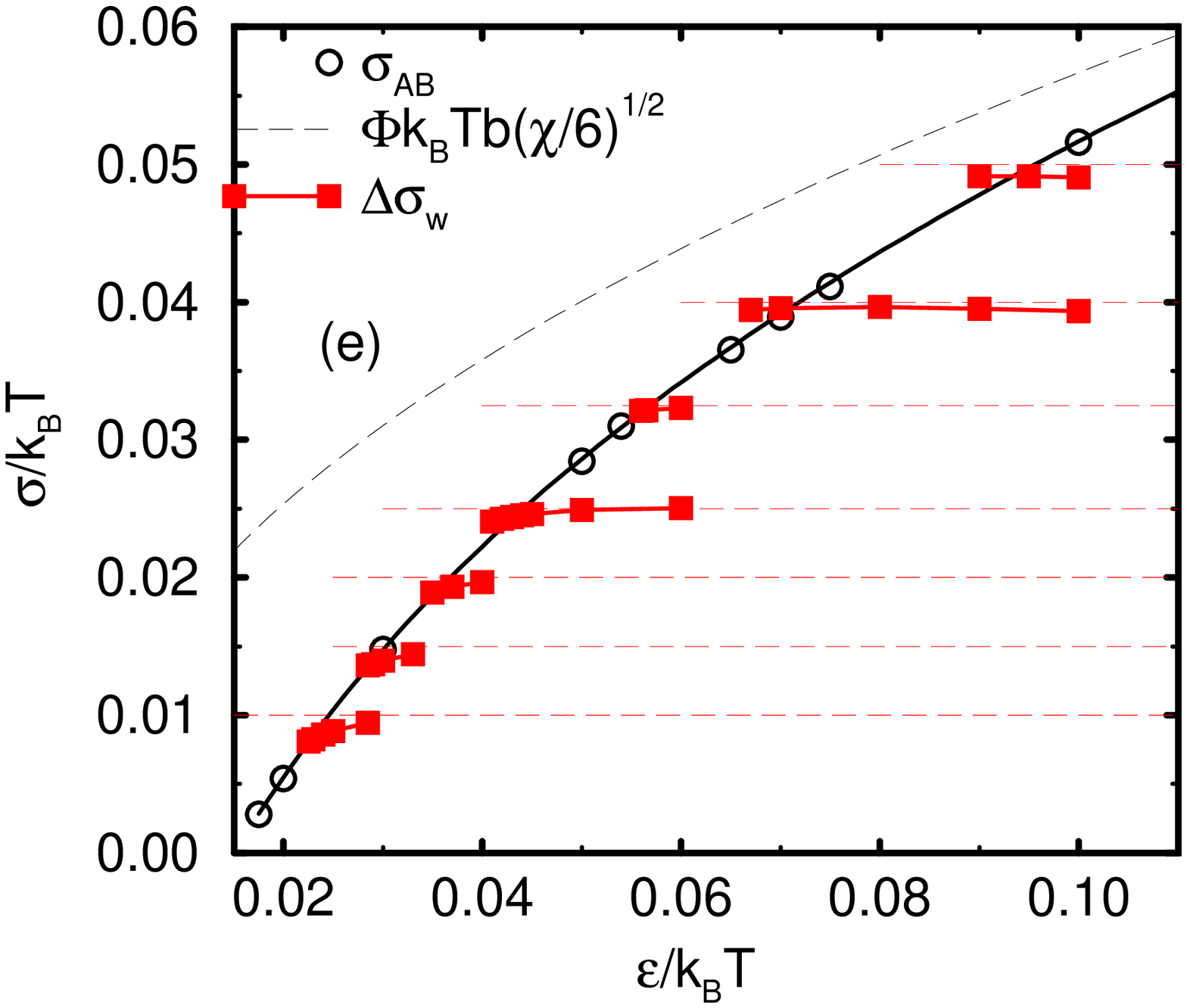}
       \setlength{\epsfxsize}{7cm}
       \hspace*{1cm}
       \epsffile{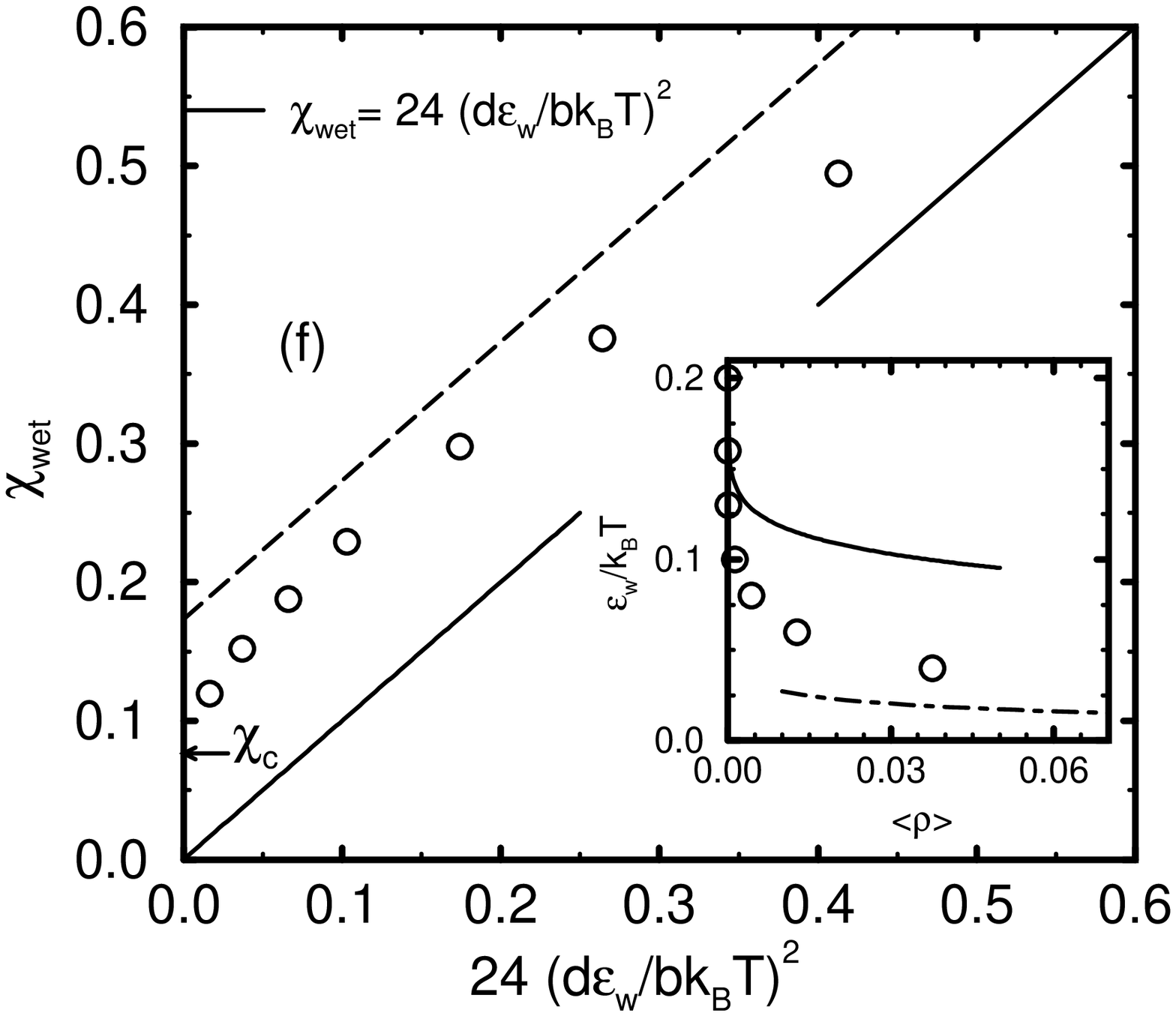}}
    \end{minipage}%
    \hfill%
    \begin{minipage}[b]{160mm}%
       \caption{Locating the wetting transition:\newline
       ({\bf a}) Mean field free energy at $\epsilon/k_BT=0.0575$ for different film thicknesses $D$.
	 	 The inset presents the same data vs the ratio of the thickness of the wetting layer $l = \rho D/2$
		 and the chain extension $R_g$. 
		 \newline
       ({\bf b}) Probability distribution of the composition  at $\epsilon/k_BT=0.0695$ in a cubic $48^3$ 
		 container. The horizontal dashed line marks the bulk interfacial tension $\sigma_{AB}=0.0387 k_BT$
		 at this temperature. The inset presents the probability distribution for small A contents at different 
		 temperatures.
		 \newline
       ({\bf c}) Surface order parameter $\phi_{\rm wall}$ and surface layer
                susceptibility as a function of temperature for fixed
		$\epsilon_{\rm w}=0.16$ at the bulk coexistence
		chemical potential difference $\Delta \mu=0$. Open
		symbols represent the simulation data for $L=96$ and
		$D=80$, whereas filled symbols refer to $L=D=128$.
		Lines are only guides to the eye. Around
		$\epsilon=0.065(3)$ we find evidence for a wetting
		{\em spinodal}.\newline 
	({\bf d}) Surface order parameter as a function of the monomer-wall interaction strength $\epsilon_w$
		at fixed $\epsilon/k_BT=0.07$. The bars do {\em not} represent statistical errors but the variance 
		of the distribution of $\phi_{\rm wall}$. The horizontal dashed line shows the naive estimates in the
		strong segregation limit, whereas the solid lines takes account of compressibility effects.
		The inset displays the dependence of $\Delta \sigma_{\rm wall}$ on the film width $D$.\newline
	({\bf e}) Interfacial tension $\sigma_{AB}$ and wall free energy difference $\Delta
		\sigma_{\rm w}$ as a function of the interaction strength $\epsilon$. Symbols 
		denote results of the Monte Carlo simulations. Horizontal dashed lines present
		our estimate in the strong segregation limit (SSL). The dashed line displays the 
		low temperature estimate of the interfacial tension, whereas the solid line
		$\sigma/k_BT = 0.1792 (\epsilon/k_BT)^{1/2} \left(1.126 - 0.0222/(\epsilon/k_BT) 
			      +8.04 \; 10^{-5}/(\epsilon/k_BT)^2 \right)$
		shows a fit to the MC data.\newline
	({\bf f}) Dependence of the inverse wetting temperature $\chi_{\rm wet}$
		on the monomer-wall interaction strength $\epsilon_w$. circles denote our MC estimates,
		the solid line shows the behavior in the strong segregation limit for infinite chain length,
		whereas the dashed line incorporates the effect of finite chain length on the bulk interfacial
		tension to first order in $1/\chi N$.
                }
       \label{fig:wet}
    \end{minipage}%
\end{figure}

\begin{figure}[htbp]
   \begin{minipage}[t]{160mm}%
       \setlength{\epsfxsize}{13cm}
       \epsffile{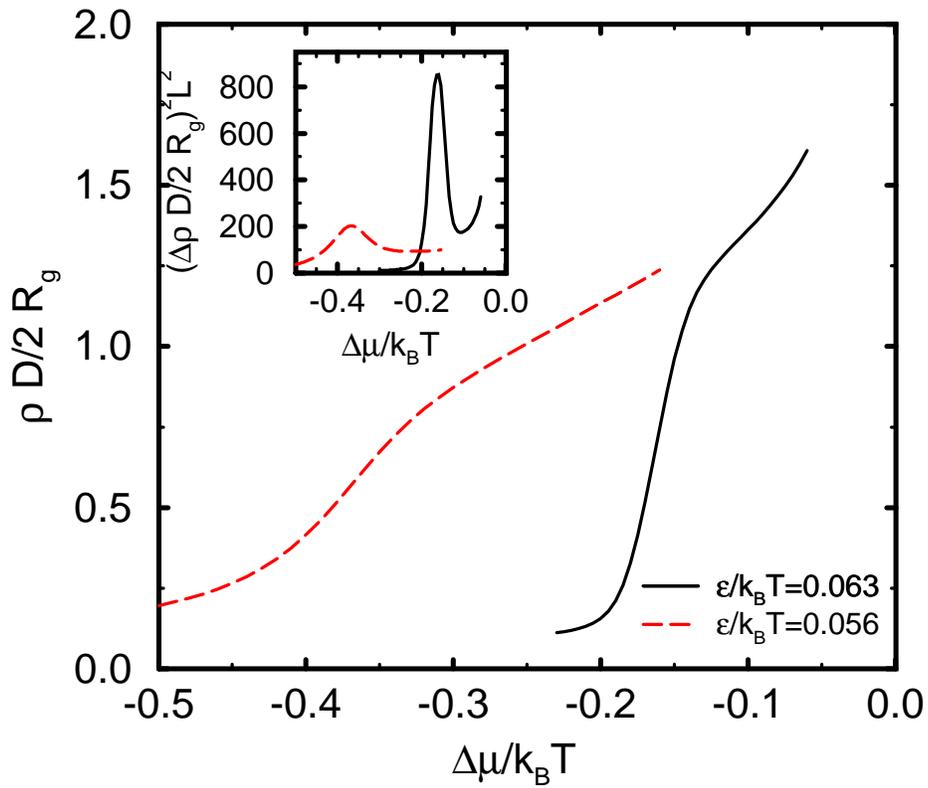}
    \end{minipage}%
    \hfill%
    \begin{minipage}[b]{160mm}%
       \caption{Adsorption isotherms slightly above the wetting transition temperature $\epsilon/k_BT_{\rm wet} =  0.071$. 
		The dependence of the layer thickness in units of the radius of gyration $\rho D/2 R_g$  exhibits
		a turning point, indicative of a first order transition rounded by finite size effects. The inset
		presents the thickness fluctuations. From the location of the peak we have estimated the location of 
		the prewetting line. Results are obtained via multihistogram analysis.
	       }
       \label{fig:pre}
    \end{minipage}%
\end{figure}

The free energy $F$ as a function of the composition $\rho$ is accessible in the simulations via the probability 
distribution $P(\rho)$: $F(\rho)/k_BT \sim -\ln( P(\rho))$. The MC results for system geometry
$48 \times 48 \times 48$ in the vicinity of the wetting transition of the semi-infinite system 
($\epsilon/k_BT=0.07$, $\epsilon_w/k_BT=0.0695$, $\Delta \mu=0$) are presented in Figure \ref{fig:wet}.
Unlike the SCF calculations, the composition in the simulation is not necessarily laterally homogeneous
nor symmetric around the middle of the film. Rather the system chooses a configuration which
minimizes the free energy at fixed overall composition. We can estimate the relative stability of 
the different conformations by some simple free energy considerations\cite{BREV,BERG2}.

If the lateral system extension $L$ is large, the systems will phase separate laterally into
A-rich and B-rich domains. Those domains are separated via two interfaces, which run perpendicular to the walls
For simplicity we approximate
the free energy costs of these perpendicular interfaces by $\sigma_{AB}(D-2l)L$ and assume that
$\Delta \sigma_{\rm w} \approx \sigma_{AB} \ll \gamma$, i.e.,\ the thickness of the wetting layer
$l$ is allowed to adjust as to minimize the overall free energy. 
The excess free energy of such a configuration is given by:
\begin{equation}
\Delta F = \min_{\alpha} \left(
\alpha 2 L^2 \sigma_{AB} + 2L(D-2l)\Delta \sigma_{\rm w} \right)
\qquad \mbox{with} \qquad
DL \frac{\langle\rho\rangle-\rho}{2\langle\rho\rangle-1} = \alpha L (D-2l)
\end{equation}
Minimization leads to $\Delta F = 4 L \sqrt{\sigma_{AB}\Delta \sigma_{\rm w} DL {[\langle\rho\rangle-\rho]}/{[2\langle\rho\rangle-1]} }$.
Comparing this result to the lateral homogeneous situation $\Delta F = 2L^2 \sigma_{AB}$ (independent of $\rho$), 
we find that for ${[\langle\rho\rangle-\rho]}/{[2\langle\rho\rangle-1]} < \Delta \sigma_{\rm w}L/4D \sigma_{AB}$ 
lateral phase separation occurs. Indeed for $\rho>0.7$ the excess free energy in the simulations decreases strongly
and upon approaching $\rho=1$ the two interfaces annihilate.

For small compositions $\rho$, no lateral phase separation occurs. However, unlike the situation in the
SCF calculations, the thickness of the wetting layers $l_r,l_l$ on both sides need not to be identical;
only the average thickness $\bar{l} = \frac{l_r+l_l}{2} = \frac{D(\rho-\langle\rho\rangle +1)}
{2(2\langle \rho \rangle-1)}$ is fixed by the constraint on $\rho$.
Let $g(l)L^2$ denote the potential of a single interface at a distance $l$ from the wall. Then
the excess free energy of the system with two interfaces is given by:
$\Delta F/L^2 = \min_{\Delta l} \left((g(\bar{l}+\Delta l)+g(\bar{l}-\Delta l)\right)$.
If the wetting transition is first order, than $g(l)$ has a minimum $g(l_0)=0$ at a finite distance
from the wall, is convex around its maximum of height $\gamma$, and  vanishes for large $l$. For 
$\bar{l}$ around the maximum, the system chooses rather $l_l \approx l_0$ and $l_r \approx 2\bar{l}-l_0$ 
than $l_l=l_r=\bar{l}$ which yields $\Delta F/2L^2 \approx g(l^*)/2$ with $l^*=2\bar{l}-l_0$\cite{COMMENT4}.
Therefore, the hump in the free energy corresponds to a conformation in which one interface is at 
$l_0$ (``dry'' state) and the height of the hump is $\gamma L^2$. If one allows for lateral variation of the wetting layer
thickness, the hump in the free energy can be reduced further to $O(1/L)$. We expect that the maximum of the free energy 
is shifted to smaller values of the composition and its height is at most $\gamma$, i.e., half the value in the
SCF calculations. From the probability distribution $P(\rho)$ we can estimate the free energy $\gamma$, which controls
the lifetime of the ``dry'' state. Using the simulation data in Figure \ref{fig:wet}, we obtain the rough estimate
$\gamma/k_BT > 0.0019(10)$ at $\epsilon/k_BT=0.0695$. Gratifyingly, this is of the same order of magnitude as the 
mean field result $\gamma/k_BT = 0.0027(8)$ at $\epsilon/k_BT=0.0575$

\pagestyle{empty}

\begin{table}[htbp]
\begin{tabular}{|c|c|c|c|c|c|c|c|c|}
\hline
 & $k_BT_c^{\rm bulk}/\epsilon$& $k_BT_c/\epsilon$ & $k_BT_{\rm wet}/\epsilon$ & $\rho_c$ & $\sigma(k_BT/\epsilon=50)/k_BT$ & $\rho(k_BT/\epsilon=50)$ & $w(k_BT/\epsilon=50)$ & $\xi(k_BT/\epsilon=50)$\\
\hline
\hline
MC  & 69.3(3) & 55.5(6) & 14.1(7)  &  0.686(5)  &  0.0054(1)  &  0.9369 &  10.71  & 6.21    \\
SCF & 84.8    & 76.7    & 17.2     &  0.69      &  0.0075     &  0.9569 &  6.91   & 4.72    \\  
\end{tabular}
\caption{
Bulk properties as measured in the Monte Carlo simulations and
self-consistent field calculations.  Length scales are measured in
units of the lattice spacing $u$, and the interfacial tension in units
of $k_BT/u^2$. The interfacial width in the Monte Carlo simulations
includes additional broadening due to capillary waves ($L=64$), whereas
the SCF result refers to the intrinsic width. The mean field correlation length
is given by $\xi = \frac{R_g}{\left(3(1-2\chi N \rho(1-\rho)\right)^{1/2}}$ with $\chi=2 {\tt z} \epsilon/k_BT$
} 
\label{tab:resbulk}
\end{table}

\begin{table}[htbp]
\begin{tabular}{|c|c|c|c|c|c|c|}
\hline
$D$   &   $\Delta \mu/k_BT$   &  $\rho_-$ (B-rich) & $\rho_+$ (A-rich) & $l$ & $\xi_{\|}(\sigma_{AB}/\sigma)^{1/2}$ & $\sigma/k_BT$ \\
\hline
\hline
36    &  -0.6575  & 0.5096  & 0.8605 &  9.20 & 11.3 & \\
48    &  -0.3944  & 0.4376  & 0.9185 & 10.29 & 14.3 & 0.0133(4) \\
48$^*$&  -0.3937  & 0.4379  & 0.9187 & 10.29 &      & \\
48$^{\dag}$ & -0.3938 & 0.4421 & 0.9143 & 10.41 &   &\\
60    &  -0.2682  & 0.4097  & 0.9301 & 11.90 & 18.0 & 0.0115(2)   \\
80    &  -0.1670  & 0.3747  & 0.9359 & 14.26 & 24.2 & 0.0094(2)\\
112   &  -0.0996  & 0.3300  & 0.9381 & 17.10 & 33.3 & 0.0074(1) \\
$\infty$ & 0      & 0.0631  & 0.9369 & $\infty$ & $\infty$ & 0.0054(1) \\
\end{tabular}
\caption{Simulation data for $\epsilon=0.02,\epsilon_{\rm w}=0.16$ and lateral system size $L=96$ (if not
indicated otherwise: ${}^*$ $L=160$,   ${}^{\dag}$ $L_x=164 \times L_y=48 \times D=48$).
The errors for the coexistence value of the chemical potential difference $\Delta \mu$ are less than 1$\%$,
the uncertainties for the coexistence composition are $2\%$ for the B-rich phase and $1\%$ for the A-rich phase.
The errors of the effective interfacial tension are extracted from the
regression of the simulation data in Figure 12.
        }
\label{tab:resmc}
\end{table}

\begin{table}[htbp]
\begin{tabular}{|c|c|c|c|c|}
\hline
$D$   &   $\Delta \mu/k_BT$   &  $\rho_-$ (B-rich) & $\rho_+$ (A-rich) & $l$\\
\hline
\hline
37    &  -0.6182  & 0.381  & 0.948 &  6.84 \\
49    &  -0.3875  & 0.356  & 0.954 &  8.39 \\
61    &  -0.2764  & 0.338  & 0.956 &  9.84 \\
81    &  -0.1803  & 0.302  & 0.958 & 11.47 \\
113   &  -0.1134  & 0.258  & 0.958 & 13.29 \\
137   &  -0.0874  & 0.233  & 0.957 & 14.22 \\
\end{tabular}
\caption{Results of the self-consistent field calculations for the same interaction strength $\epsilon/k_BT=0.02, \epsilon_{\rm w}/k_BT=0.16$ as
the Monte Carlo simulation.
        }
\label{tab:resscf}
\end{table}

\newpage
\vfill

\begin{figure}[htbp]
    \begin{minipage}[t]{160mm}%
    figures available upon request
    \end{minipage}%
    \hfill%
    \begin{minipage}[b]{160mm}%
     \caption{
                Configuration snapshots of dewetting in thin polymer layers.
		The left row of snapshots displays the spinodal dewetting 
		process in an ultrathin polymer layer above the first order
		wetting transition ($\epsilon/k_BT=0.0695$). The middle row presents the behavior of a thicker
		layer at the same temperature; the layer remains stable.
		The right sequence shows the behavior of the thin layer above the
		prewetting critical temperature ($\epsilon/k_BT=0.03$). The layer does not dewet; however 
		there are strong thermal fluctuations.
                }
       \label{fig:dewet1}
    \end{minipage}%
\end{figure}

\newpage

\begin{figure}[htbp]
   \begin{minipage}[t]{160mm}%
       \setlength{\epsfxsize}{8cm}
       \mbox{
       \epsffile{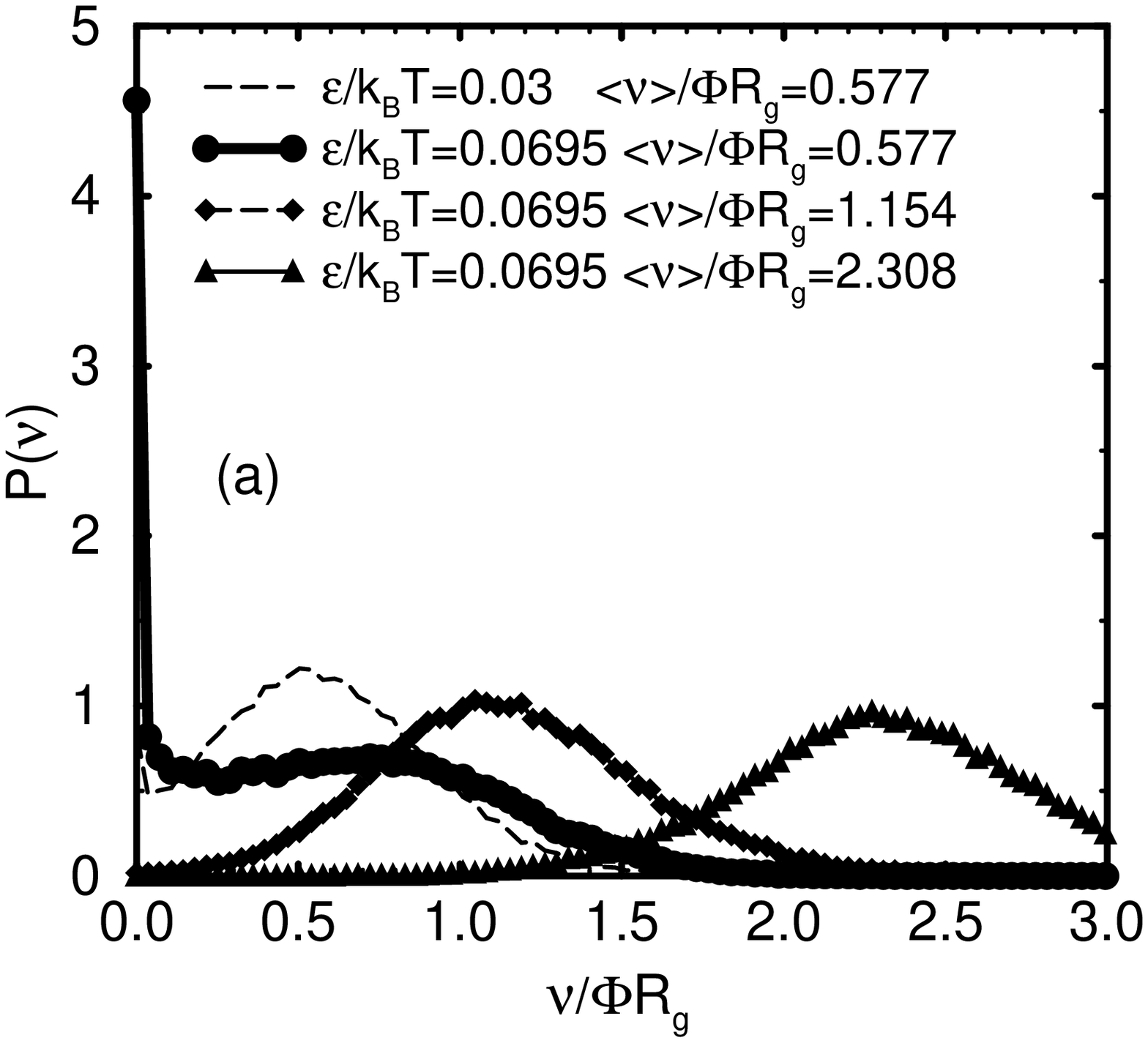}
       \hspace*{1cm}
       \setlength{\epsfxsize}{8cm}
       \epsffile{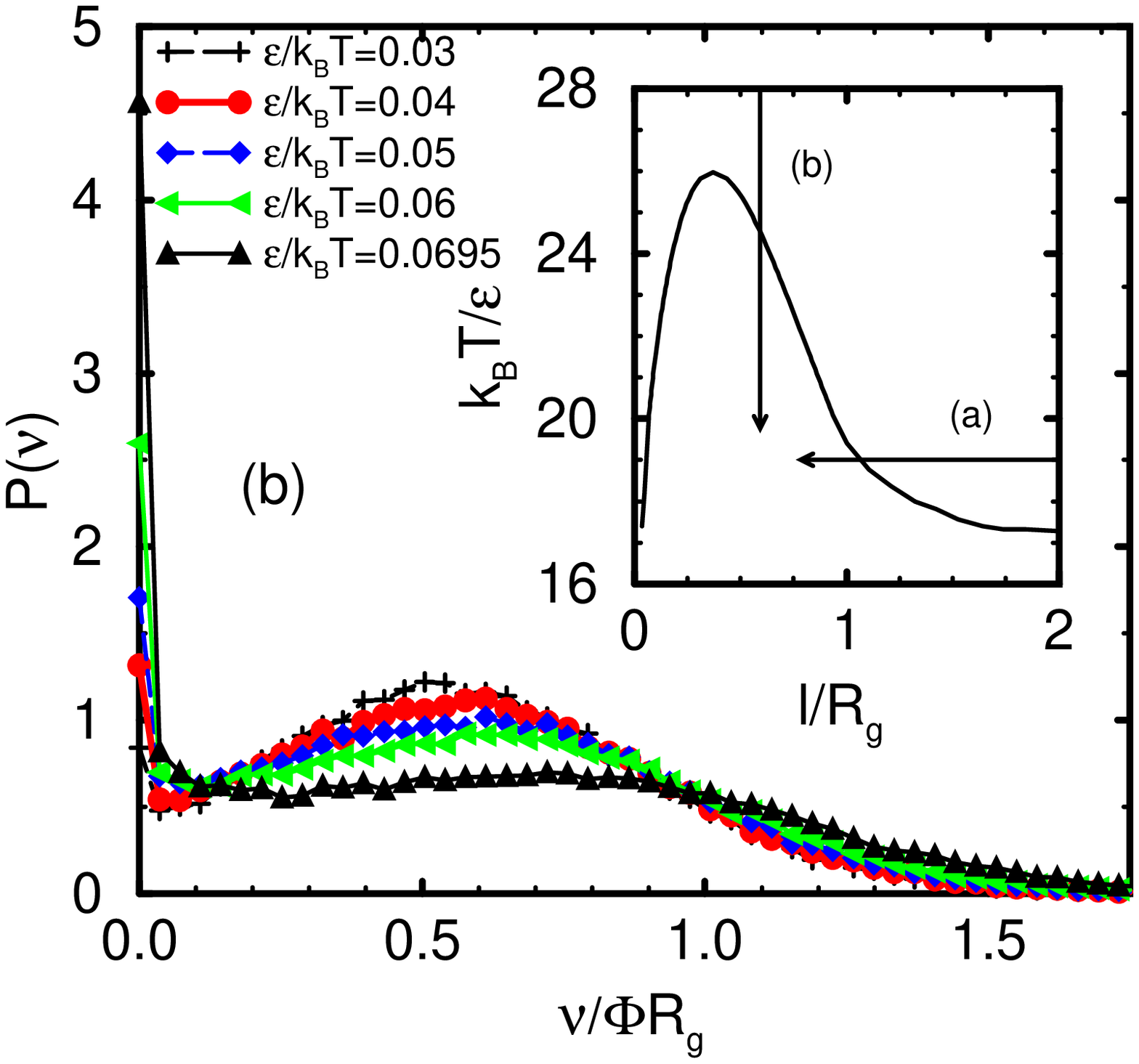}
       }
    \end{minipage}%
    \hfill%
    \begin{minipage}[b]{160mm}%
       \caption{ Subbox analysis of the local $A$-monomer density distribution:\newline
		({\bf a}) Dependence on the layer thickness above the wetting
			  transition ($\epsilon/k_BT=0.0695$). Upon reducing the layer thickness the distribution
			  becomes bimodal; this indicates prewetting. The distribution of a thin film above
			  the prewetting critical point ($\epsilon/k_BT=0.03$) is also shown for comparison.
			  \newline
                ({\bf b}) Temperature dependence of the local lateral A-monomer density.
			  At high temperature the distribution is nearly Gaussian; whereas 
			  it becomes bimodal at low temperatures but above the wetting transition.\newline
			  The inset illustrates the different path to coexistence in ({\bf a}) and ({\bf b}).
			  The solid curve, depicting the thickness of the coexisting layers, is
			  obtained from the SCF calculations.
                }
       \label{fig:dewet2}
    \end{minipage}%
\end{figure}

\begin{figure}[htbp]
    \begin{minipage}[t]{160mm}%
       \setlength{\epsfxsize}{7cm}
       \mbox{\epsffile{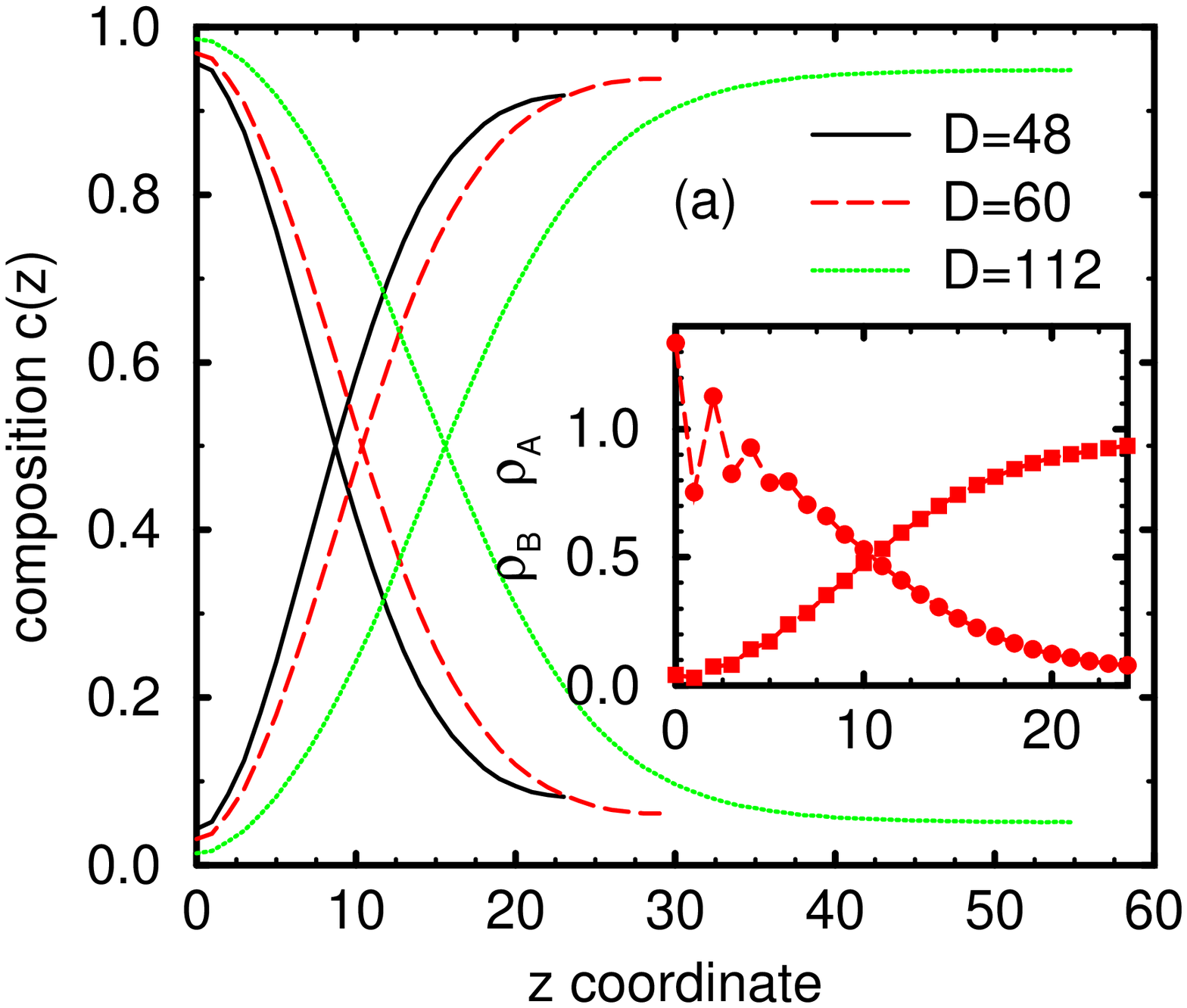}
       \setlength{\epsfxsize}{7cm}
       \epsffile{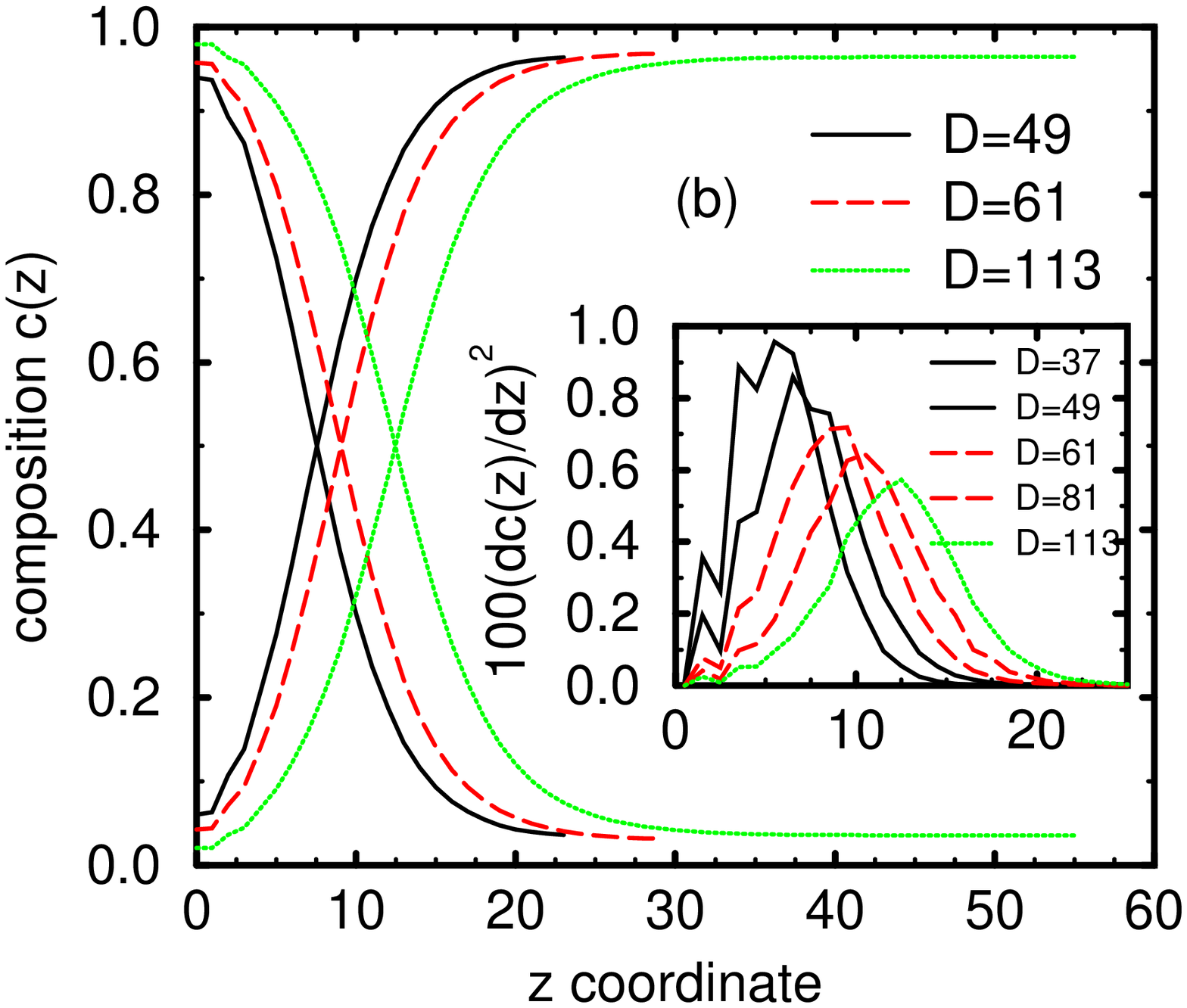}}
       \setlength{\epsfxsize}{7cm}
       \mbox{\epsffile{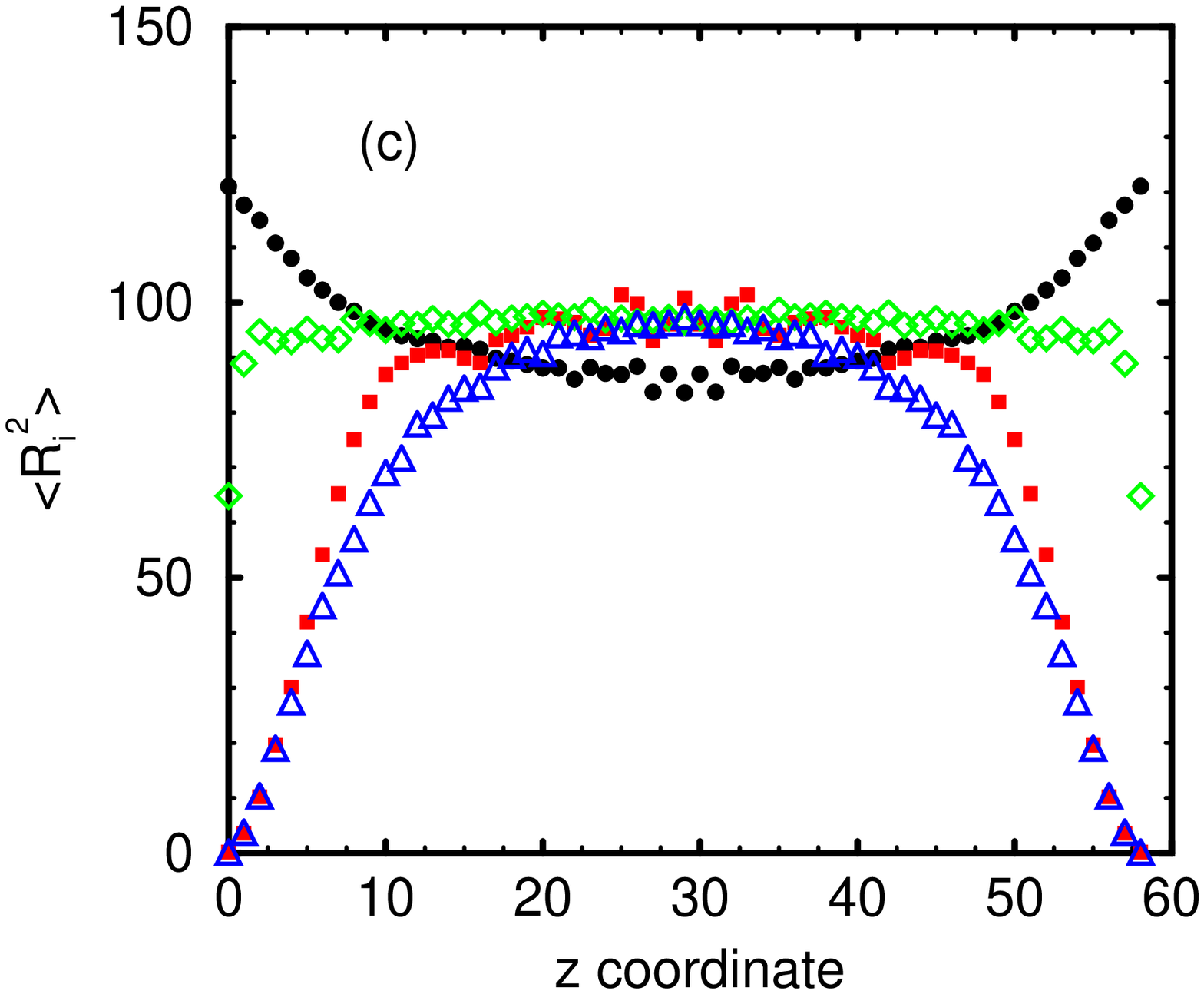}
       \setlength{\epsfxsize}{7cm}
       \epsffile{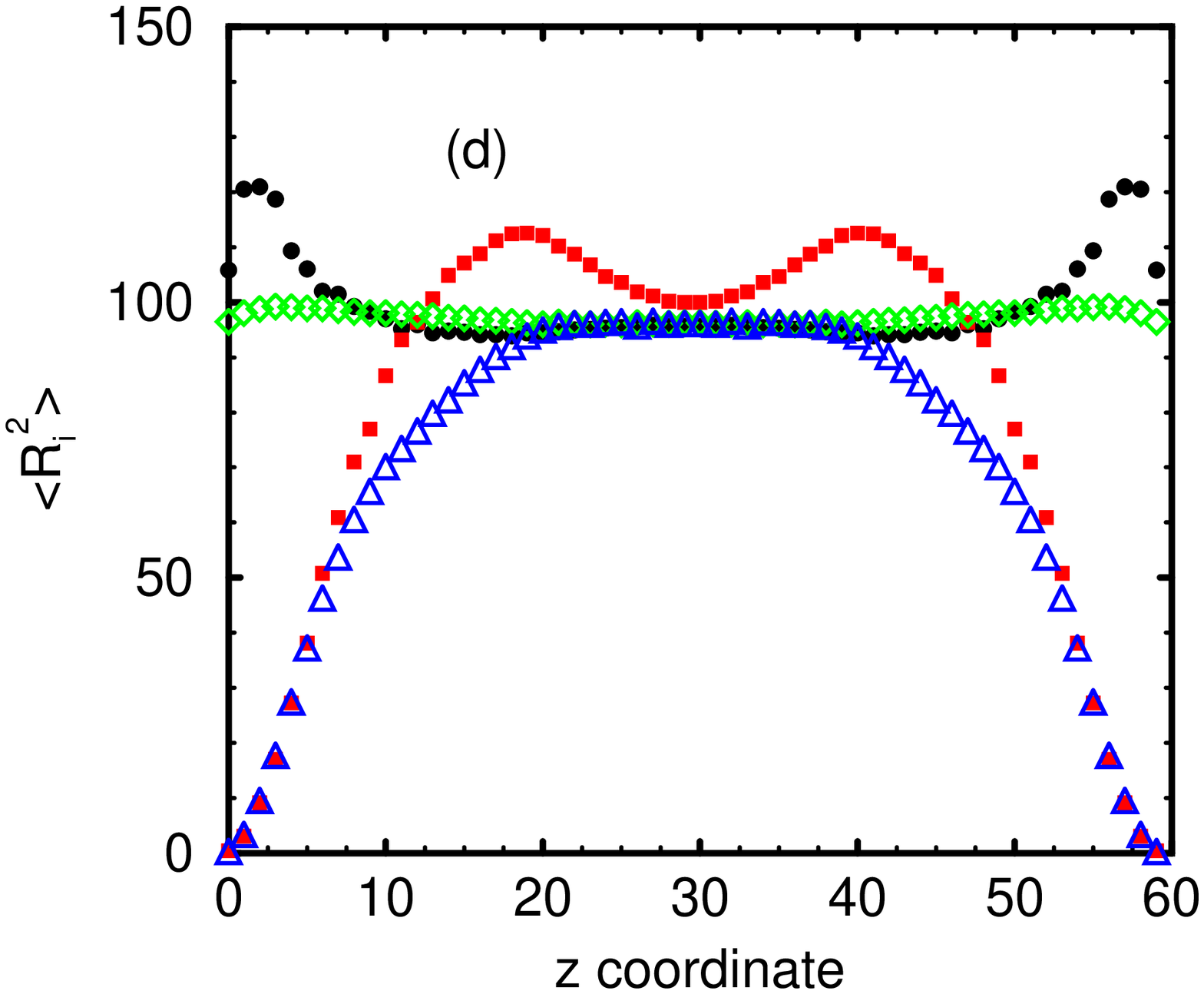}}
       \mbox{\epsffile{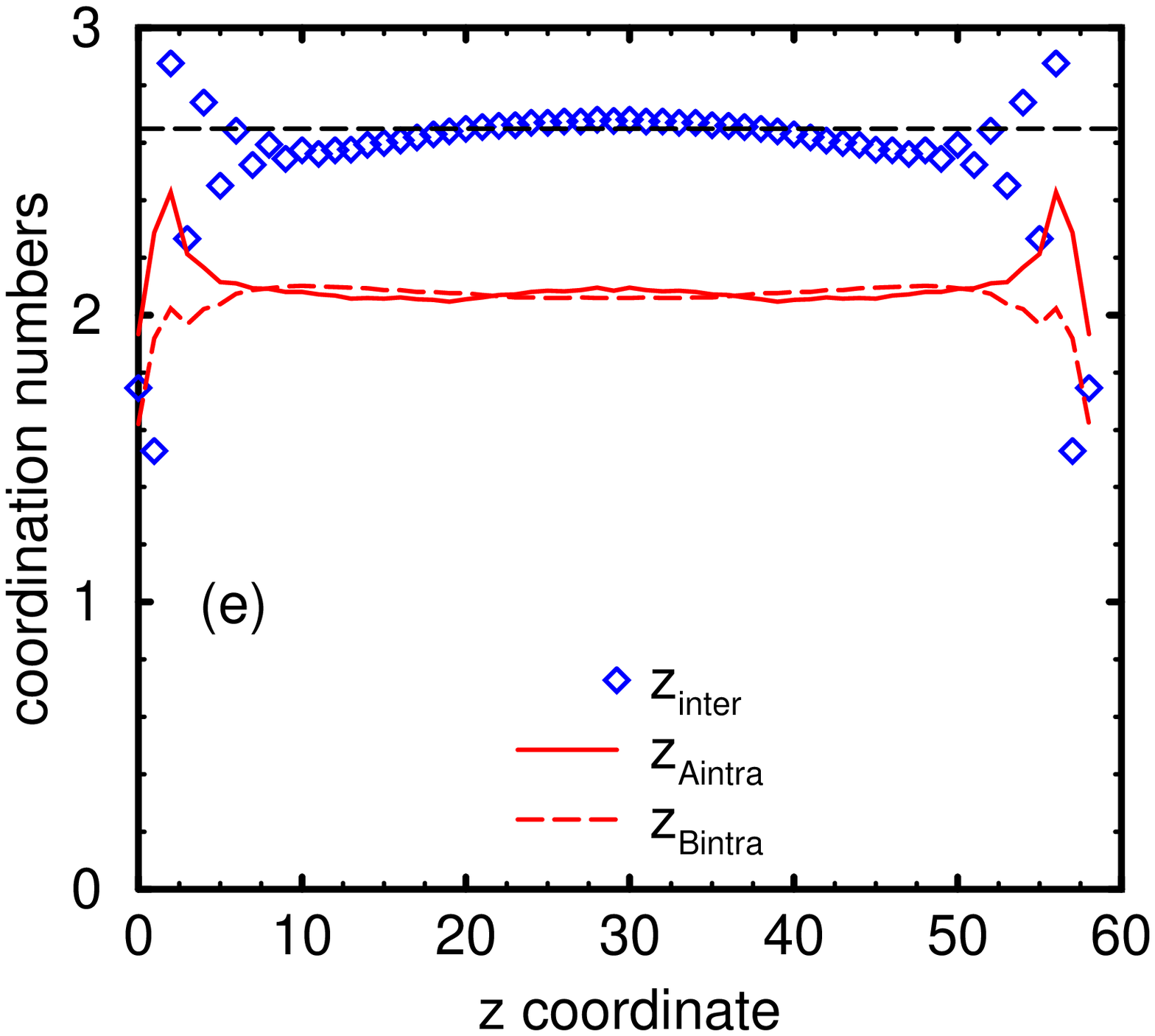}
       \setlength{\epsfxsize}{7cm}
       \epsffile{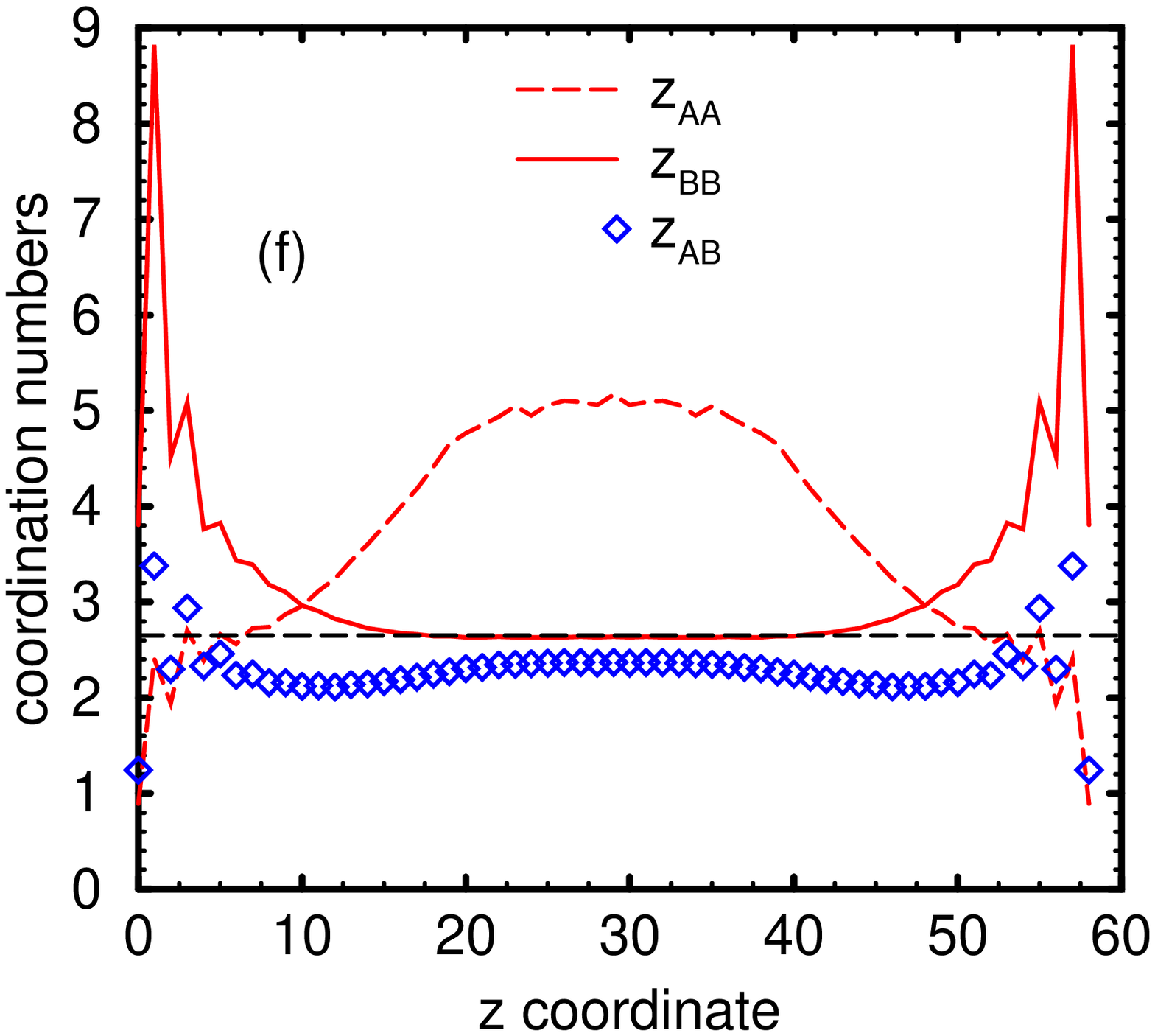}}
    \end{minipage}%
    \hfill%
    \begin{minipage}[b]{160mm}%
       \caption{
       Profiles across the film:\newline
       ({\bf a}) Monte Carlo results for the composition profiles $c(z)=\rho_A(z)/(\rho_A(z)+\rho_B(z))$
		 at $\epsilon/k_BT=0.02$, $\epsilon_{\rm w}/k_BT=0.16$ 
		 and thickness $D$ as indicated in the figure. The inset shows the A- and B-monomer densities near the 
		 wall for $D=60$.\newline
       ({\bf b}) Composition profiles at the interaction as in ({\bf a}) calculated in the self-consistent field theory.
		 The inset displays the squared composition gradient.
		 \newline
       ({\bf c}) Chain extension for $D=60$ as function of the distance from the wall. Filled circles represent 
		 $\langle R_{\|}^2\rangle$ of A-chains, filled squares denote $\langle R_{\perp}^2\rangle$ of A-chains,
		 and diamonds, triangles correspond to $\langle R_{\|}^2\rangle$, $\langle R_{\perp}^2\rangle$ of B-chains
		 respectively.\newline
       ({\bf d}) Chain extensions for $D=61$ in the mean field calculations. Symbols are the same as in ({\bf c}).\newline
       ({\bf e}) MC results for the total number of intermolecular contacts and intrachain contacts as a function of the 
                 distance from the wall ($D=60$) \newline
       ({\bf f}) Intermolecular contacts between like and unlike monomer species as a function of the distance from 
		 the wall ($D=60$) in the MC simulations. Rather pronounced non-random mixing effects are observed.
                }
       \label{fig:profile}
    \end{minipage}%
\end{figure}

\begin{figure}[htbp]
    \begin{minipage}[t]{160mm}%
       \setlength{\epsfxsize}{8cm}
       \mbox{\epsffile{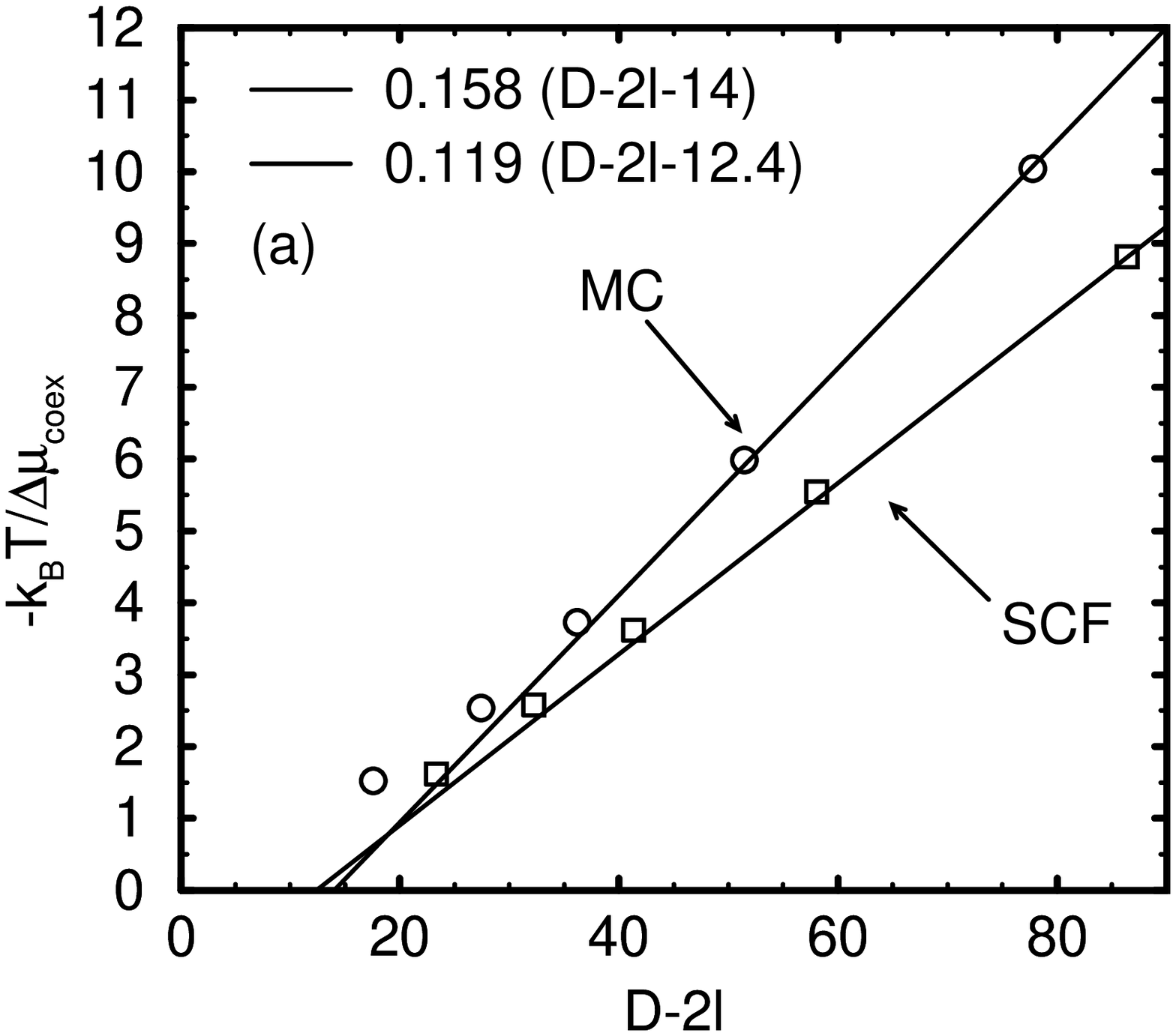}
       \setlength{\epsfxsize}{8cm}
       \epsffile{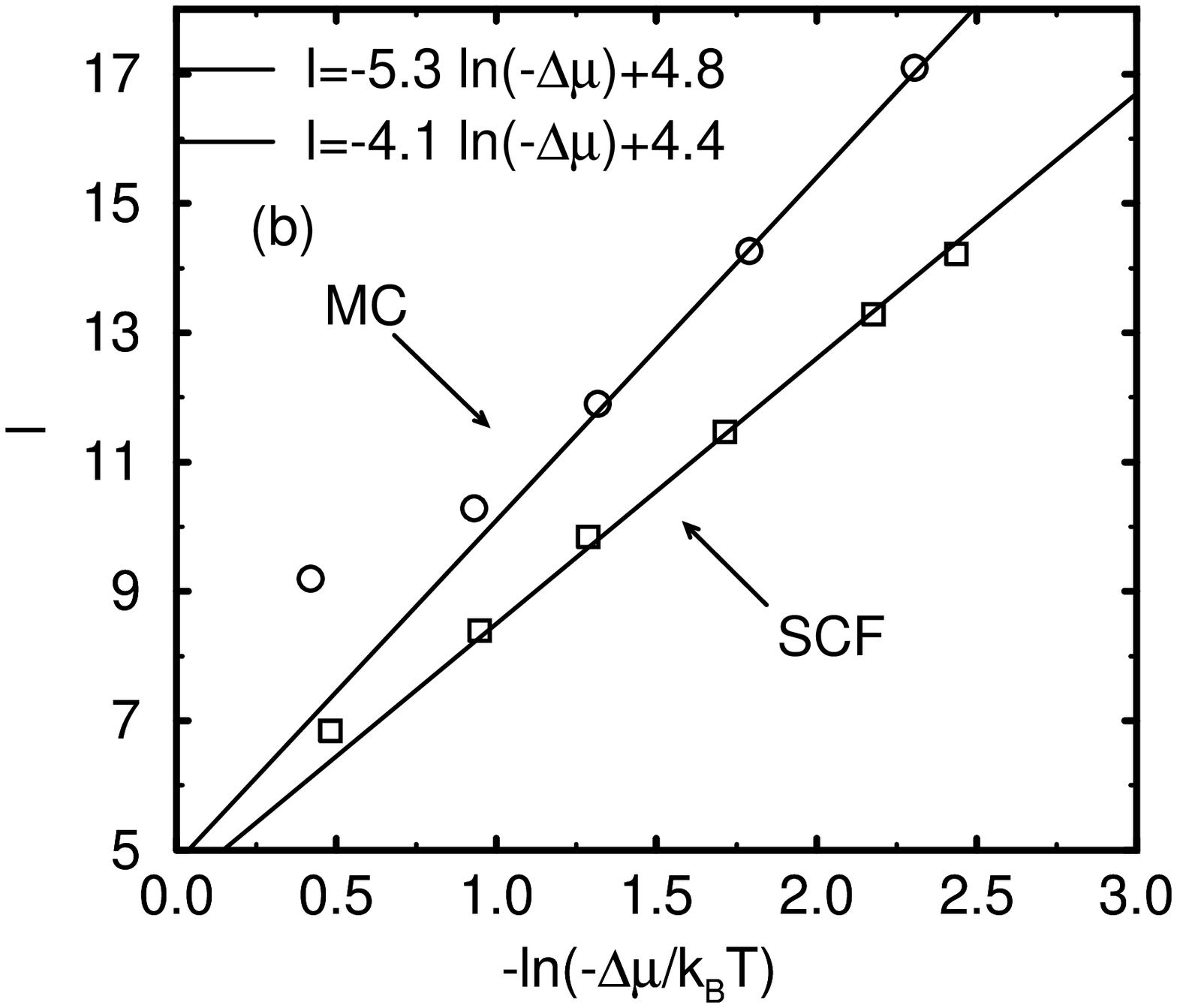}}
    \end{minipage}%
    \hfill%
    \begin{minipage}[b]{160mm}%
       \caption{
       Kelvin equation and thickness of the wetting layer:\newline
       ({\bf a}) MC (circles) and SCF (squares) results for the coexistence chemical potential difference $\Delta \mu$ at 
		 $\epsilon=0.02$, $\epsilon_{\rm w}=0.16$. The solid lines correspond to eq.\ (5), where the slope is
		 fixed by the independently measured interfacial tension.\newline
       ({\bf b}) Thickness $l$ of the wetting layer in the MC simulations (circles) and SCF calculations (squares). 
		 The slope is a measure for the wall-interface interaction range $1/\lambda$.\newline
                }
       \label{fig:kelvin}
    \end{minipage}%
\end{figure}

\begin{figure}[htbp]
    \begin{minipage}[t]{160mm}%
       \setlength{\epsfxsize}{13cm}
       \mbox{\epsffile{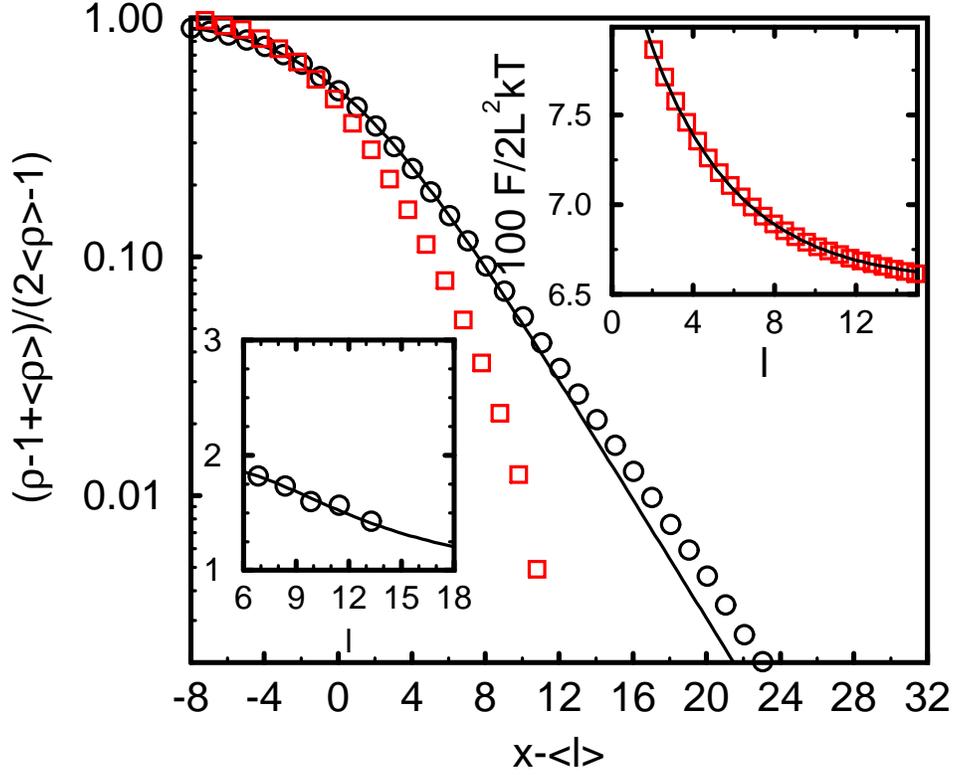}}
    \end{minipage}%
    \hfill%
    \begin{minipage}[b]{160mm}%
       \caption{Interfacial profiles in the self-consistent field calculation:
                Circles denote the interfacial profile in the bulk ($64 \times 64 \times 128$ 
		geometry and periodic boundary conditions), and the
		solid line presents a tanh-profile with width $w=6.91$. There
		are deviations in the wings of the profile. Squares represent
		the profile in the confined system $D=49$.\newline
                The right inset presents the 
		effective interface-wall potential in the self-consistent field calculations for $D=49$, 
		$\epsilon/k_BT=0.02$ and $\epsilon_{\rm w}/k_BT=0.16$. The solid line is a fit with an exponential 
		decaying interaction $g(l)=0.0656+0.0209 \exp(-0.231 l)$.
		(This corresponds to a parallel correlation length of 
		$\xi_{\|}(\sigma_{AB}/\sigma)^{1/2} \approx 43$)\newline
		The left inset displays the position dependence of the effective interfacial tension $\Sigma$.
		The solid line corresponds to $\Sigma/\Sigma_{AB} = 1 +2.4 (0.21 l) \exp(-0.21 l)$
                }
       \label{fig:mfpot}
    \end{minipage}%
\end{figure}

\begin{figure}[htbp]
    \begin{minipage}[t]{160mm}%
       \setlength{\epsfxsize}{13cm}
       \mbox{\epsffile{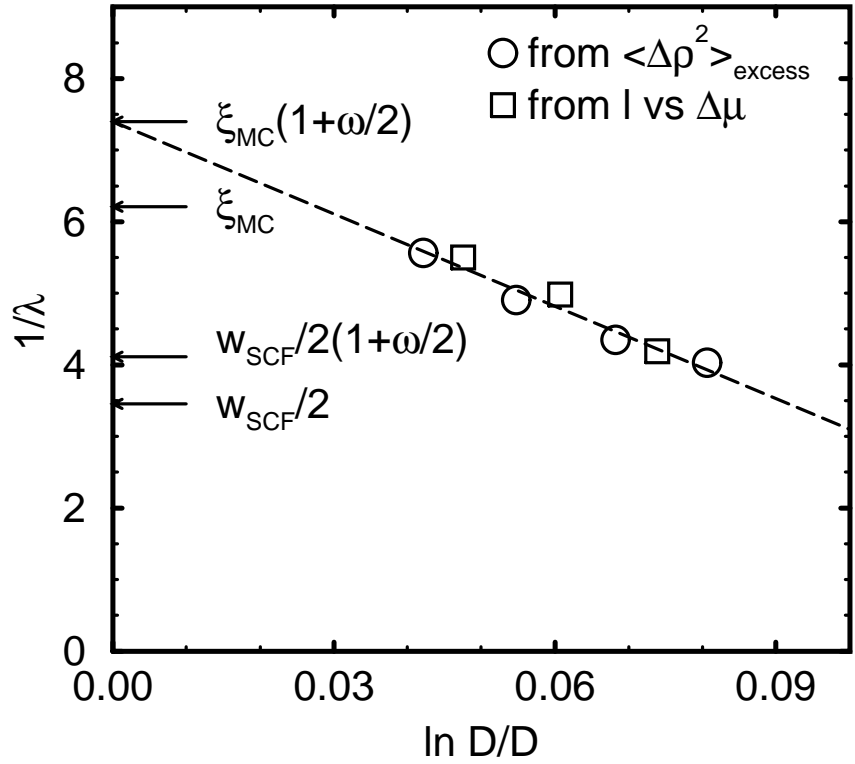}}
    \end{minipage}%
    \hfill%
    \begin{minipage}[b]{160mm}%
       \caption{Effective interaction range of the interface-wall potential extracted from the Monte Carlo data.
		The circles are estimates according to eq.\ (35), the squares correspond to the slope of thickness $l$ 
		of the wetting layer vs the logarithm of the coexistence potential $\Delta \mu$. The arrows denote
		possible limiting values for the interaction range.
                }
       \label{fig:lambda}
    \end{minipage}%
\end{figure}

\begin{figure}[htbp]
    \begin{minipage}[t]{160mm}%
       \setlength{\epsfxsize}{13cm}
       \mbox{\epsffile{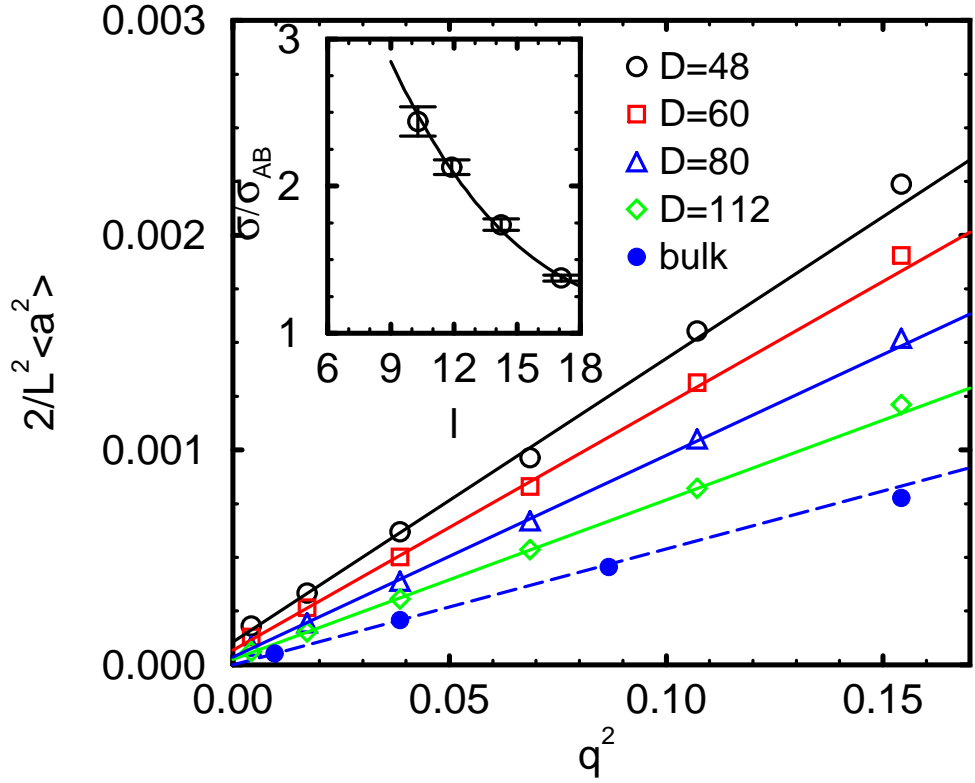}}
    \end{minipage}%
    \hfill%
    \begin{minipage}[b]{160mm}%
       \caption{Fluctuation spectrum of the AB-interface for different system sizes $D$. 
		The solid lines corresponds to fits in the range $0 < q^2 <0.12$.
		The dashed line corresponds to the
		capillary fluctuations with $\sigma_{AB}=0.054$, the value measured independently via reweighting 
		techniques. 
		The filled circles represent the capillary fluctuation spectrum in the bulk system 
		($64 \times 64 \times 128$ geometry
                and periodic boundary conditions).
		The inset shows the ratio of the effective interfacial tension and
		its bulk value on the thickness $l$ of the wetting layer. The solid line corresponds to 
		$\sigma/\sigma_{AB} = 1 +  8.72(0.272 l) \exp(-0.272 l)$. 
                }
       \label{fig:cap}
    \end{minipage}%
\end{figure}


\begin{thebibliography}{99}
\bibitem{COMMENT1}   There are corrections to the coefficient of the next to leading term $O(1/D)$,
		     due to the shift of the chemical potential and corrections to the interfacial tension.
\bibitem{COMMENT2}   The restriction of interfacial fluctuations due to the presence of the wall
		     gives rise to an additional repulsion, which, decays Gaussian\cite{SCHICK}, however.
\bibitem{COMMENT3}   The coordination number {\tt z}, as determined via the intermolecular pair correlation 
		     function\cite{M4} is slightly temperature dependent: ${\tt z}=2.65 + 0.9(\epsilon/k_BT-0.064)$
\bibitem{COMMENTPRE} Indeed, the prewetting line $\Delta \mu_{\rm pre}(T)$ in the semi-infinite system approaches the 
		     coexistence curve $\Delta \mu_{\rm coex}(T)$ for $D=48$ very closely. However, due to the
		     small wall separation $D$, the two interfaces interact with each other across the film. This
		     interaction across the film modifies the dependence of the free energy on the layer thickness $l$
		     such that only a single layer thickness is stable on the A-poor side of the phase diagram.
\end{thebibliography}
\end{document}